\documentclass[preprint,12pt, 3p, sort&compress]{elsarticle}
\usepackage{fullpage}
\usepackage[T1]{fontenc}
\usepackage{amssymb,amsmath,amsthm,amsfonts,amsbsy,mathrsfs}
\usepackage{calligra}
\usepackage[caption=false]{subfig}
\usepackage{graphicx,color}
\usepackage{multirow}
\usepackage{booktabs}
\usepackage[shortlabels]{enumitem}

\usepackage[colorlinks   = true, urlcolor     = cyan, linkcolor    = blue, citecolor   = red]{hyperref}
\usepackage{bm}
\usepackage{pict2e}
\usepackage{epsfig}
\usepackage{epstopdf}
\usepackage{comment}
\usepackage{accents}
\usepackage{ulem}
\usepackage{url}
\usepackage[ruled,vlined]{algorithm2e}
\usepackage{tikz}
\usepackage{pgfplots}
\usepackage{pgfplotstable}
\usepgfplotslibrary{external}
\usetikzlibrary{external}
\usetikzlibrary{shapes.geometric, arrows}
\pgfplotsset{
    compat=newest,
    table/header=false,
    title style={font=\small},
    tick label style={font=\scriptsize},
    label style={font=\scriptsize},
    legend style={font=\scriptsize},
    legend cell align=left
}

\usepackage{tabularx,multirow,booktabs}
\newcolumntype{L}{>{\raggedright\arraybackslash}X}
\newcolumntype{C}{>{\centering\arraybackslash}X}
\newcolumntype{R}{>{\raggedleft\arraybackslash}X}

\journal{arXiv}

\newcommand{\rz}{\mathbb{R}}
\newcommand{\gz}{\mathbb{Z}}

\newcommand{\nz}{\mathbb{N}}

\newcommand{\bfe}{{\bf e}}
\newcommand{\bff}{{\bf f}}

\newcommand{\bfp}{{\bf p}}
\newcommand{\bfq}{{\bf q}}
\newcommand{\bfr}{{\bf r}}

\newcommand{\bft}{{\bf t}}

\newcommand{\bfx}{{\bf x}}
\newcommand{\bfy}{{\bf y}}

\newcommand{\bfz}{{\bf z}}

\newcommand{\bfI}{{\bf I}}

\newcommand{\bfQ}{{\bf Q}}
\newcommand{\bfR}{{\bf R}}

\newcommand{\bfX}{{\bf X}}
\newcommand{\bfY}{{\bf Y}}
\newcommand{\bfZ}{{\bf Z}}

\newcommand{\beq}{\begin{equation}}
\newcommand{\eeq}{\end{equation}}
\newcommand{\beqs}{\begin{eqnarray}}
\newcommand{\eeqs}{\end{eqnarray}}
\newcommand{\beql}{\begin{equation} \label}

\newcommand{\half}{\frac{1}{2}}

\newcommand{\calC}{{\cal C}}
\newcommand{\calD}{{\cal D}}

\newcommand{\calF}{{\cal F}}
\newcommand{\calG}{{\cal G}}

\newcommand{\calK}{{\cal K}}

\newcommand{\calN}{{\cal N}}
\newcommand{\calO}{{\cal O}}
\newcommand{\calP}{{\cal P}}

\newcommand{\calV}{{\cal V}}

\newcommand{\calZ}{{\cal Z}}

\newcommand{\hamil}{\mathfrak{H}}

\newcommand{\abs}[1]{\lvert#1\rvert}
\newcommand{\norm}[2]{\lVert#1\rVert_{#2}}
\newcommand{\innprod}[3]{\langle#1,#2\rangle_{#3}}

\newcommand{\Lpspc}[3]{\textsf{L}^{#1}_{#2}(#3)}

\newcommand{\pd}[2]{\frac{\partial#1}{\partial#2}}
\newcommand{\hpd}[3]{\frac{\partial^{#3}#1}{\partial#2^{#3}}}

\let\oldFootnote\footnote
\newcommand\nextToken\relax

\renewcommand\footnote[1]{%
    \oldFootnote{#1}\futurelet\nextToken\isFootnote}

\newcommand\isFootnote{%
    \ifx\footnote\nextToken\textsuperscript{,}\fi}

\DeclareMathOperator{\arctantwo}{arctan2}
\definecolor{mycolor1}{rgb}{1.00000,0.00000,1.00000}

\begin{document}

%%%%%%%%%%%%%%%%%%%%%%%%%%%%%%%%%%%%%%%%%%%%%%%%%%%%%%%%%%%%%%%%%%%%%%%%%%%%%%%%%%%%%%%%%%%%%%%%%%%%%%%%%%%%%%%%%%%%%%%%%%%%%%%%%%%%
%%%%%%%%%%%%%%%%%%%%%%%%%%%%%%%%%%%%%%%%%%%%%%%%%%%%%%%%%%%%%%%%%%%%%%%%%%%%%%%%%%%%%%%%%%%%%%%%%%%%%%%%%%%%%%%%%%%%%%%%%%%%%%%%%%%%

\begin{frontmatter}
\title{Density functional theory method for twisted geometries with application to torsional deformations in group-IV nanotubes}

\author[uclamse]{Hsuan Ming Yu}
\ead{kingforever10@ucla.edu}
\author[uclamse]{Amartya S.\ Banerjee\corref{cor1}}
\ead{asbanerjee@ucla.edu}
\cortext[cor1]{Corresponding author}
\address[uclamse]{Department of Materials Science and Engineering, University of California, Los Angeles, CA 90095, U.S.A}
\begin{abstract}
We present a real-space formulation and implementation of Kohn-Sham Density Functional Theory suited to twisted geometries, and apply it to the study of torsional deformations of X (X = C, Si, Ge, Sn) nanotubes.  Our formulation is based on higher order finite difference discretization in helical coordinates, uses ab intio pseudopotentials, and naturally incorporates rotational (cyclic) and screw operation (i.e., helical) symmetries. We discuss several aspects of the computational method, including the form of the governing equations, details of the numerical implementation, as well as its convergence, accuracy and efficiency properties. 

The technique presented here is particularly well suited to the first principles simulation of quasi-one-dimensional structures and their deformations, and many systems  of interest can be investigated using small simulation cells containing just a few atoms. We apply the method to systematically study the properties of single-wall zigzag and armchair group-IV nanotubes in the range of (approximately) $1$ to $3$ nm radius, as they undergo twisting. For the range of deformations considered, the mechanical behavior of the tubes is found to be largely consistent with isotropic linear elasticity, with the torsional stiffness, $k_{\text{twist}}$, varying as the cube of the nanotube radius. Furthermore, for a given tube radius, $k_{\text{twist}}$ is seen to be highest for carbon nanotubes and the lowest for those of tin, while nanotubes of silicon and germanium are found to have intermediate values of this quantity close to each other. We also describe different aspects of the variation in electronic properties of the nanotubes as they are twisted. In particular, we find that akin to the well known behavior of armchair carbon nanotubes, armchair nanotubes of silicon, germanium and tin also exhibit bandgaps that vary periodically with imposed rate of twist, and that the periodicity of the variation scales in an inverse quadratic manner with the tube radius. These examples highlight the utility of the proposed method in the accurate and efficient computational characterization of important nanomaterials from first principles.
\end{abstract}
 
\end{frontmatter}
\section{Introduction}
\label{sec:introduction}
Over the past few decades, the synthesis and characterization of novel nanomaterials and nanostructures has blossomed into a major scientific and technological endeavor \citep{bhushan2017springer, cao2004nanostructures, nanotech_applications_1, nanotech_gov_website}. Such materials are usually associated with shapes and structures that are quite different from crystalline materials, and they often display properties that are radically distinct from the bulk phase. Consequently, a variety of computational techniques employing different physical theories have been developed over the years, to aid in their design and discovery \citep{Zhigilei2012, musa2011computational, Hutter_abinitio_MD, chelikowsky2019introductory, comp_nano}.

A defining feature of the aforementioned class of materials is that they are of limited spatial extent along one or more dimensions. This often makes it possible to sustain unusual and/or large modes of deformation in such systems, without incurring material failure. Since a variety of material properties of nanostructures, including, e.g., optical, electronic and transport behavior are often strongly coupled to distortions in the material's structure, engineering the response of these systems through the application of mechanical strains constitutes an active and important area of scientific research today \citep{pereira2009strain, ghassemi2012field, fei2014strain, roldan2015strain, schlom2014elastic, li2014elastic}. In particular, inhomogeneous strain fields --- such as those associated with overall torsion (i.e., twisting) or flexure  (i.e., bending) of the nanostructure, as well as those arising from localized deformations such as wrinkles or corrugations, have often been used to elicit fascinating electro-mechanical responses in such systems \citep{hall2006experimental, wei2011strain, falvo1997bending, yang2015tuning}. A persistent issue however, is that there appears to be a paucity of systematic and efficient computational techniques that can model these systems as they are undergoing such deformations, especially from first principles. We view the current contribution as an important step in addressing this gap in the literature and present a real-space formulation and implementation of Kohn-Sham Density Functional Theory (KS-DFT) that is suited to twisted geometries. 

Systems associated with intrinsic twist are quite common among nanomaterials, with chiral carbon nanotubes \citep{endo2013carbon}, nanocoils \citep{chen2003mechanics} and inorganic nanoassemblies  \citep{ma2017chiral} constituting well known examples. Twisting is particularly relevant as a mode of deformation for quasi-one-dimensional systems such as nanotubes, nanoribbons, nanowires and nanorods  \citep{James_OS}, and can be an important route to engineering the properties of these materials through the imposition of strain. In particular, imposition of twist naturally gives rise to so-called \textit{helical potentials} in achiral nanostructures, which can then cause  these materials to display unusual transport properties and fascinating light-matter interactions \citep{aiello2020chirality}. Twisted geometries also have found relevance recently in the context of quasi-two-dimensional systems such as graphene bilayers \citep{bistritzer2011moire, cao2018unconventional, cao2018correlated, carr2019exact}, which are associated with strong electronic correlations and superconductivity, as well as the use of screw dislocations to engineer growth processes \citep{zhao2020supertwisted, bierman2008dislocation, jin2010new}. We anticipate that the simulation technique discussed in this work will have broad relevance to most of the materials systems described above, while being particularly consequential for the computational study of quasi-one-dimensional systems and their deformations, from first principles. 

\begin{figure}[ht]
\centering
{\includegraphics[trim={5cm 1.7cm 10.0cm 1.8cm}, clip, width=0.65\textwidth]{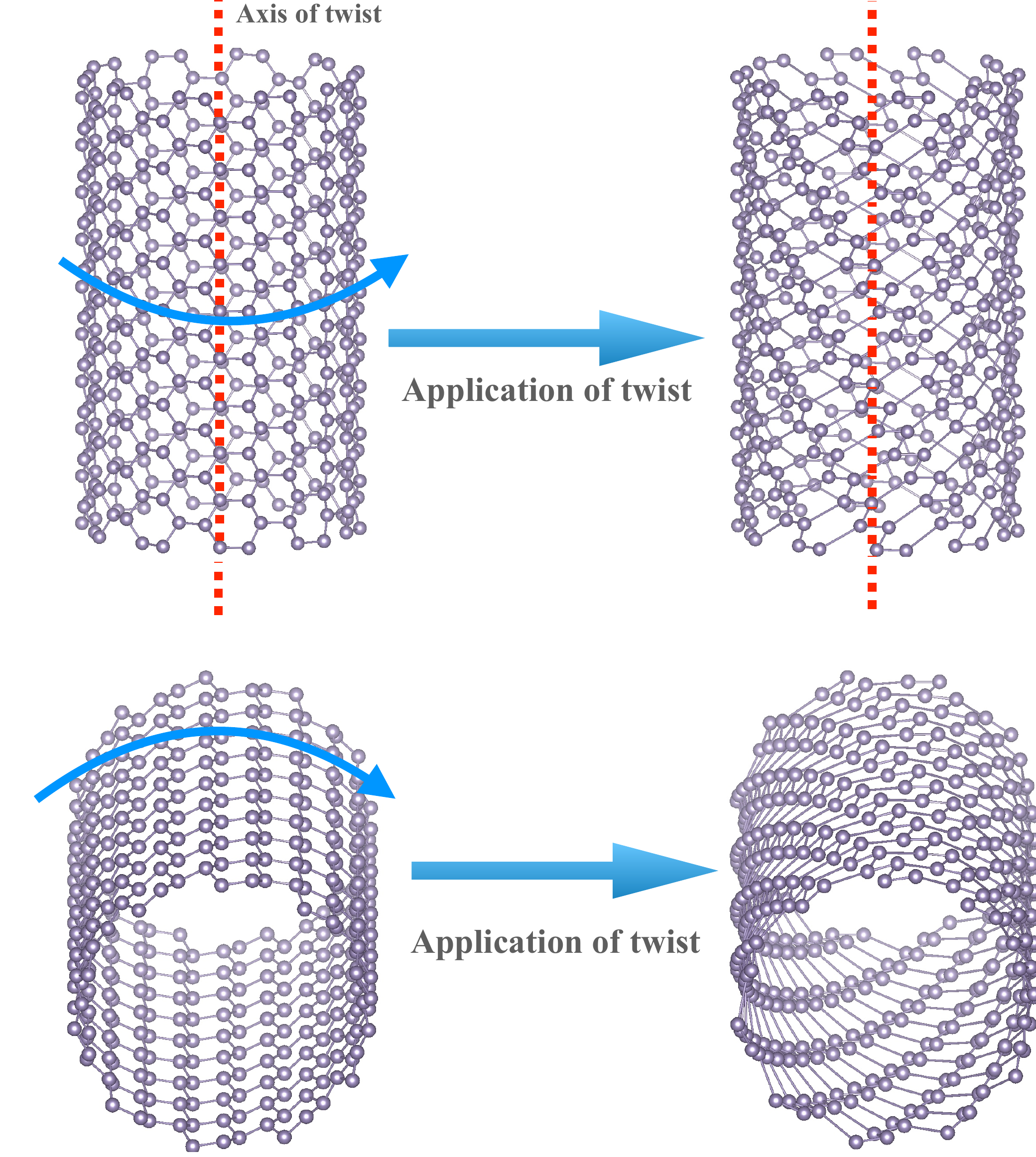}}
\caption{Depiction of a prototypical twisted geometry --- a nanotube being subjected to torsional deformation. Two views are shown.}
\label{fig:Nanotube_Twisted}
\end{figure}

The vast majority of first principles calculations being carried out today use KS-DFT, as implemented using the pseudopotential plane-wave method \citep{Kresse_abinitio_iterative, Gonze_ABINIT_1, Quantum_Espresso_1, CASTEP_1}. While this is a powerful computational technique for the study of periodic systems (such as crystals) and their homogeneous deformations, it is 
fundamentally unsuitable for modeling systems subjected to inhomogeneous strain fields (such as those associated with bending or torsion), that break periodic symmetry. Indeed, modeling such systems by use of the plane-wave method can result in the use of uncontrolled approximations and/or performance and convergence {(with respect to discretization parameters)} issues that can render the calculations infeasible. For example, plane-wave calculations of  a quasi-one-dimensional system that is undergoing twisting (Figure \ref{fig:Nanotube_Twisted}) will usually involve making the system artificially periodic along the direction of the twist axis --- thus resulting in a supercell containing a very large number of atoms, as well as the inclusion of a substantial amount of vacuum padding in the directions orthogonal to the twist axis, so as to minimize interactions between periodic images. Together, these conditions can make such calculations extremely challenging even on high performance computing platforms, if not altogether impractical.

It has been pointed out in the literature however, that the aforementioned computational issues  related to the study of twisted or bent nanostructures can be avoided by making use of the  connections of such inhomogeneous strain states with non-periodic symmetries \citep{James_OS, Dumitrica_James_OMD, Dumitrica_Bending_Graphene, CNT_Dumitrica, ma2015thermal, Pekka_Efficient_Approach, Pekka_CNT_Bending, Pekka_GNR_Bending, Pekka_Revised_Periodic, cai2008torsion}. Specifically, as long as edge effects are unimportant in a system under study, cyclic symmetries can be used to simulate bent nanostructures, while helical symmetries can be used to simulate systems with twist. A key ingredient for such an approach is the availability of efficient computational methods that can adequately handle such non-periodic symmetries. Following this line of thought, we have been developing systematic first principles simulation techniques suited to the study of systems with non-periodic symmetries \citep{My_PhD_Thesis}. In particular, we have  developed \textit{ab initio} methods that explicitly incorporate cyclic symmetries, and used this methodology to simulate bending in nanoribbons \citep{banerjee2016cyclic} and sheets of two-dimensional materials \citep{ghosh2019symmetry}. More recently, we have rigorously formulated and implemented a novel first principles computational technique that explicitly accounts for helical symmetries \citep{banerjee2021ab}. {We view the present contribution as a follow up of this  most recent development, and focus on the computational and application aspects of the simulation technique in this work, in contrast to our earlier contribution, which was largely concerned with the mathematical aspects. In particular, salient features of the current contribution are as follows. We present in this work a self-contained, intuitive derivation of the governing equations for systems associated with twisted geometries and make connections with helical symmetries, while also allowing for the possibility that such systems may have inherent cyclic symmetries. We describe the details of our computational strategy, including discretization choices in real and reciprocal space, numerical linear algebra issues and choice of eigensolvers. We touch upon specific aspects of our MATLAB based numerical implementation. We then discuss various features of the simulation method, including its convergence, accuracy, consistency, computational efficiency and parallel scaling properties. Finally, we apply the method to the study of torsional deformations of an important class of nanomaterials (i.e., nanotubes from Group IV of the periodic table\footnote{In modern IUPAC convention this group is also referred to as Group 14. Elsewhere, this group is also referred to as Group IVa or the Carbon group.}) and investigate the electro-mechanical response of these systems.} Notably, the present contribution subsumes our earlier work on KS-DFT for cylindrical geometries \citep{ghosh2019symmetry}, and many of the results in that former contribution can be derived as special cases of the results presented here for twisted geometries (by considering simulations with zero twist). Together, the present contribution, and our earlier body of work extends symmetry adapted molecular dynamics and tight-binding based computational methods developed in the literature for studying bent and/or twisted nanomaterials, to the realm of first principles calculations.

The numerical technique described here employs finite difference discretization in helical coordinates\footnote{We are aware of chemistry literature based on Linear Combination of Atomic Orbitals (LCAO) methods \citep{d2009single, dovesi2017crystal17, Mintmire_White1, CNT_1, CRYSTAL, CNT_4}, which have explored the use of helical and cyclic symmetries for studying nanostructures of interest. The connection of such symmetries with deformation modes in nanostructures does not appear to have been explored by these authors, as far as we can tell, and at any rate, these methods are quite distinct from the real space technique presented here.} which allows us to set up a computational domain in an annular region of space. In turn, this enables us to carry out simulations of systems associated with twisted geometries, while employing small unit cells containing just a few atoms. With this setup in hand, we were able to carry out an extensive series of simulations involving zigzag and armchair nanotubes of carbon, silicon, germanium and tin, with radii approximately in the range of $1$ to $3$ nanometers. This enabled us to compare and contrast the properties of these different materials, and also allowed us to extend some well-known qualitative and quantitative features of the electro-mechanical properties of carbon nanotubes, to the broader class of Group IV nanotubes. We would like to point out that these studies would not have been possible without the use of a specialized computational method such as the one presented here. We anticipate that the rich repository of simulation data produced by our method can be utilized for the development of efficient, accurate, interpretable machine learning models  \citep{yadav2021interpretable}, in the near future.\footnote{After the submission of this manuscript, we were made aware of recent work \citep{sharma2021real} wherein the techniques presented by us here as well as our  earlier contribution \citep{banerjee2021ab} have been implemented into an efficient C/C++ framework.}

The rest of this work is organized as follows. We derive the governing equations of our method in Section \ref{sec:formulation}. We discuss implementation aspects in Section \ref{sec:implementation}. Results from the computational method are presented in Section \ref{sec:simulation_results}. Finally, Section \ref{sec:conclusions} summarizes the work and mentions ongoing and future research directions.
\section{Formulation} \label{Section:Formulation}
\label{sec:formulation}
In this section, we describe our formulation of Kohn-Sham density functional theory for
twisted geometries. We first lay out the notation used in the rest of the paper. In what follows, $\bfe_{\bfX}, \bfe_{\bfY}, \bfe_{\bfZ}$ will denote the standard orthonormal basis of $\rz^3$. The Cartesian coordinates of a point $\bfp \in \rz^3$ will be denoted as $(x_{\bfp},y_{\bfp},z_{\bfp})$, i.e.,  $\bfx = x_{\bfp}\,\bfe_{\bfX} + y_{\bfp}\,\bfe_{\bfY} + z_{\bfp}\,\bfe_{\bfZ}$. The corresponding helical coordinates (introduced later in Section \ref{subsec:helical_coordinates}) and cylindrical coordinates of the point will be denoted as $(r_{\bfp},\theta_{1\,{\bfp}},\theta_{2\,{\bfp}})$ and $(r_{\bfp},\vartheta_{\bfp},z_{\bfp})$  respectively. The coordinates of a generic point will be denoted as $(x,y,z)$, $(r,\theta_1,\theta_2)$ and $(r,\vartheta, z)$  in Cartesian, helical and cylindrical coordinates respectively. Vectors and matrices will be denoted in boldface, with vectors typically denoted using lower case letters (e.g., $\bfp)$ and matrices using uppercase (e.g. ${\bfQ}$). The symbol $\cdot$ will be often used as a generic placeholder instead of specifying a variable explicitly (e.g. $f(\cdot)$ instead of $f(x)$ or $f(y)$). The notation $\Lpspc{2}{}{\Omega}$ will be used to denote the space of square integrable functions over a domain $\Omega$. The inner product over such a space will be denoted as $\innprod{\cdot}{\cdot}{\Lpspc{2}{}{\Omega}}$. An overbar will be used to denote complex conjugation (e.g. $\overline{f(\bfx)}$). Finally, $\lvert \cdot \rvert$ will be used to denote the absolute value of a scalar, and $\norm{\cdot}{}$ will be used to denote the norm of a vector or function.
\subsection{System specification: Computational domain, atomic configuration and symmetries}
\label{subsec:system_specification}
We consider a nanostructure aligned along $ \bfe_{\bfZ}$, the axis of twist, as the prototypical system of interest.  In order to avoid quantum finite-size effects and/or mechanical constraints at the edges due to the imposition of twist \citep{koskinen2016quantum, Pekka_Efficient_Approach} , we will assume that the structure is infinite in extent along $\bfe_{\bfZ}$. For the sake of simplicity, we will also assume that the structure is of limited spatial extent along  $\bfe_{\bfX}$ and  $\bfe_{\bfY}$, i.e., it is a quasi-one-dimensional system. The large majority of nanomaterials for which twisted geometries might be relevant as deformation modes, are included within the scope of the above set of assumptions. These conditions imply that the system can be embedded in a cylinder with axis $\bfe_{\bfZ}$ (or annular cylinder, if the system is tubular), of infinite height and finite radius, and we will refer to this region of space as the \textit{global simulation domain}.

For most quasi-one-dimensional systems of interest, the infinite extent along $\bfe_{\bfZ}$ is related  to periodicity along this axis. Additionally, for many such systems, including for example, the tubular structures considered in this work, there may be rotational symmetries about the same axis. Let the atoms of the untwisted structure have positions: 
\begin{align}
\mathcal{S}_{\text{untwisted}} = \{\bfp_1,  \bfp_2, \bfp_3,\ldots \}\,.
\end{align}
The above assumptions on periodicity and rotational symmetry imply that there is a periodic group consisting of translations along $\bfe_{\bfZ}$: 
\begin{align}
\calG_{\text{periodic}} = \bigg\{\big(\,{\bfI} \,| \,m\tau\,\bfe_{\bfZ}\,\big): m \in \gz\bigg\}\,,
\end{align}
a cyclic group of order $\mathfrak{N}$ about  $\bfe_{\bfZ}$ (consisting of rotations through multiples of the angle $\displaystyle\Theta = \frac{2\pi}{\mathfrak{N}}$):
\begin{align}
\calG_{\text{cyclic}} = \bigg\{\big({\,{\bfR}}_{n\Theta} \,|\,\mathbf{0}\,\big): n=0,1,\ldots,\mathfrak{N}-1 \bigg\}\,,
\end{align}
and a finite collection of points:
\begin{align}
\calP = \big\{\bfr_k \in \rz^3: k = 1,2,\ldots,M\big\}\,,
\label{eq:points_in_P}
\end{align}
such that the entire structure $\mathcal{S}_{\text{untwisted}}$ can be described as the action of the composite group:
\begin{align}
\calG_{\text{untwisted}} = \bigg\{\big(\,{\bfR}_{n\Theta} \,|\,m\tau\,\bfe_{\bfZ}\,\big) : m \in \gz, n=0,1,\ldots,\mathfrak{N}-1\bigg\}\,,
\end{align}
on the points in $\calP$, i.e.,
\begin{align}
\mathcal{S}_{\text{untwisted}} =\bigcup_{\substack{{\Upsilon} \in \calG_{\text{untwisted}},\\  k = 1,2,\ldots,M }}\!{\Upsilon} \circ \bfr_k\, = \bigcup_{\substack{m \in \gz,\\  n=0,1,\ldots,\mathfrak{N}-1 \\ k = 1,2, \ldots, M}}\!{\bfR}_{n\Theta}\,\bfr_k + m\tau\,\bfe_{\bfZ}\,.
\label{eq:S_untwisted}
\end{align}
In the above equations, a symbol of the form $\big(\,{\bfQ}\,|\,\bft \,\big)$ denotes an isometry with rotation ${\bfQ} \in \text{SO}(3)$ and translation $\bft \in \rz^3$. Its action on a point $\bfx \in \rz^3$ can be written as:
\begin{align}
\big(\,{\bfQ}\,|\,\bft \,\big) \circ \bfx ={\bfQ}\,\bfx + \bft \,.
\end{align}
Additionally, ${\bfR}_{n\Theta}$ denotes the following rotation matrix with axis $\bfe_{\bfZ}$: 
\begin{align}
{\bfR}_{n\Theta} = 
\begin{pmatrix}
\cos (n\Theta)  & -\sin (n\Theta)   & 0 \\
\sin (n\Theta)   & \quad\cos (n\Theta)   & 0 \\
0 & 0 & 1
\end{pmatrix}\,,\Theta = \frac{2\pi}{\mathfrak{N}}\,,
\end{align}
${\bfI}$ denotes the identity matrix and $\mathbf{0}$ denotes the zero vector. The scalar $0 < \tau <\infty$ is the fundamental period of the group $\calG_{\text{periodic}}$. We will refer to the points in $\calP$ as the \textit{simulated atoms}. We will use $Z_k $ to denote the valence nuclear charge of the simulated atom located at position $\bfr_k$.

Now let us  suppose that the structure $\mathcal{S}_{\text{untwisted}}$ is subjected to a uniform twist of $2\pi \alpha$ radians per $\tau$ bohr along the axis $\bfe_{\bfz}$, so as to result in the structure $\mathcal{S}_{\text{twisted}}$ with the atomic positions:
\begin{align}
\mathcal{S}_{\text{twisted}} = \{\bfq_1,  \bfq_2, \bfq_3,\ldots \}\,.
\end{align}
Since we are dealing with structures that extend to infinity along $\bfe_{\bfZ}$, we may obtain the deformed (twisted) configuration by prescribing a mapping of the form $\bfq = \displaystyle{{\bfR}_{\frac{2\pi \alpha z_{\bfp}}{\tau}}} \bfp$, to the undeformed one. Here, $\alpha \in [0,1)$ is a scalar twist parameter, $\tau$ can be re-identified as the \textit{pitch of the twist}, and $\beta = {\frac{2\pi \alpha}{\tau}}$, is the \textit{rate of twist}. Furthermore,
\begin{align}
{{\bfR}_{\frac{2\pi \alpha z_{\bfp}}{\tau}}} = \begin{pmatrix}
\cos (\frac{2\pi \alpha z_{\bfp}}{\tau})  & -\sin (\frac{2\pi \alpha z_{\bfp}}{\tau})  & 0 \\
\sin (\frac{2\pi \alpha z_{\bfp}}{\tau})  & \quad\cos (\frac{2\pi \alpha z_{\bfp}}{\tau})  & 0 \\
0 & 0 & 1
\end{pmatrix}
= \begin{pmatrix}
\cos (\beta  z_{\bfp})  & -\sin (\beta  z_{\bfp})  & 0 \\
\sin (\beta  z_{\bfp})  & \quad\cos (\beta  z_{\bfp})  & 0 \\
0 & 0 & 1
\end{pmatrix}\,,
\end{align} 
denotes a rotation matrix with axis $\bfe_{\bfZ}$ for which the (twist) angle depends on the coordinate along $\bfe_{\bfZ}$. At the atomic level, this implies \citep{James_OS, Dumitrica_James_OMD, banerjee2021ab} that the deformed structure may be obtained from the undeformed one by replacing the group of translations $\calG_{\text{periodic}}$ used to generate $\mathcal{S}_{\text{untwisted}}$, by a group of screw transformations (or helical isometries), i.e.:
\begin{align}
\calG_{\text{helical}} = \bigg\{\big(\,{\bfR}_{{2\pi m \alpha}} \,|\,m\tau\,\bfe_{\bfZ}\,\big) : m \in \gz\bigg\}\,.
\end{align}
Here $\bfR_{2\pi m\alpha}$ denotes the following rotation matrix with axis $\bfe_{\bfZ}$: 
\begin{align}
{\bfR_{2\pi m\alpha}} = 
\begin{pmatrix}
\cos (2\pi m\alpha)  & -\sin (2\pi m\alpha)  & 0 \\
\sin (2\pi m\alpha)  & \quad\cos (2\pi m\alpha)  & 0 \\
0 & 0 & 1
\end{pmatrix}\,.
\end{align}
In other words, by replacing the composite group $\calG_{\text{untwisted}}$ with:
\begin{align}
\calG_{\text{twisted}} = \bigg\{\big(\,{{\bfR}}_{(2\pi m\alpha + n\Theta)} \,|\,m\tau\,\bfe_{\bfZ}\,\big) : m \in \gz, n=0,1,\ldots,\mathfrak{N}-1\bigg\}\,,
\end{align}
we may generate the structure with the prescribed amount of twist as:
\begin{align}
\mathcal{S}_{\text{twisted}} =\bigcup_{\substack{{\Upsilon} \in \calG_{\text{twisted}},\\  k = 1,2,\ldots,M }}\!{\Upsilon} \circ \bfr_k =  \bigcup_{\substack{m \in \gz,\\  n=0,1,\ldots,\mathfrak{N}-1 \\ k = 1,2, \ldots, M}}\!{\bfR}_{(2\pi m\alpha+ n\Theta)}\,\bfr_k + m\tau\,\bfe_{\bfZ}\,.
\label{eq:S_twisted}
\end{align}
In the above equations, $\bfR_{(2\pi m\alpha + n\Theta)}$ denotes the following rotation matrix with axis $\bfe_{\bfZ}$: 
\begin{align}
{\bfR_{(2\pi m\alpha + n\Theta)}} = 
\begin{pmatrix}
\cos (2\pi m\alpha  + n\Theta)  & -\sin (2\pi m\alpha + n\Theta )  & 0 \\
\sin (2\pi m\alpha + n\Theta)  & \quad\cos (2\pi m\alpha + n\Theta)  & 0 \\
0 & 0 & 1
\end{pmatrix}\,.
\end{align}
Note that in this formulation, the structure continues to maintain its cyclic symmetries even after twisting. Also note that the formula in eq.~\ref{eq:S_twisted} (and similarly, eq.~\ref{eq:S_untwisted}) is meant to be species preserving in the sense that an atom in the simulated set $\calP$ has the same atomic number as its images under the isometries in $\calG_{\text{untwisted}}$ (or $\calG_{\text{untwisted}}$).\footnote{More specifically, if the atom at $\bfq_k \in \mathcal{S}_{\text{twisted}}$ has atomic number $Z$, then the simulated atom at $\bfr_{k'}$ which satisfies $\bfq_k =  \Upsilon \circ \bfr_{k'}$ for some $\Upsilon \in \calG_{\text{twisted}}$ also has atomic number $Z$. Similarly also for $\mathcal{S}_{\text{untwisted}}$ and $\calG_{\text{untwisted}}$.} Also note that by virtue of the above definitions, the group $\calG_{\text{twisted}}$ serves as a physical symmetry group for the structure $\mathcal{S}_{\text{twisted}}$ in the sense that the action of any $\Upsilon \in  \calG_{\text{twisted}}$ on all the points in $\mathcal{S}_{\text{twisted}}$ leaves it invariant (and similarly for $\calG_{\text{untwisted}}$  and $\mathcal{S}_{\text{untwisted}}$).

The group $\calG_{\text{twisted}}$ will play a central role in the rest of this work.  Note that this group  subsumes the group $\calG_{\text{untwisted}}$ in the sense that the latter can be recovered by simply setting $\alpha = 0$ in the former. In what follows, we will simplify notation a bit and simply use $\calG$ to denote this group. Further, we will use the notation:
\begin{align}
\Upsilon_{m,n}  = \big(\,{\bfR}_{(2\pi m\alpha + n\Theta)} \,|\,m\tau\,\bfe_{\bfZ}\,\big)\,,
\end{align}
to denote group elements from $\calG$. The action of $\Upsilon_{m,n}$ on a generic point in space is to rotate it about axis $\bfe_{\bfZ}$ by angle $2\pi m\alpha + n\Theta$ while also translating it by $m\tau$ along the same axis.

\begin{figure}[ht]
\centering
\subfloat[Front view]
{\includegraphics[trim={1cm 0.45cm 2cm 0.5cm}, clip, width=0.49\textwidth]{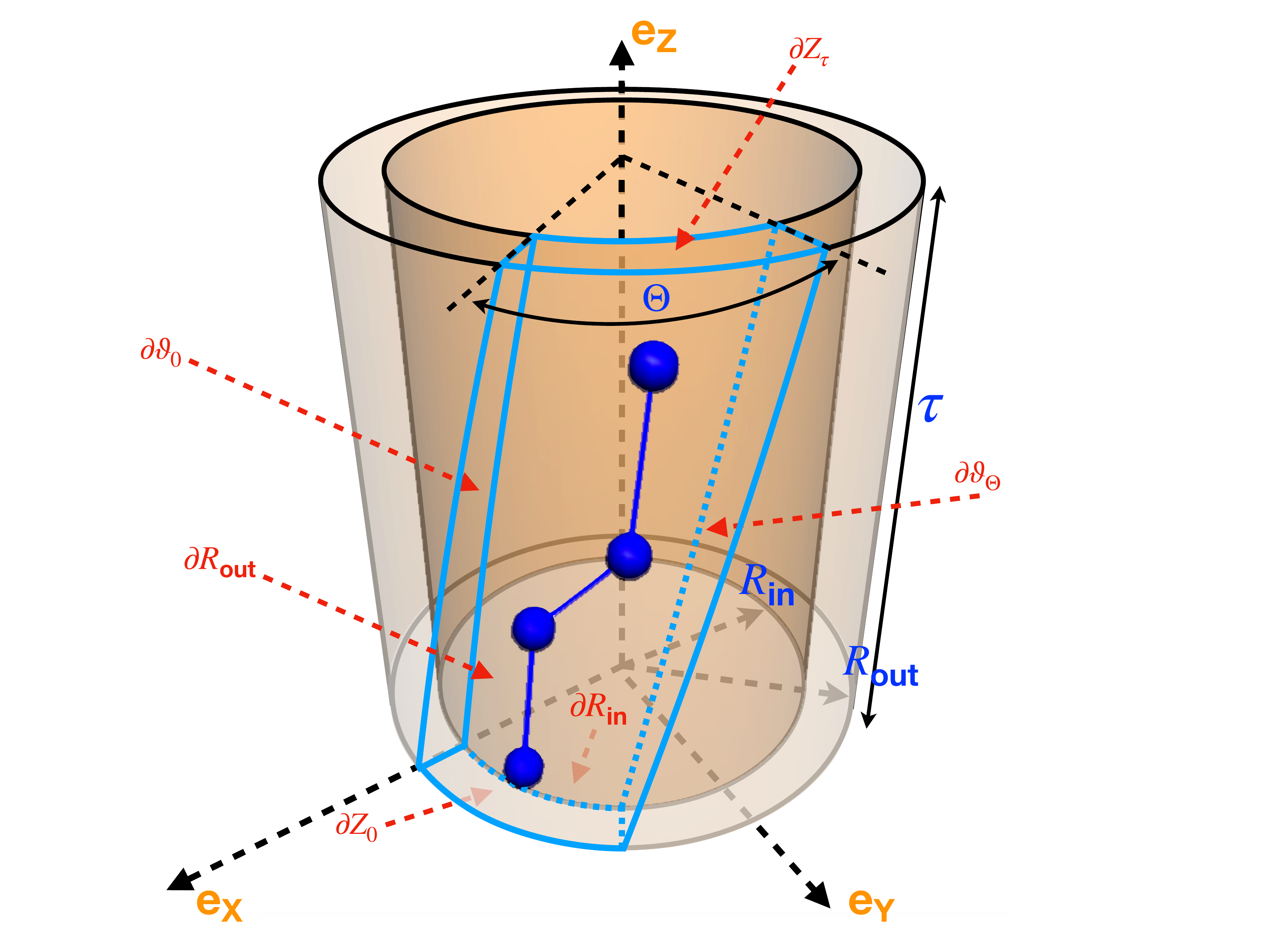}}$\;$
\subfloat[Top view]
{\includegraphics[trim={1cm 0.46cm 2cm 0.5cm}, clip, width=0.49\textwidth]{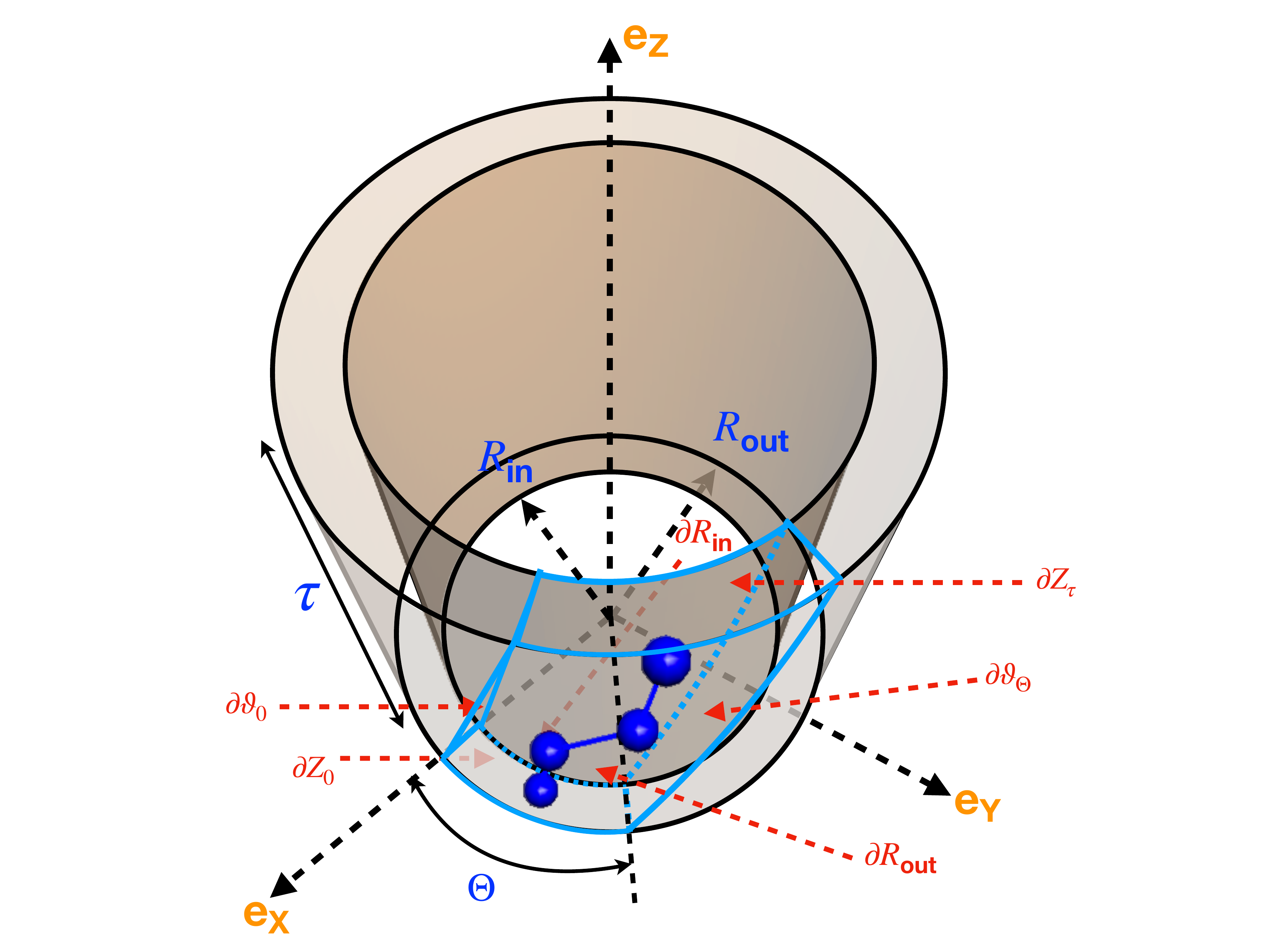}}
\caption{Illustration of the symmetry adapted unit cell or fundamental domain $\calD$ (domain boundary lines in blue) of the twisted structure. The region $\calD$ also serves as the computational domain for the calculations presented in this work. A few contained atoms as well as various bounding surfaces of the domain are also shown. For a tubular structure, the parameter $\Theta = 2\pi/\mathfrak{N}$ relates to the cyclic symmetry of the structure. The parameter $\tau$ is related to the pitch of the applied twist.}
\label{fig:FD_Domain}
\end{figure}

In subsequent sections, we will describe how the Kohn-Sham problem for the entire twisted structure as posed on the global simulation domain, can be appropriately reformulated as a problem over a \textit{fundamental domain} (or \textit{symmetry adapted unit cell}), such that only the simulated atoms and the symmetry group $\calG$ are involved in the resulting equations. This symmetry adapted computational domain has to be a regular region of space with sufficiently smooth boundaries that encompasses the simulated atoms and can be used to tile the global simulation domain by the action of the group $\calG$. Furthermore, this region should be minimal in the sense that the above tiling operation should not produce intersecting volumes. In the context of the twisted tubular structures considered in this work, if the simulated atoms have radial coordinates lying between $R_{\text{in}}$ and $R_{\text{out}}$, the following region serves as an appropriate fundamental domain (expressed using cylindrical coordinates):
\begin{align}
\calD =  \big\{(r,\vartheta,z) \in \rz^3:R_{\text{in}} \leq r\leq R_{\text{out}}, \frac{2\pi\alpha z}{\tau} \leq \vartheta \leq \frac{2\pi\alpha z}{\tau} + \Theta, 0 \leq z \leq \tau\big\}\,.
\label{eq:fundamental_domain}
\end{align}
The boundaries of the fundamental domain defined above can be expressed as:
\begin{align}
\partial\calD= \partial R_{\text{in}} \bigcup \partial R_{\text{out}} \bigcup \partial\vartheta_{0} \bigcup \partial\vartheta_{\Theta}\bigcup\partial\calZ_0\bigcup\partial\calZ_{\tau}\,.
\label{eq:calD_boundary}
\end{align}
Here $\partial R_{\text{in}}$ and  $\partial R_{\text{out}}$ denote boundaries related to the radial direction (i.e., the surfaces $r = R_{\text{in}}$ and $r = R_{\text{out}}$ respectively), $\partial\vartheta_{0}$ and $\partial\vartheta_{\Theta}$ denote ($z$-dependent) bounding surfaces related to the angular direction (i.e., $\vartheta = \frac{2\pi\alpha z}{\tau}$ and $\vartheta = \frac{2\pi\alpha z}{\tau} + \Theta$ respectively), and finally, $\partial\calZ_0$ and $\partial\calZ_{\tau}$ denote boundaries related to the $\bfe_{\bfZ}$ direction (i.e., the surfaces $z = 0$ and $z = \tau$ respectively). Note that for no applied twist, the region $\calD$ is simply an annular cylindrical sector, i.e., 
\begin{align}
\calD^{\,\alpha = 0} =  \big\{(r,\vartheta,z) \in \rz^3:R_{\text{in}} \leq r\leq R_{\text{out}},0 \leq \vartheta \leq \Theta, 0 \leq z \leq \tau\big\}\,,
\end{align}
and the boundaries $\partial\vartheta_{0}$ and $\partial\vartheta_{\Theta}$ are then vertical surfaces perpendicular to the $\bfe_{\bfY}-\bfe_{\bfZ}$ plane. Figure \ref{fig:FD_Domain} shows two views of the fundamental domain used for the simulations used in this work and also highlights the boundaries described above.
%\begin{figure}[ht]
%\begin{subfigure}{0.49\textwidth}
%\includegraphics[trim={1cm 0.5cm 2cm 0.5cm}, clip, width=\linewidth]{./FD_Front.pdf}
%\caption{Front view}
%\end{subfigure}
%\hspace*{\fill}
%\begin{subfigure}{0.49\textwidth}
%\includegraphics[trim={1cm 0.5cm 2cm 0.5cm}, clip, width=\linewidth]{./FD_Front.pdf}
%\caption{Top view}
%\end{subfigure}
%\caption{Illustration of the symmetry adapted unit cell or fundamental domain $\calD$ (domain boundary lines in blue) of the twisted structure. The region $\calD$ also serves as the computational domain for the calculations presented in this work. A few contained atoms as well as various bounding surfaces of the domain are also shown. For a tubular structure, the parameter $\Theta = 2\pi/\mathfrak{N}$ relates to the cyclic symmetry of the structure. The parameter $\tau$ is the related to the pitch of the applied twist.}
%\label{fig:FD_Domain}
%\end{figure} 
In what follows, we will formulate suitable versions of the equations of Kohn-Sham theory as posed on the simulation cell $\calD$ and also elaborate on the conditions that have to be applied on the bounding surfaces that make up $\partial D$. Our derivation of the governing equations presented here is largely heuristic, and a more nuanced, mathematically rigorous discussion is available in \citep{banerjee2021ab}.
\subsection{Governing equations}
\label{subsec:gov_eqn}
\subsubsection{Helical Bloch theorem and block-diagonalization of Hamiltonian}
As described above (eq.~\ref{eq:S_twisted}), the atomic positions of the twisted structure can be described as the orbit of a discrete group of isometries (i.e., the group $\calG$). Due to the presence of such symmetries in the system, it follows under fairly general hypotheses \citep{My_PhD_Thesis, banerjee2016cyclic, ghosh2019symmetry, banerjee2021ab} that the ground state electron density for such a system is invariant under the same symmetry group.  Furthermore, the Kohn-Sham Hamiltonian for the system commutes with the symmetry operations of the group  \citep{Hammermesh, McWeeny}. Consequently, the eigenstates of the Hamiltonian can be labeled using irreducible representations of the group $\calG$, and they transform under action of the group in the same manner as the irreducible representations themselves do \citep{My_PhD_Thesis, McWeeny,Hammermesh, banerjee2021ab}. Since the group $\calG$ is Abelian, results from group representation theory\citep{ Folland_Harmonic, Barut_Reps} imply that the complex irreducible representations are one dimensional. These are the so called \textit{complex characters} of $\mathcal{G}$, which, keeping in mind that $\calG$ is the direct product of the groups $\calG_{\text{helical}}$ and $\calG_{\text{cyclic}}$, can be expressed as (for $m \in \gz, n \in\{0,1,2,\ldots,\mathfrak{N} - 1\}$):
\begin{align}
\label{eq:dual_group}
\widehat{\mathcal{G}} =\bigg\{e^{2\pi i \big({m \eta + \frac{n\nu}{ \mathfrak{N}} }\big)}:  \eta \in \big[-\half,\half \big); \nu \in   \big\{0,1,2,\ldots,\mathfrak{N} - 1\big\} \bigg\}\,.
\end{align}
In other words, for each value of $\eta \in $ and $\nu$ as defined above, the character $\hat{\zeta} \in \widehat{\mathcal{G}}$ is a complex valued map on the group\footnote{ $\widehat{\mathcal{G}}$ is often referred to as the \textit{dual group} of $\calG$ in the mathematics literature \citep{Folland_Harmonic, Barut_Reps}.}, that assigns the value $e^{2\pi i \big({m \eta + \frac{n\nu}{ \mathfrak{N}} }\big)}$ to the group element $\Upsilon_{m,n} \in \mathcal{G} $. Since  any character $\hat{\zeta} \in \widehat{\mathcal{G}}$ can be labeled using the pair $(\eta,\nu)$, these can be also used to label the eigenstates of the Kohn-Sham Hamiltonian, and other quantities related to its spectrum. Accordingly, we will use $\lambda_j(\eta,\nu)$, $\psi_j(\bfx; \eta,\nu)$ and $g_j(\eta,\nu)$ to explicitly indicate the labels for the eigenvalues, the eigenvectors, and the electronic occupations, respectively. Collections of the eigenvalues, eigenvectors and  electronic occupations will be denoted using $\Lambda$, $\Psi$ and $\mathfrak{G}$ respectively, i.e.:
\begin{align}
\nonumber
\Lambda &=\bigg\{\lambda_j(\eta,\nu): \eta \in \big[-\half,\half \big); \nu \in   \big\{0,1,2,\ldots,\mathfrak{N} - 1 \big\}; j=1,2,\ldots,\infty \bigg\}\,,\\\nonumber
\Psi &=\bigg\{\psi_j(\cdot;\eta,\nu): \eta \in \big[-\half,\half \big); \nu \in   \big\{0,1,2,\ldots,\mathfrak{N} - 1 \big\}; j=1,2,\ldots,\infty \bigg\}\,,\\
\mathfrak{G} &=\bigg\{g_j(\eta,\nu): \eta \in \big[-\half,\half \big); \nu \in   \big\{0,1,2,\ldots,\mathfrak{N} - 1 \big\}; j=1,2,\ldots,\infty \bigg\}\,.
\end{align}
Mathematical properties of the characters and the above discussion lead to a number of important  considerations that are worth mentioning at this point. First, as a consequence of the orthogonality relations obeyed by the characters \citep{Folland_Harmonic, Barut_Reps} the eigenstates associated with distinct characters are orthogonal to each other. This can be used to cast the Hamiltonian (which commutes with the symmetry operations in $\calG$)  in a symmetry adapted basis \citep{McWeeny}, such that it appears \textit{block-diagonal} \citep{My_PhD_Thesis, banerjee2021ab}. Since the blocks associated with distinct characters can be dealt with   independently of each other and are of reduced dimension compared to the full Hamiltonian (within any finite dimensional approximation, e.g.), this implies that the  problem of diagonalizing the Hamiltonian is greatly simplified. Second, the fact that the eigenstates of the Hamiltonian transform under symmetry operations in the same manner as the characters, implies that they obey a \textit{Helical Bloch theorem} \citep{My_PhD_Thesis, banerjee2021ab, koskinen2016quantum, Dumitrica_Tight_Binding1}, i.e., for any $\Upsilon_{m,n} \in \calG$:
\begin{align}
\psi_j(\Upsilon_{m,n}^{-1}\circ \bfx;\eta,\nu) = e^{2\pi i \big({m \eta + \frac{n\nu}{ \mathfrak{N}} }\big)} \psi_j(\bfx;\eta,\nu)\,,
\label{eq:Bloch_theorem_1}
\end{align}
or equivalently:
\begin{align}
\label{eq:Bloch_theorem_2}
\psi_j(\Upsilon_{m,n} \circ \bfx;\eta,\nu) = e^{-2\pi i \big({m \eta + \frac{n\nu}{ \mathfrak{N}} }\big)} \psi_j(\bfx;\eta,\nu)\,.
\end{align}
These relations can be used to deduce the conditions that need to be applied to the boundary surfaces of the fundamental domain while formulating the Kohn-Sham problem.  Finally, in order to write down quantities that depend on all eigenstates cumulatively, we need to account for contributions from each $\zeta \in \widehat{\mathcal{G}}$. This amounts to integrating the eigenstate dependent quantities against a suitable integration measure over $\widehat{\mathcal{G}}$, i.e., by forming sums of the form $\displaystyle \frac{1}{\mathfrak{N}} \sum_{\nu=0}^{\mathfrak{N}-1}$ , along with integrals in $\eta$. As an example, if we intend to compute the sum of the occupation numbers over all the electronic states in the system, we need to evaluate:
\begin{align}
{s} = \int_{\mathfrak{I}}\frac{1}{\mathfrak{N}}\sum_{\nu = 0}^{\mathfrak{N}-1}\sum_{j=1}^{\infty}\,g_j(\eta,\nu)\,.
\end{align}
Here and henceforth, $\mathfrak{I}$ is used to denote the set $\big[-\half,\half \big)$.
\subsubsection{Electronic free energy functional and Kohn-Sham equations for twisted structure}
\label{subsubsec:KS_eqn}
In what follows, we will consider the (twisted) system of interest to be one in which the effects of spin can be ignored, and for which the electronic temperature is set at $T_{\text{e}}$. This implies that the electronic occupations can be expressed in terms of the Kohn-Sham eigenvalues as:
\begin{align}
g_j(\eta,\nu) = f_{T_{\text{e}}}\big(\lambda_j(\eta, \nu)\big)\,,
\end{align}
with $f_{T_{\text{e}}}\big(\cdot)$ denoting the Fermi-Dirac function, i.e.,
\begin{align}
\label{eq:Fermi_Dirac}
f_{T_e}(y) = \frac{1}{1 + \exp\big({\frac{y\,-\,\lambda_{\text{F}}}{k_{\text{B}} T_{\text{e}}}}\big)}\,.
\end{align}
Here $\lambda_{\text{F}}$ and $k_{\text{B}}$ denote the system's Fermi level and the Boltzmann constant respectively.  In order to motivate the correct form of the various terms of the governing equations for the twisted structure, we will often refer to the simpler, more well known expressions of these quantities for finite (or isolated) systems. We will denote these finite system relevant quantities (scalar fields, energies, etc.) with a $^\circ$ superscript. 

For a finite system \citep{ghosh2019symmetry, ghosh2017sparc_1}, the electron density can be expressed in terms of the Kohn-Sham eigenvectors and the electronic occupations as: 
\begin{align}
\displaystyle \rho^{\circ}(\bfx) = 2\sum_{j=1}^{\infty}\,g_j^{\circ}\,\lvert \psi_j^{\circ}(\bfx)\rvert^2\,.
\end{align} 
Following the discussion above, this expression has to be modified for our case as:
\begin{align}
\rho(\bfx) = 2\int_{\mathfrak{I}}\frac{1}{\mathfrak{N}}\sum_{\nu = 0}^{\mathfrak{N}-1}\sum_{j=1}^{\infty}\,g_j(\eta,\nu)\,\lvert \psi_j(\bfx; \eta, \nu)\rvert^2\,d\eta\,.
\label{eq:electron_density_expression}
\end{align}
Note that the factor of $2$ in the expressions above is due to ignoring electronic spin. Further note that due to the Helical Bloch conditions obeyed by the Kohn-Sham eigenvectors (eq.~\ref{eq:Bloch_theorem_2}), the expression above is invariant under the symmetry operations in $\calG$, as is required of the ground state electron density.

$\bullet$ \underline{Electronic free energy:} To derive the governing equations of Kohn-Sham theory for our system, we take recourse to an energy minimization approach \citep{banerjee2021ab, ghosh2019symmetry, ghosh2017sparc_2}. The relevant quantity in this case, since the system is of an extended nature, is the \textit{ground state electronic free energy per unit fundamental domain}. We denote this quantity here as  $\calF(\mathfrak{G}, \Psi, \calP, \calD, \calG)$ to emphasize its dependence on the electronic occupation numbers, the eigenstates, the positions of the simulated atoms, the fundamental domain and the symmetry group $\calG$. Within the pseudopotential \citep{troullier1991efficient, chelikowsky2019introductory} and Local Density Approximations \citep{KohnSham_DFT}, we may express it as:
\begin{align}
\nonumber
\calF(\mathfrak{G}, \Psi, \calP, \calD, \calG) = T_{\text{kin}}&(\mathfrak{G}, \Psi, \calP, \calD, \calG) \,+\,E_{\text{xc}}(\rho, \calD)\,+\,K(\mathfrak{G}, \Psi, \calP, \calD, \calG)\\
&+ E_{\text{el}}(\rho, \calP, \calD, \calG)\,-\,T_{\text{e}}\,S(\mathfrak{G})\,.
\label{eq:electronic_free_energy}
\end{align}
The terms on the right-hand side of the above expression represent (per unit fundamental domain) the kinetic energy of the electrons, the exchange correlation energy, the nonlocal pseudopotential energy, the electrostatic energy and the electronic entropy contribution, respectively. We now elaborate on each of these quantities.

$\bullet$ \underline{Kinetic energy:} The first term on the right hand side of the expression above is the electronic kinetic energy per unit fundamental domain. For an isolated system (placed in $\rz^3$), this term can be expressed \citep{ghosh2017sparc_1, ghosh2019symmetry} in terms of the Kohn-Sham eigenstates and the occupations as:
\begin{align}
T_{\text{kin}}^{\circ} = \sum_{j=1}^{\infty} {2}\,g_j^{\circ}\,\innprod{{-\half}\Delta \psi_j^{\circ}}{\psi_j^{\circ}}{\Lpspc{2}{}{{\rz^3}}} = \sum_{j=1}^{\infty} {2}\,g_j^{\circ} \int_{\rz^3}{-\half}\Delta\psi_j^{\circ}(\bfx)\,\overline{\psi_j^{\circ}(\bfx)}\,d\bfx\,.
\end{align}
For the system at hand, this is modified to read:
\begin{align}
T_{\text{kin}}&(\mathfrak{G}, \Psi, \calP, \calD, \calG) = \int_{\mathfrak{I}}\frac{1}{\mathfrak{N}}\sum_{\nu = 0}^{\mathfrak{N}-1}\bigg(\sum_{j=1}^{\infty}{2}\,g_j(\eta,\nu)\big\langle{{-\half}\Delta \psi_j(\cdot; \eta,\nu)},{\psi_j(\cdot; \eta,\nu)}\big\rangle_{\Lpspc{2}{}{{D}}}\,\bigg)\,d\eta\,.
\end{align}
$\bullet$ \underline{Exchange-correlation energy:} The second term represents the exchange correlation energy per unit fundamental domain and is expressible using the Local Density Approximation (LDA) \citep{KohnSham_DFT} as:
\begin{align}
E_{\text{xc}}(\rho, {\calD}) = \int_{{\calD}}\varepsilon_{\text{xc}}[\rho(\bfx)]\,\rho(\bfx)\,d\bfx\,.
\end{align}
Note that the above formulation does not preclude the use of more sophisticated exchange correlation functionals such as the Generalized Gradient Approximation \citep{GGA_made_simple_Perdew}. Since the use of such functionals has little bearing on the subsequent discussion, we do not consider them further in this work.

$\bullet$ \underline{Nonlocal pseudopotential energy:} The third term on the right hand side of eq.~\ref{eq:electronic_free_energy} represents the nonlocal pseudopotential energy per unit fundamental domain. For a finite system consisting of $M^{\circ}$ atoms located at the points $\displaystyle \{\bfr_k^{{\circ}} \in \rz^3\}_{k=1}^{M^{\circ}}$, the non-local pseudopotential operator in Kleinman-Bylander form \citep{kleinman1982efficacious} can be written as:
\begin{align}
\calV_{\text{nl}}^{\,\circ} = \sum_{k=1}^{M^{\circ}}\sum_{p \in \calN_k}\gamma_{k,p}\,\chi_{k,p}(\cdot;\bfr_k^{\circ})\,\overline{\chi_{k,p}(\cdot;\bfr_k^{\circ})}\,,
\label{eq:Kleinman_Bylander_form}
\end{align}
in terms of the projection functions $\displaystyle \{\chi_{k,p}(\cdot;\bfr_k)\}_{p=1}^{\calN_k}$ and the corresponding normalization constants $\displaystyle\{\gamma_{k,p}\}_{p=1}^{\calN_k}$ associated with the $k^{\text{th}}$ atom (located at $\bfy_k$). The nonlocal pseudopotential energy in that case has the form:
\begin{align}
K^{\circ} = 2\, \sum_{k=1}^{M^{\circ}}\sum_{p \in \calN_k}\gamma_{k,p}\sum_{j=1}^{\infty}g_j \bigg\lvert \innprod{\chi_{k,p}(\cdot;\bfr_k)}{\psi_j^{\circ}(\cdot)}{\Lpspc{2}{}{{\rz^3}}} \bigg\rvert^2
\label{eq:Finite_Vnl_Expression}
\end{align}
To obtain the analogous expression for the twisted structure, we consider the contributions from the atoms located within the fundamental domain and all the electronic states in the system \citep{banerjee2021ab} to get the nonlocal pseudopotential energy per unit fundamental domain as:
\begin{align}
\nonumber
&K(\mathfrak{G}, \Psi, \calP, \calD, \calG)\\
&= 2\,\sum_{k=1}^{M}\sum_{p \in \calN_k}\gamma_{k,p}\sum_{j=1}^{\infty}\int_{\mathfrak{I}}\frac{1}{\mathfrak{N}}\sum_{\nu = 0}^{\mathfrak{N}-1}\bigg(g_j(\eta, \nu)\bigg\lvert{\big\langle{{\chi}_{k,p}(\cdot;\eta, \nu;\bfr_k)},{\psi_j(\cdot; \eta, \nu)}\big\rangle}_{\Lpspc{2}{}{{\calC}}}\bigg\rvert^2\bigg)\,d\eta\,.
\end{align}
Here, the overlaps of the  orbitals with the atom centered projectors are carried out over the global simulation domain $\calC$, since the latter can have support extending beyond the fundamental domain. With the aid of the Helical  Bloch Theorem (eq.~\ref{eq:Bloch_theorem_2}) and by using the properties of the projection functions ${\chi}_{k,p}$, the integral implicit in the above expression can be reduced to the fundamental domain \citep{ghosh2019symmetry, banerjee2021ab}, so that a more computationally convenient expression for the nonlocal pseudopotential energy per unit fundamental domain reads as:
\begin{align}
\nonumber
&K(\mathfrak{G}, \Psi, \calP, \calD, \calG)\\
&= 2\,\sum_{k=1}^{M}\sum_{p \in \calN_k}\gamma_{k,p}\sum_{j=1}^{\infty}\int_{\mathfrak{I}}\frac{1}{\mathfrak{N}}\sum_{\nu = 0}^{\mathfrak{N}-1}\bigg(g_j(\eta, \nu)\bigg\lvert{\big\langle{\hat{\chi}_{k,p}(\cdot;\eta, \nu;\bfr_k)},{\psi_j(\cdot; \eta, \nu)}\big\rangle}_{\Lpspc{2}{}{{\calD}}}\bigg\rvert^2\bigg)\,d\eta\,.
\end{align}
The functions $\hat{\chi}_{k,p}$ in the equation above can be expressed as:
\begin{align}
\label{eq:chi_hat_vnl}
\hat{\chi}_{k,p}(\bfx;\eta, \nu;\bfr_k) = \sum_{m\in \gz}\sum_{n=0}^{\mathfrak{N}-1}{\chi}_{k;p}\big( \Upsilon_{m,n}\circ \bfx;\bfr_k\big)\,e^{i2\pi (m\eta + \frac{n\nu}{\mathfrak{N}})}\,.
\end{align}

$\bullet$ \underline{Electrostatic interaction energy:} The fourth term on the right hand side of eq.~\ref{eq:electronic_free_energy} represents the electrostatic interaction energy per unit fundamental domain. This includes the Coulombic attraction between the electrons and the nuclei, as well as the mutual repulsion between the electrons themselves. To express this term, it is useful to introduce the net electrostatic potential $\Phi$, which also appears in the Kohn-Sham equations (as part of the effective potential). To see how this can be done, we consider first a finite system placed in $\rz^3$, with nuclei located at the points $\displaystyle \{\bfr_k^{\circ} \in \rz^3\}_{k=1}^{M^{\circ}}$. For this example system, the net electrostatic potential $\Phi^{\circ}$, can be expressed in terms of the total charge of the (finite) system as:
\begin{align}
\label{eq:ES_phi_formula_finite} 
\Phi^{^{\circ}}(\bfx) = \int_{\rz^3} \frac{\rho^{^{\circ}}(\bfy) + b^{^{\circ}}(\bfy)}{\norm{\bfx - \bfy}{\rz^3}}\,d\bfy\,.
\end{align}
Here, $\rho^{^{\circ}}$ represents the electron density and $ b^{\text{finite}}$ represents the total nuclear pseudocharge. The latter can be expressed in terms of the individual nuclear pseudocharges $\big\{b_k(\bfx;\bfr_k^{\circ})\big\}_{k=1}^{M^{\circ}}$ as:
\begin{align}
b^{{\circ}}(\bfx) = \sum_{k=1}^{M^{\circ}}b_k(\bfx;\bfr_k^{\circ})\,, 
\end{align}
Note that for each atom, the pseudocharge (typically a smooth, radially symmetric, compactly supported function) integrates to its valence nuclear charge, i.e.,
\begin{align}
\int_{\rz^3}b_k(\bfx;\bfr_k^{\circ})\,d\bfx = Z_k\,.
\end{align}
The connection between the potential $\Phi^{\circ}$ and the electrostatic interaction energy is that we may express the latter as:
\begin{align} 
\nonumber
{E}_{\text{el}}^{\,\circ} = \max_{\widetilde{\Phi}^{\circ} }&\bigg\{-\frac{1}{8\pi}\int_{\rz^3}\abs{\nabla \widetilde{\Phi}^{\circ} }^2\,d\bfx + \int_{\rz^3}(\rho^{\circ}  + b^{\circ})\,\widetilde{\Phi}^{\circ}\,d\bfx\bigg\} \\ \label{eq:ES_finite}  &+ E_{\text{sc}}^{\,\circ}(\bfr_1^{\circ},\bfr_2^{\circ},\ldots,\bfr_k^{\circ})\,,
\end{align}
and the scalar field $\widetilde{\Phi}^{^{\circ}}$ which attains the maximum in the above problem is precisely the one presecribed using eq.~\ref{eq:ES_phi_formula_finite}. Note that the constant term $E_{\text{sc}}^{\,\circ}(\bfr_1^{\circ},\bfr_2^{\circ},\ldots,\bfr_k^{\circ})$ is added as a correction  for self-interactions and possible overlaps of the nuclear pseudocharges \citep{suryanarayana2014augmented}. It is independent of $\widetilde{\Phi}^{\circ}$ and does not play a role in the above optimization problem.

With the above discussion in mind, we may now introduce the net electrostatic potential for the twisted structure using the electron density (eq.~\ref{eq:electron_density_expression}) and the net nuclear pseudocharge associated with the system, in a manner that is analogous to eq.~\ref{eq:ES_phi_formula_finite}, i.e.,
\begin{align}
\Phi(\bfx) = \int_{\calC} \frac{\rho(\bfy) +b(\bfy,\calP,\calG)}{\norm{\bfx - \bfy}{\rz^3}}\,d\bfy\,,
\label{eq:ES_phi_formula_helical} 
\end{align}
The net nuclear pseudocharge at any point in the global simulation domain can be expressed using the pseudocharges of the atoms in the fundamental domain as:
\begin{align}
b(\bfx, \calP,\calG) = \sum_{m \in \gz}\sum_{n = 0}^{\mathfrak{N}-1} \sum_{k = 1}^{M}b_{k}(\bfx; \Upsilon_{m,n} \circ \bfr_k)\,,
\label{eq:Net_pseudocharge}
\end{align}
Note that since the electron density is group invariant, as is the net nuclear pseudocharge (by construction), the total electrostatic potential  for the twisted structure is group invariant as well. Thus, it suffices to compute this quantity over the fundamental domain, in addition to specifying boundary conditions that are consistent with the group invariance of the function. Following eq.~\ref{eq:ES_finite}, we now write the electrostatic interaction energy per unit fundamental domain as:
\begin{align}
\nonumber
E_{\text{el}}(\rho, \calP, \calD, \calG) =  \max_{\widetilde{\Phi}}&\bigg\{-\frac{1}{8\pi}\int_{\calD}\abs{\nabla \widetilde{\Phi}}^2\,d\bfx + \int_{\calD}\bigg(\rho(\bfx) + b(\bfx,\calP,\calG)\bigg)\widetilde{\Phi}(\bfx)\,d\bfx\bigg\}\\ 
&+ E_{\text{sc}}(\calP,\calG, \calD)\,.
\label{eq:ES_helical} 
\end{align}
The scalar field $\widetilde{\Phi}$ which attains the maximum in the above problem, is the same one specified in eq.~\ref{eq:ES_phi_formula_helical}. The constant (i.e., $\widetilde{\Phi}$-independent)  term $E_{\text{sc}}(\calP,\calG, \calD)$ accounts for self-interaction corrections and possible overlaps between pseudocharges. We omit the details of this term here for the sake of brevity, and cite  references \citep{banerjee2016cyclic, ghosh2017sparc_1, suryanarayana2014augmented} for relevant details.

$\bullet$ \underline{Electronic entropy:} Finally, the last term on the right hand side of eq.~\ref{eq:electronic_free_energy} deals with the contribution of the electronic entropy to the free energy. Using Fermi-Dirac smearing, for a finite system at electronic temperature $T_{\text{e}}$, the electronic entropy can be represented :
\begin{align}
S^{\circ} = -2\,k_{\text{B}}\sum_{j=1}^{\infty} g_j^{\circ} \log(g_j^{\circ}) + (1 - g_j^{\circ})\,\log(1 - g_j^{\circ})\,.
\end{align}
Analogously, the corresponding term for the twisted structure reads as:
\begin{align}
S(\mathfrak{G}) = -2\,k_{\text{B}} {\int_{\mathfrak{I}}\bigg[\frac{1}{\mathfrak{N}}\sum_{\nu = 0}^{\mathfrak{N}-1}}\sum_{j=1}^{\infty} g_j(\eta, \nu) \log\big(g_j(\eta, \nu)\big) + \big(1 - g_j(\eta, \nu)\big)\,\log\big(1 - g_j(\eta, \nu)\big)\, \bigg]{d\eta}\,.
\end{align}
$\bullet$ \underline{Kohn-Sham Equations:} With the expressions for the various energy terms in place, we write the electronic ground-state energy for the twisted structure as the following minimization problem:
\begin{align}
{\calF}_{\substack{\text{Ground}\\ \text{State}}}(\calP , \calD, \calG) = \text{inf.}_{{\Psi,\mathfrak{G}}}\,{\calF}(\mathfrak{G}, \Psi, \calP, \calD, \calG)\,,
\end{align}
with the added constraints that:  
\begin{enumerate}
\item the orbitals in $\Psi$ are helical Bloch states, namely, they obey eq.~\ref{eq:Bloch_theorem_2} and are orthonormal over the fundamental domain for each $\zeta \in \widehat{\calG}$, i.e.:
\begin{align}
\label{eq:orthonormality}
\innprod{\psi_j(\cdot;\eta, \nu)}{\psi_{j'}(\cdot;\eta, \nu)}{\Lpspc{2}{}{\calD}} = \delta_{j,{j'}}\,,
\end{align}
and,
\item the number of electrons per unit fundamental domain is a fixed number, i.e., 
\begin{align}
\label{eq:electrons_per_unit_cell}
\int_{\calD} \rho(\bfx)\,d\bfx = \frac{2}{\mathfrak{N}}\sum_{\nu = 0}^{\mathfrak{N}-1} \int_{\mathfrak{I}}\sum_{j=1}^{\infty} g_j(\eta, \nu) = N_{\text{e}}\,.
\end{align}
\end{enumerate}
The Euler-Lagrange equations corresponding to the above variational problem are the Kohn-Sham equations for the twisted structure, as posed on the fundamental domain.  For $j\in\nz$, $\eta \in\mathfrak{I}$ and $\nu = 0,1,\ldots \mathfrak{N}-1$, we may express them as:
\begin{align}
\hamil^{\text{KS}}\,\psi_j(\cdot;\eta, \nu) &= \lambda_j(\eta, \nu)\,\psi_j(\cdot;\eta, \nu)\,,
\label{eq:Euler_Lagrange}
\end{align}
with $\hamil^{\text{KS}}$ denoting the Kohn-Sham operator, i.e.:
\begin{align}
\label{eq:KS_Operator}
\hamil^{\text{KS}} \equiv -\half \Delta + V_{\text{xc}} + \Phi + \calV_{\text{nl}}\,.
\end{align}
Here, $V_{\text{xc}}$ denotes the exchange correlation potential:
\begin{align}
V_{\text{xc}} = \frac{\delta E_{\text{xc}}(\rho, {\calD}) }{\delta \rho} = \varepsilon_{\text{xc}} + \rho \,\frac{d \varepsilon_{\text{xc}}}{d\rho}\,,
\end{align}
$\Phi$ (as introduced in eq.~\ref{eq:ES_phi_formula_helical}) denotes the net electrostatic potential arising from the electrons and the nuclear pseudocharges, and obeys the Poisson equation:
\begin{align}
-\Delta \Phi &= 4\pi\,\big(\rho + b(\cdot,\calP ,\calG)\,\big)\,,
\label{eq:Poisson_Equation}
\end{align}
while $\calV_{\text{nl}}$ denotes the non-local pseudoptential operator (specifically, its $(\eta, \nu)$ component), and can be expressed in terms of the functions $\hat{\chi}_{k,p}$ (introduced in eq.~\ref{eq:chi_hat_vnl}) as:
\begin{align}
\label{eq:vnl_twisted}
{\calV}_{\text{nl}} =  \sum_{k=1}^{M}\sum_{p \in \calN_k}\gamma_{k,p}\,\hat{\chi}_{k,p}(\cdot;\eta, \nu;\bfx_k)\,\overline{\hat{\chi}_{k,p}(\cdot;\eta, \nu;\bfx_k)}
\end{align}
Note that the use of eq.~\ref{eq:Poisson_Equation} in lieu of eq.~\ref{eq:ES_phi_formula_helical} is preferable for practical calculations since computationally inconvenient non-local integrals that appear in the latter equation are avoided  \citep{Pask2005, suryanarayana2014augmented, motamarri2012higher, ghosh2019symmetry}. Together, eqs.~\ref{eq:Euler_Lagrange} - \ref{eq:vnl_twisted}, along with eq.~\ref{eq:orthonormality} and \ref{eq:electrons_per_unit_cell} form the governing equations for our system and need to be solved self-consistently. 
\subsection{Boundary Conditions}
\label{subsec:BCs}
The unknown fields in the governing equations above are the orbitals $\psi_j(\cdot;\eta,\nu) \in \Psi$ and the electrostatic potential $\Phi$. Since these fields are posed on the fundamental domain $\calD$, we need to augment the governing equations with boundary conditions on the surfaces that make up $\partial \calD$. By using the conditions in eq.~\ref{eq:Bloch_theorem_2} on the orbitals, and observing that the symmetry operation $\Upsilon_{1,0} = \big(\,{\bfR}_{2\pi\alpha} \,|\,\tau\,\bfe_{\bfZ}\,\big)$ maps $\partial\calZ_{0}$ to $\partial\calZ_{\tau}$, while the operation $\Upsilon_{0,1} = \big(\,{\bfR}_{\Theta} \,|\,\bf0\,\big)$  maps $\partial{\vartheta_0}$ to $\partial{\vartheta_{\Theta}}$, we arrive at:
\begin{align}
\psi_j(\bfx \in \partial \calZ_{\tau},\eta, \nu) &= e^{-2 \pi i \eta}\psi_j(\bfx \in \partial \calZ_{0},\eta, \nu)\,, \\
\psi_j(\bfx \in \partial \vartheta_{\Theta},\eta, \nu) &= e^{-2 \pi i \frac{\nu}{\mathfrak{N}}}\psi_j(\bfx \in \partial \vartheta_{0} ,\eta, \nu)\,.
\end{align}
Concurrently, since the net electrostatic potential is invariant under all symmetry operations in $\calG$, it obeys the boundary conditions:
\begin{align}
\Phi(\bfx \in \partial \calZ_{\tau}) &= \Phi(\bfx \in \partial \calZ_{0})\,, \\
\Phi(\bfx \in \partial \vartheta_{\Theta}) &= \Phi(\bfx \in \partial \vartheta_{0})\,.
\end{align}
The above equations leave the boundary conditions on the surfaces $\partial R_{\text{in}}$ and $\partial R_{\text{out}}$ unspecified. As far as the wavefunctions are concerned, we may enforce Dirichlet boundary conditions on these surfaces, by appealing to the decay of the electron density along the radial direction \citep{banerjee2021ab, ghosh2019symmetry}. This gives us:
\begin{align}
\psi_j(\bfx \in \partial R_{\text{in}},\eta, \nu) &= \psi_j(\bfx \in \partial R_{\text{out}},\eta, \nu) = 0\,.
\end{align}
On the other hand, the electrostatic potential $\Phi$ may not decay to zero quickly along the radial direction. Therefore, it is more prudent to set $\Phi(\bfx \in \partial R_{\text{in}})$ and $\Phi(\bfx \in \partial R_{\text{out}})$ by direct evaluation of eq.~\ref{eq:ES_phi_formula_helical} by using a modified version of the Ewald summation technique \citep{Dumitrical_Ewald}. In practical calculations however, this correction may be sometimes ignored \citep{banerjee2021ab} in favor of Dirichlet boundary conditions on those surfaces. 
\subsection{Other quantities of interest at self-consistency}
\label{subsec:other_quantities}
At the end of the self consistent field iterations, a number of other quantities may be computed from the converged electronic states. For instance, we may obtain a more accurate estimate {(i.e., one that is less sensitive to self-consistency errors)} of the {Kohn-Sham ground  state electronic free energy} (per unit fundamental domain) by using the Harris-Foulkes functional \citep{harris1985simplified,foulkes1989tight} instead of eq.~\ref{eq:electronic_free_energy}. This can be written for the twisted structure, using quantities expressed over the fundamental domain as:
\begin{align}
\nonumber
{\calF}^{\text{HF}}(\Lambda, \Psi, \calP, {\calD}, \calG) &=  2\int_{\mathfrak{I}}\frac{1}{\mathfrak{N}}\sum_{\nu = 0}^{\mathfrak{N}-1}\sum_{j=1}^{\infty}\lambda_j(\eta, \nu)\,g_j(\eta, \nu)\,d\eta + E_{\text{xc}}(\rho, {\calD}) \\\nonumber &-\int_{ {\calD}}V_{\text{xc}}(\rho(\bfx))\rho(\bfx)\,d\bfx
+\half\int_{{\calD}}\bigg(b(\bfx, \calP, \calG) - \rho(\bfx) \bigg)\Phi(\bfx)\,d\bfx \\&+ E_{\text{sc}}(\calP ,\calG, {\calD}) - T_{\text{e}}\,S(\Lambda)\,.
\label{eq:Harris_Foulkes}
\end{align}
Note that the first term on the right hand side of the above equation is the electronic band energy.

For ab initio molecular dynamics or structural relaxation calculations, atomic forces need to be calculated. The Hellmann-Feynman forces on the atom located at $\bfr_k$ in the fundamental domain can be computed about the ground-state as:
\begin{align}
\nonumber
\mathbf{f}_k &= -\pd{{\calF}(\mathfrak{G}, \Psi, \calP, {\calD}, \calG)}{\bfr_k}\bigg\lvert_{\substack{\text{Ground}\\ \text{State}}}
\\\nonumber &= \sum_{m \in \gz}\sum_{n = 0}^{\mathfrak{N}-1}(\bfR_{2\pi m \alpha + n\Theta})^{-1} \int_{{\calD}}\nabla b_{k}\big(\bfx;(\Upsilon_{m,n}\circ \bfx_k \big)\Phi(\bfx)\,d\bfx - \pd{E_{\text{sc}}(\calP,\calG, {\calD})}{\bfr_k}\\\nonumber
&-4\sum_{j=1}^{\infty}\Bigg(\int_{\mathfrak{I}} \frac{1}{\mathfrak{N}}\sum_{\nu = 0}^{\mathfrak{N}-1} g_j(\eta, \nu)\sum_{p \in \mathcal{N}_k}\gamma_{k;p}\,\text{Re.}\Bigg\{\bigg[\int_{{\calD}} \hat{\chi}_{k,p}(\bfx;\eta, \nu;\bfr_k)\,\overline{\psi_j(\bfx;\eta, \nu)}\,d\bfx\bigg]\\
&\quad\quad\quad\quad\quad\quad\quad\quad\quad\quad\;\,\quad\quad\quad\quad\quad\quad\times\bigg[\int_{{\calD}} \psi_j(\bfx;\eta, \nu)\,\overline{\pd{\hat{\chi}_{k,p}(\bfx;\eta, \nu;\bfr_k)}{\bfr_k}}\,d\bfx\bigg]
\Bigg\}\Bigg)\,d\eta\,.
\label{eq:Hellman_Feynman_Force}
\end{align}
Note that since the forces are derivatives of a free energy which is invariant with respect to the symmetry operations in $\calG$, it follows that the force on an atom $\Upsilon_{m,n}\circ\bfr_{k}$ located outside the fundamental domain can be evaluated in terms of the force on its counterpart in the fundamental domain as $(\bfR_{2\pi m \alpha + n\Theta})^{-1}\bff_{k}$  \citep{Dumitrica_James_OMD}. Thus, to perform structural relaxations on the twisted structure, it suffices to concentrate on the atoms in the fundamental domain and drive their forces to zero.

Finally, the electronic density of states which often offers useful information about the electronic properties of a material under study, can be computed at an electronic temperature $T_{\text{e}}$ as \citep{Defranceschi_LeBris}:
\begin{align}
\aleph_{T_{\text{e}}}(E) =2\,\int_{\mathfrak{I}} \frac{1}{\mathfrak{N}}\sum_{\nu = 0}^{\mathfrak{N}-1}\bigg(\sum_{j=1}^{\infty} f^\prime_{T_{\text{e}}}\big(E - \lambda_j(\eta,\nu)\big)\bigg)\,d\eta\,,
\end{align}
with $f^\prime_{T_{\text{e}}}(\cdot)$ denoting the derivative of the Fermi-Dirac function.
\section{Implementation}
\label{sec:implementation}
We now discuss different numerical and computational aspects of the implementation of the above methodology. 
\subsection{Use of helical coordinates}
\label{subsec:helical_coordinates}
The equations in Section \ref{Section:Formulation} above are expressed in a manner that do not make any explicit reference to a coordinate system. For numerical implementation purposes however, it is useful to introduce a coordinate system that is commensurate with the geometry of the twisted structure and the symmetries of the system. The helical coordinate system, introduced in \citep{My_PhD_Thesis, banerjee2021ab} is well suited for these purposes. If a point $\bfp$ in the global simulation domain $\calC$ has Cartesian coordinates $(x_{\bfp},y_{\bfp},z_{\bfp})$ and cylindrical coordinates $(r_{\bfp}, \vartheta_{\bfp}, z_{\bfp})$, then the corresponding helical coordinates  $(r_{\bfp},\theta_{1\,{\bfp}},\theta_{2\,{\bfp}})$ are defined as:
\begin{align}
\nonumber
r_{\bfp} &= \sqrt{x_{\bfp}^2 + y_{\bfp}^2}\,,\,\theta_{1\,{\bfp}} =\frac{z_{\bfp}}{\tau}\,,\\
\theta_{2\,{\bfp}} &= \frac{1}{2\pi}\arctantwo(y_{\bfp}, x_{\bfp})-\alpha \frac{z_{\bfp}}{\tau} = \frac{\vartheta_{\bfp}}{2\pi} - \alpha \frac{z_{\bfp}}{\tau}\,.
\label{eq:helical_coordinates}
\end{align}
The helical coordinates reduce to the usual cylindrical coordinates when the twist parameter of the system is $0$ and the pitch $\tau$ is set to unity. The inverse relations:
\begin{align}
x_{\bfp} = r_{\bfp}\cos\big(2\pi(\alpha\theta_{1\,{\bfp}}+\theta_{2\,{\bfp}})\big)\,,\, y_{\bfp} = r_{\bfp}\sin\big(2\pi(\alpha\theta_{1\,{\bfp}}+\theta_{2\,{\bfp}})\big)\,,\,z_{\bfp} = \tau\,\theta_{1\,{\bfp}}\,,
\label{eq:inverse_helical_coordinates}
\end{align} 
map the helical coordinates of $\bfp$ to their Cartesian counterparts. 

The coordinate transformations introduced above can be used to map the curvilinear coordinate system associated with the twisted structure, to a rectilinear one in which computations are simpler to set up. Specifically, the relations in eq.~\ref{eq:inverse_helical_coordinates} above map the cuboid $(R_{\text{in}},R_{\text{out}})\times(0,1)\times(0,1/\mathfrak{N})$ to the fundamental domain $\calD$. In particular, the bounding surfaces of the fundamental domain can be described in helical coordinates as $r= R_{\text{in}}$ (for $\partial R_{\text{in}}$), $r= R_{\text{out}}$ (for $\partial R_{\text{out}}$), $\theta_{1} = 0$ (for $\partial \calZ_0$), $\theta_{1} = 1$ (for $\partial \calZ_{\tau}$), $\theta_{2} = 0$ (for $\partial \vartheta_0$) and $\theta_{2} = 1/{\mathfrak{N}}$ (for $\partial \vartheta_{\Theta}$). Furthermore, the symmetry operation $\Upsilon_{m,n}$ maps the helical coordinates of a point $\bfp$ from $(r_{\bfp},\theta_{1\,{\bfp}},\theta_{2\,{\bfp}})$ to $(r_{\bfp},\theta_{1\,{\bfp}} + m,\theta_{2\,{\bfp}} + \frac{n}{\mathfrak{N}})$.

In order to express the equations in Section \ref{subsec:gov_eqn} in helical coordinates, we need the   the Laplacian operator, the Cartesian gradient and the integral of a function (over the fundamental domain) expressed in helical coordinates. For a function $f(r,\theta_1,\theta_2)$ these take the form \citep{banerjee2021ab}:
\begin{align}
\Delta f &= \hpd{f}{r}{2}+\frac{1}{r}\pd{f}{r}+\frac{1}{\tau^2}\hpd{f}{\theta_1}{2}-
\frac{2\alpha}{\tau^2}\frac{\partial^2 f}{\partial \theta_1  \partial\theta_2} +
\frac{1}{4\pi^2}\bigg(\frac{1}{r^2}+\frac{4\pi^2\alpha^2}{\tau^2}\bigg)\hpd{f}{\theta_2}{2}\,,\\\nonumber
\nabla f &= \bigg(\pd{f}{r} \cos{\big(2\pi(\alpha\theta_1 + \theta_2)\big)} - \pd{f}{\theta_2}\frac{\sin{\big(2\pi(\alpha\theta_1 + \theta_2)\big)}}{2\pi r}\bigg)\bfe_{\bfX}\\\nonumber
&+ \bigg(\pd{f}{r} \sin{\big(2\pi(\alpha\theta_1 + \theta_2)\big)} - \pd{f}{\theta_2}\frac{\cos{\big(2\pi(\alpha\theta_1 + \theta_2)\big)}}{2\pi r}\bigg)\bfe_{\bfY}\\
&+ \bigg(\frac{1}{\tau}\big(\pd{f}{\theta_1} - \alpha \pd{f}{\theta_2}\big)\bigg)\bfe_{\bfZ}\\
\int_{\bfx \in \calD}\!f(\bfx)\,d\bfx &= \int_{r=R_{\text{in}}}^{r=R_{\text{out}}}\int_{\theta_1 = 0}^{\theta_1 = 1}\int_{\theta_2 = 0}^{\theta_2 = \frac{1}{\mathfrak{N}}}f(r,\theta_1,\theta_2)\,2\pi\tau r\,dr\,d{\theta_1}\,d{\theta_2}\,.
\end{align}
Upon expressing the Kohn-Sham orbitals as $\psi_j(r, \theta_1,\theta_2; \eta, \nu)$, the above expressions allow the governing equations and boundary conditions to be expressed in helical coordinates exclusively. For numerical implementation purposes however, it is convenient to work with functions that are completely invariant under symmetry operations instead of being invariant upto a Bloch phase, as the orbitals are. To this end, we write:
\begin{align}
 \psi_j(r, \theta_1,\theta_2; \eta, \nu) = e^{-2\pi i(\eta \theta_1 + \nu \theta_2)} \,u_j(r, \theta_1,\theta_2; \eta, \nu)\,,
 \label{eq:Bloch_ansatz}
\end{align}
where the functions $u_j(r, \theta_1,\theta_2; \eta, \nu)$ are group invariant. In terms of these auxiliary functions, the governing equations over the  fundamental domain are:
\begin{align}
\nonumber
&-\frac{1}{2}\Delta u_j(r,\theta_1,\theta_2;\eta,\nu) -\frac{2i\pi}{\tau^2}\left(\nu\alpha-\eta\right) \pd{ u_j(r,\theta_1,\theta_2;\eta,\nu)}{\theta_1}\\\nonumber
&- 2i\pi\bigg(\frac{\alpha}{\tau^2}\left(\eta-\nu\alpha\right)-\frac{\nu}{4\pi^2 r^2}\bigg)\pd{u_j(r,\theta_1,\theta_2;\eta,\nu)}{\theta_2} \\\nonumber 
&+\bigg(\frac{\nu^2}{2r^2} - \frac{2\pi^2}{\tau^2}\Big\{\nu\alpha\left(2\eta-\nu\alpha\right)-\eta^2\Big\} + V_{\text{xc}}(r,\theta_1,\theta_2) + \Phi(r,\theta_1,\theta_2)\bigg) u_j(r,\theta_1,\theta_2;\eta,\nu)\\\label{eq:final_gov_eqn_1}
&+e^{2\pi i \big({\eta\theta_1 + {\nu}\theta_2 }\big)}\,\calV_{\text{nl}}\,e^{-2\pi i \big({\eta\theta_1 + {\nu}\theta_2 }\big)}u_j(r,\theta_1,\theta_2;\eta,\nu) 
=\lambda_j(\eta,\nu)\,u_j(r,\theta_1,\theta_2;\eta,\nu)\,,\\\label{eq:final_gov_eqn_2}
&-\frac{1}{2}\Delta\Phi(r,\theta_1,\theta_2) = \rho(r,\theta_1,\theta_2) + b(r,\theta_1,\theta_2; \calP,\calG)\\\label{eq:final_gov_eqn_3}
&\rho(r,\theta_1,\theta_2) = 2\int_{\mathfrak{I}}\frac{1}{\mathfrak{N}}\sum_{\nu = 0}^{\mathfrak{N}-1}\sum_{j=1}^{\infty}\,g_j(\eta,\nu)\,\lvert u_j(r,\theta_1,\theta_2;\eta,\nu) \rvert^2\,d\eta\,,\\
&g_j(\eta,\nu) = f_{T_{\text{e}}}\big(\lambda_j(\eta, \nu)\big)\,,\,\frac{2}{\mathfrak{N}}\sum_{\nu = 0}^{\mathfrak{N}-1} \int_{\mathfrak{I}}\sum_{j=1}^{\infty} g_j(\eta, \nu) = N_{\text{e}}
\label{eq:final_gov_eqn_4}
\end{align}
The boundary conditions\footnote{{The use of $R_{\text{in}} > 0$ is well justified for tubes with large enough radii (based on wavefunction decay effects or the nearsightedness principle \citep{prodan2005nearsightedness}, for example), such as the ones considered in this work. This has the added benefit of being able to computationally avoid the coordinate singularities in the Laplacian at the origin, without incurring any performance or accuracy issues.}} are:
\begin{align}
\nonumber
u_j(r = R_{\text{out}},\theta_1,\theta_2;\eta,\nu) &= u_j(r = R_{\text{out}},\theta_1,\theta_2;\eta,\nu) = 0\,,\\\nonumber
u_j(r,\theta_1 = 0 ,\theta_2;\eta,\nu) &= u_j(r,\theta_1 = 1 ,\theta_2;\eta,\nu)\,,\\
u_j(r,\theta_1,\theta_2 = 0;\eta,\nu) &= u_j(r,\theta_1,\theta_2 = \frac{1}{\mathfrak{N}};\eta,\nu)\,.\\\nonumber
\quad\quad\quad\quad\quad\quad\quad\quad\Phi(r = R_{\text{out}},\theta_1,\theta_2) = \phi_{R_{\text{out}}}\,&,\,\Phi(r = R_{\text{in}},\theta_1,\theta_2) = \phi_{R_{\text{in}}}\,,\\\nonumber
\Phi(r,\theta_1 = 0 ,\theta_2) &= \Phi(r,\theta_1 = 1 ,\theta_2)\,,\\
\Phi(r,\theta_1,\theta_2 = 0) &= \Phi(r,\theta_1,\theta_2 = \frac{1}{\mathfrak{N}})\,.
\end{align}

\subsection{Approximation of infinite series in governing equations}
\label{subsec:infinite_sums}
The governing equations as posed above, contain series sums over infinite numbers of terms which need to be truncated for the purposes of numerical implementation. Such infinite sums not only appear explicitly while summing over an infinite number of electronic states (eqs.~\ref{eq:final_gov_eqn_3}, \ref{eq:final_gov_eqn_4}), but also implicitly in the calculation of quantities such as the net pseudocharge (eqs.~\ref{eq:final_gov_eqn_2}, \ref{eq:Net_pseudocharge}) and the nonlocal pseudopotential operator (eqs.~\ref{eq:final_gov_eqn_1}, \ref{eq:vnl_twisted}, \ref{eq:chi_hat_vnl}). We now describe our strategies for dealing with such quantities.

In order to truncate sums involving an infinite number of electronic states, we may assume --- as is commonly done in the literature \citep{Gavini_higher_order, banerjee2018two}, that the electronic occupation numbers decay to zero beyond the lowest $N_{\text{s}}$ electronic states. In effect, this implies that sums over the index $j$ in equations \ref{eq:final_gov_eqn_3} -  \ref{eq:final_gov_eqn_4}  run from $1$ to $N_{\text{s}}$ (instead of $1$ to $\infty$), and that a set of $N_{\text{s}}$ eigenvalue problems for each value of $\eta$ and $\nu$, as posed in eq.~\ref{eq:final_gov_eqn_1}, have to be considered. In practical calculations when the electronic temperature is less than a few thousand Kelvin, the number of states $N_{\text{s}}$ can be related to the number of electrons per unit cell $N_{\text{e}}$ as $\displaystyle N_{\text{s}} = \big\lceil c_{\text{s}} \times \frac{N_{\text{e}}}{2} \big\rceil$, with the constant $c_{\text{s}}$ chosen to be between $1.05$ and $1.20$ \citep{banerjee2018two}.

The infinite sums involved in calculation of the net pseudocharge and the non-local pseudopotential operator both arise due to summations over individual atoms in the fundamental domain, as well as  their repeated images under the group $\calG$ (eqs.~\ref{eq:Net_pseudocharge}, \ref{eq:vnl_twisted}, \ref{eq:chi_hat_vnl}). However, we observe that the functions being summed in these cases are  always centered about the atoms in question, and they have the property of being compactly supported in a small spherical region of space around the atom (i.e., the functions $b_{k}(\cdot)$ in eq.~\ref{eq:Net_pseudocharge} and ${\chi}_{k;p}(\cdot)$ in eq.~\ref{eq:chi_hat_vnl} all have this property). Thus, the contribution of such terms to the fundamental domain is zero beyond a few terms of the series expressed in eqs.~\ref{eq:Net_pseudocharge} and \ref{eq:chi_hat_vnl}, and the infinite summations in these expressions can be reduced to a set of values of $m$ and $n$ that are near zero.\footnote{This typically involves $m = \pm 1, \pm 2, \pm 3$, etc., from the helical symmetry operations, and $n = 1,2,3, \mathfrak{N} - 1, \mathfrak{N} - 2, \mathfrak{N} - 3$, etc. from the cyclic symmetry operations.}
\subsection{Discretization Strategy}
\label{subsec:discretization}
The equations above need to be discretized in real space (i.e., over the fundamental domain $\calD$) as well as in reciprocal space (i.e., over the set $\widehat{\calG}$). We now describe our strategies for addressing each of these issues. 
\subsubsection{Real space discretization of the fundamental domain}
\label{subsubsec:real_space_discretization}
We use a higher order finite difference scheme \citep{chelikowsky1994finite, Chelikowsky_Saad_1,  Chelikowsky_Saad_2, kikuji2005first, ghosh2017sparc_1, ghosh2017sparc_2, ghosh2019symmetry, banerjee2021ab} for real space discretization. Since helical coordinates have the property of ``unwrapping'' the fundamental domain $\calD$ to a cuboid, a convenient meshing of the computational domain can be attained by choosing equispaced points in the $r$, $\theta_1$ and $\theta_2$ directions. Accordingly, we choose $\mathscr{N}_{r}$, $\mathscr{N}_{\theta_1}$ and $\mathscr{N}_{\theta_2}$ grid points along these directions (respectively), and observe that the corresponding mesh spacings $h_{r}, h_{\theta_1}, h_{\theta_2}$ satisfy:
\begin{align}
\mathscr{N}_{r}\,h_{r} = R_{\text{out}} - R_{\text{in}} \,,\,\mathscr{N}_{\theta_1}\,h_{\theta_1} = 1\,,\,\mathscr{N}_{\theta_2}\,h_{\theta_2} = \frac{1}{\mathfrak{N}}\,.
\end{align}
We will often refer to the quantity $h = \text{Max.}\bigg(h_r, \tau h_{\theta_1}, 2\pi\big(\frac{R_{\text{in}} + R_{\text{out}}}{2}\big)h_{\theta_2}\bigg)$ as the overall mesh spacing for a particular level of discretization. We index each finite difference node using a triplet of natural numbers:
\begin{align}
(\mathsf{i}, \mathsf{j}, \mathsf{k}) \in \{1,2,\ldots,\mathscr{N}_{r}\} \times \{1,2,\ldots,\mathscr{N}_{\theta_1}\} \times \{1,2,\ldots,\mathscr{N}_{\theta_2}\}\,,
\end{align}
and we use $f^{(\mathsf{i}, \mathsf{j}, \mathsf{k})}$ to denote the value a function $f$ at the grid point $\mathsf{i}, \mathsf{j}, \mathsf{k}$. The grid point with indices $(\mathsf{i}, \mathsf{j}, \mathsf{k})$ is associated with the radial coordinate $r_{\mathsf{i}} = R_{\text{in}} + (\mathsf{i} - 1) * h_r$, $\theta_1$ coordinate $\theta_{1_{\mathsf{j}}} =  (\mathsf{j} - 1) * h_{\theta_1}$ and $\theta_2$ coordinate $\theta_{2_{\mathsf{k}}} =  (\mathsf{k} - 1) * h_{\theta_2}$. The total number of real space grid points is $N_{\calD} = \mathscr{N}_{r} \times \mathscr{N}_{\theta_1} \times \mathscr{N}_{\theta_2}$.

To discretize equations \ref{eq:final_gov_eqn_1} - \ref{eq:final_gov_eqn_4} using the finite difference scheme, we require expressions for first and second order derivatives, as well as a quadrature rule to compute integrals over the fundamental domain (e.g., to evaluate the action of $\calV_{\text{nl}}$ on a given function). The expressions for the first order derivatives are:
 \begin{align}
\frac{\partial f}{\partial r} \bigg|^{(\mathsf{i},\mathsf{j},\mathsf{k})} &\approx  \sum_{p=1}^{n_o} \bigg({w}_{p,r}^{\text{first}}  \big( f^{(\mathsf{i}+p,\mathsf{j},\mathsf{k})} - f^{(\mathsf{i}-p,\mathsf{j},\mathsf{k})}\big) \bigg)\,,\nonumber \\
\frac{\partial f}{\partial \theta_1} \bigg|^{(\mathsf{i},\mathsf{j},\mathsf{k})} &\approx  \sum_{p=1}^{n_o} \bigg(  {w}_{p,\theta_1}^{\text{first}} \big( f^{(\mathsf{i},\mathsf{j}+p,\mathsf{k})} - f^{(\mathsf{i},\mathsf{j}-p,\mathsf{k})}\big) \bigg)\,,\nonumber \\
\frac{\partial f}{\partial \theta_2} \bigg|^{(\mathsf{i},\mathsf{j},\mathsf{k})} &\approx \sum_{p=1}^{n_o} \bigg({w}_{p,\theta_2}^{\text{first}} \big( f^{(\mathsf{i},\mathsf{j},\mathsf{k}+p)} - f^{(\mathsf{i},\mathsf{j},\mathsf{k}-p)}\big) \bigg)\,.
\label{eq:FD_expressions}
\end{align}
The expressions for the second order derivatives are:
\begin{align}
\nonumber 
\frac{\partial^2 f}{\partial r^2} \bigg|^{(\mathsf{i},\mathsf{j},\mathsf{k})} &\approx  \sum_{p=0}^{n_o} \bigg({w}_{p,r}^{\text{second}}  \big( f^{(\mathsf{i}+p,\mathsf{j},\mathsf{k})} + f^{(\mathsf{i}-p,\mathsf{j},\mathsf{k})}\big) \bigg) \,,  \\\nonumber 
\frac{\partial^2 f}{\partial \theta_1^2} \bigg|^{(\mathsf{i},\mathsf{j},\mathsf{k})} &\approx  \sum_{p=0}^{n_o} \bigg(  {w}_{p,\theta_1}^{\text{second}} \big( f^{(\mathsf{i},\mathsf{j}+p,\mathsf{k})} + f^{(\mathsf{i},\mathsf{j}-p,\mathsf{k})}\big) \bigg) \,, \\\nonumber 
\frac{\partial^2 f}{\partial \theta_2^2} \bigg|^{(\mathsf{i},\mathsf{j},\mathsf{k})} &\approx \sum_{p=0}^{n_o} \bigg({w}_{p,\theta_2}^{\text{second}} \big( f^{(\mathsf{i},\mathsf{j},\mathsf{k}+p)} + f^{(\mathsf{i},\mathsf{j},\mathsf{k}-p)}\big) \bigg) \,,\\\nonumber 
\frac{\partial^2 f}{\partial \theta_1\partial\theta_2} \bigg|^{(\mathsf{i},\mathsf{j},\mathsf{k})} &\approx
\sum_{p=1}^{n_o} {w}_{p,\theta_2}^{\text{first}} \bigg[ 
\bigg\{ \sum_{p'=1}^{n_o} {w}_{p',\theta_1}^{\text{first}} \big( f^{(\mathsf{i},\mathsf{j}+p',\mathsf{k}+p)} - f^{(\mathsf{i},\mathsf{j}-p',\mathsf{k}+p)}\big) \bigg\}\\
&\quad\quad\quad\quad\;\,- \bigg\{ \sum_{p'=1}^{n_o} {w}_{p',\theta_1}^{\text{first}} \big( f^{(\mathsf{i},\mathsf{j}+p',\mathsf{k}-p)} - f^{(\mathsf{i},\mathsf{j}-p',\mathsf{k}-p)}\big)\bigg\}\bigg]\,.
\label{eq:FD_second_derivatives}
\end{align}
In the above expressions, $n_o$ denotes half the finite difference order, $s$ denotes $r$, $\theta_1$ or $\theta_2$, and the finite difference weights $w_{p,s}^{\text{second}}$ and ${w}_{p,s}^{\text{first}}$ can be expressed as \citep{mazziotti1999spectral}:
 \begin{align}
w_{0,s}^{\text{second}} & =  - \frac{1}{h_s^2} \sum_{q=1}^{n_o} \frac{1}{q^2} \,, \,\,
\nonumber \\\nonumber
w_{p,s}^{\text{second}}  & =  \frac{2 (-1)^{p+1}}{h_s^2\,p^2} \frac{(n_o!)^2}{(n_o-p)! (n_o+p)!} \,\,\text{for}\,\,p=1, 2, \ldots, n_o\,,\\
{w}_{p,s}^{\text{first}} & =  \frac{(-1)^{p+1}}{h_s\,p} \frac{(n_o!)^2}{(n_o-p)! (n_o+p)!}\,\,\text{for}\,\,p=1, 2, \ldots, n_o\,.
\label{eq:FD_weights}
\end{align}
Finally, the expression for evaluating integrals over the fundamental domain is:
\begin{align}
\int_{\bfx \in \calD}f(\bfx)\,d\bfx \approx  h_r h_{\theta_1} h_{\theta_2} \sum_{\mathsf{i}=1}^{\mathscr{N}_r} \sum_{\mathsf{j} =1}^{\mathscr{N}_{\theta_1}} \sum_{\mathsf{k}=1}^{\mathscr{N}_{\theta_2}} 2\pi\tau r_{\mathsf{i}}\,f^{(\mathsf{i}, \mathsf{j}, \mathsf{k})}\,.
\end{align}
\subsubsection{Reciprocal space discretization}
\label{subsubsec:rec_space_discretization}
As is evident from the governing equations, many quantities of interest (including the electron density, for example) involve accumulating sums from each of the characters $\zeta \in \widehat{\calG}$. Since this is equivalent to computing sums of the form $\displaystyle \frac{1}{\mathfrak{N}}\sum_{\nu =0}^{\mathfrak{N}-1}$ and integrals over the set ${\mathfrak{I}}$, we need a suitable scheme for discretizing the set $\widehat{\calG}$. Accordingly, we perform quadratures over the set $\widehat{\calG}$ using:
\begin{align}
\frac{1}{\mathfrak{N}}\sum_{\nu =0}^{\mathfrak{N}-1}\int_{\mathfrak{I}} f(\eta, \nu)\,d\eta \approx \frac{1}{\mathfrak{N}}\sum_{\nu =0}^{\mathfrak{N}-1} \sum_{{b=1}}^{\calN_{\eta}}\,w_b\,f(\eta_b, \nu)\,.
\label{eq:MP_Scheme}
\end{align}
In the above expression, in accordance with the Monkhorst-Pack scheme \citep{monkhorst1976special}, the quadrature nodes $\eta_b$ are equi-spaced, while the corresponding quadrature weights $w_b$ are uniform. Effectively, the above scheme discretizes the set $\calG$ using $N_{\calK} = \calN_{\eta} \times \mathfrak{N}$ representative reciprocal space points. By use of time reversal symmetry \citep{geru2018time, ghosh2019symmetry, banerjee2021ab}, it is possible to reduce the number $N_{\calK}$ by a factor of $2$, which helps in cutting down computational wall time in practical calculations.  
\subsection{Solution strategies for the discretized equations and MATLAB implementation}
\label{subsec:discretized_solution_and_Matlab}
The governing equations for the twisted structure represent a set of coupled nonlinear eigenvalue problems. We use self consistent field (SCF) iterations accelerated via Periodic-Pulay extrapolation \citep{banerjee2016periodic} to solve them in this work. The total effective potential (i.e., $V_{\text{xc}} + \Phi$) is used as the mixing variable. Solution of the Poisson equation associated with the electrostatic field (eq.~\ref{eq:final_gov_eqn_2}) is carried out using the Generalized Minimal Residual method (GMRES) \citep{saad1986gmres}, and an incomplete LU factorization based preconditioner \citep{saad2003iterative} is used to accelerate convergence {of the GMRES iterations}. Solution to eq.~\ref{eq:final_gov_eqn_4} is carried out using a nonlinear equation root finder \citep{forsythe1977computer}.

As a consequence of the discretization choices and other simplifications outlined previously, there are $N_{\calK}$ linear eigenvalue problems, each of dimension $N_{\calD}$, that have to be solved on each SCF iteration step. Furthermore, for each of these eigenvalue problems, the lowest $N_{\text{s}}$ eigenstates have to be determined via a suitable diagonalization process. Due to our use of finite differences, the discretized Hamiltonian operators (at each value of $\eta$ and $\nu$) are non-Hermitian, even though the original infinite dimensional operators they represent are not. This is a well known issue that arises while approximating differential operators such as the Laplacian in curvilinear coordinates using finite differences \citep{gygi1995real, banerjee2016cyclic, ghosh2019symmetry}. In practice, this issue is mitigated by a combination of factors. First, as the mesh spacing $h$ is made finer, and/or the degree of the finite difference discretization $n_o$ is made larger, the discretized operators approximate their infinite dimensional counterparts more closely. Consequently, the discretized operators become more Hermitian (i.e., the norm of the difference between the operator and its Hermitian conjugate goes to zero), the eigenvalues have small imaginary components, and conventional iterative methods for obtaining the spectrum of a sparse symmetric Hamiltonian \citep{zhou2014chebyshev, saad1996solution, saad2010numerical} can be employed for diagonalization. Indeed, for the discretization parameters used to produce the results in this work, the imaginary parts of the Hamiltonian eigenvalues are small enough that they can be  ignored without any deleterious effects on the stability or quality of the simulations. Second, by choosing eigensolvers that can handle non-Hermitian problems in a robust manner, even calculations involving relatively coarse meshes (i.e., for which the Hamiltonian is well conditioned, but might have some eigenvalues with non-vanishing imaginary parts), or problems with poorly conditioned Hamiltonian matrices (which can arise if a system with a large amount of prescribed twist is being studied) can be performed.

\begin{figure}[ht]
\centering
\subfloat{
{\includegraphics[width=0.55\textwidth]{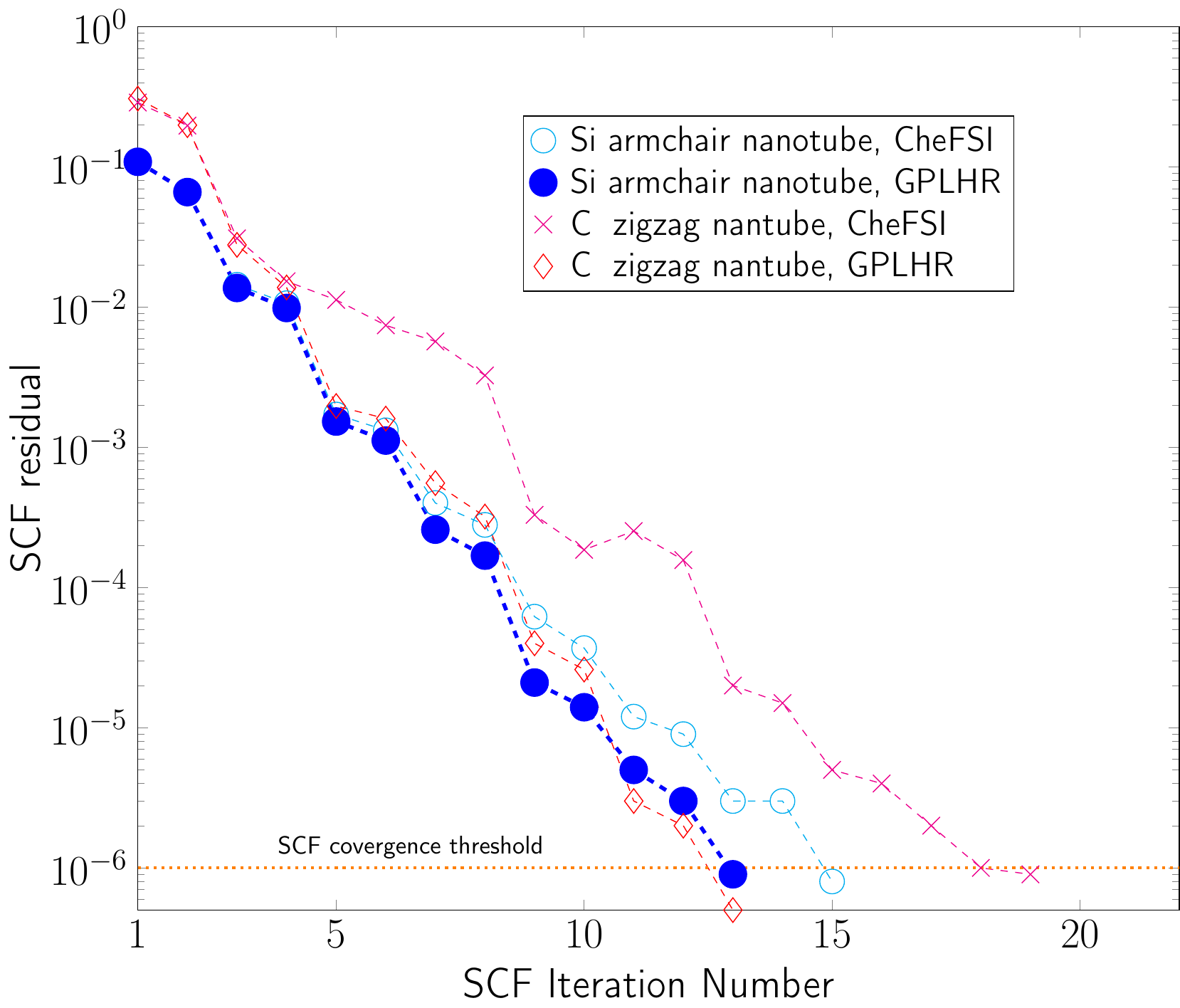}
}}
\caption{Examples of SCF convergence using CheFSI and GPLHR methods in Helical DFT. The armchair silicon nanotube has radius $0.96$ nm, and was subjected to a twist of $5.67$ degrees/nm. The zigzag carbon nanotube has radius $0.70$ nm, and was subjected to a twist of $4.27$ degrees/nm.}
\label{fig:SCF_convergence}
\end{figure}

Keeping the above factors in mind, our implementation employs a combination of the Generalized Preconditioned Locally Harmonic Residual (GPLHR) method \citep{vecharynski2015generalized}, as well as iterative diagonalization based on Chebyshev polynomial filtered subspace iterations (CheFSI) \citep{zhou2014chebyshev,Serial_Chebyshev,Parallel_Chebyshev}. Due to the ability of  GPLHR to employ preconditioners (based on incomplete LU factorization, e.g.), the method can be particularly effective in handling poorly conditioned Hamiltonian matrices --- i.e., for problems in which the CheFSI method tends to use relatively large polynomial filter orders. For such problems, we have also observed that GPLHR generally requires fewer iterations to reach {SCF} convergence, when compared to CheFSI, and between $5$ to $8$ iterations of the method are sufficient per SCF step. Nevertheless, for the systems considered in this work, we found that Chebyshev polynomial filter orders in the range $60$ to $80$ were adequate in guaranteeing stable, well converged simulations, and in this scenario the CheFSI method generally required shorter wall-times-to-solution overall. Thus, for the bulk of the simulations presented in this work, CheFSI was the method of choice. We show examples of the SCF convergence behavior for two example systems using CheFSI and GPLHR in Figure \ref{fig:SCF_convergence}.

We have implemented the above methods and algorithms in a computational package called Helical DFT. The current version of the code is largely written in MATLAB \citep{MATLAB:2019}, with certain key routines (including Hamiltonian matrix-vector products, sections containing multiple nested loops, etc.) written in C to alleviate speed and/or memory footprint issues. The code makes use of MATLAB's vectorization capabilities, and achieves parallelization by performing diagonalization of the Hamiltonian for different values of $\eta$ and $\nu$ simultaneously over multiple computational cores. Helical DFT is capable of performing structural relaxation by use of the Fast Intertial Relaxation Engine (FIRE) algorithm \citep{bitzek2006structural} as well as \textit{ab initio} molecular dynamics simulations by use of a velocity Verlet integrator \citep{verlet1967computer}. 
\section{Simulations and Results}
\label{sec:simulation_results}
\subsection{Computational Platform}
\label{subsec:comp_platform}
All simulations involving Helical DFT were run using a dedicated desktop workstation (Dell Precision 7920 Tower) or single nodes of the Hoffman2 cluster at UCLA's Institute for Digital Research and Education (IDRE). The desktop workstation has an $18$-core Intel Xeon Gold 5220 processor ($24.75$ MB cache, running at $2.2$ GHz), $256$ GB of RAM and a $1$ TB SATA Class 20 Solid State Drive (SSD). Each compute node of the Hoffman2 cluster has two $18$-core Intel Xeon Gold 6140 processors (with $24.75$ MB cache, running at $2.3$ GHz), $192$ GB of RAM and local SSD storage. MATLAB version $9.7.0$ (R2019b) was used for the simulations. Compilation of C language routines was carried out using MinGW (on the workstation) and GCC (on the Hoffman2 nodes) software suites. Interfacing between MATLAB and C language routines was carried out by means of MATLAB's MEX and Coder frameworks, while parallelization was achieved by use of using MATLAB's Parallel Computing Toolbox.
\subsection{Simulation Parameters}
\label{subsec:Sim_Params}
We used an SCF iteration convergence tolerance of $10^{-6}$ in the total effective potential (relative residual). The Periodic Pulay mixing scheme \citep{banerjee2016periodic} used a history of $7$ iterations, the mixing parameter was set at $0.2$, and Pulay extrapolation was performed on every alternate SCF step. GMRES iterations for the Poisson problem was carried out till the residual dropped below $10^{-9}$ on every SCF step. We employed an electronic temperature of $T_{\text{e}} = 315.77$ Kelvin in the Fermi-Dirac function (this corresponds to about 1 milli-Hartree of smearing), and included $2$ extra states at each value of $\eta$ and $\nu$ to accommodate fractional occupancies.  We used Troullier-Martins norm conserving pseudopotentials \citep{troullier1991efficient} and Perdew-Wang parametrization \citep{Perdew_Wang} of the Local Density Approximation \citep{KohnSham_DFT}. We used a $12^{\text{th}}$ order finite difference discretization scheme (i.e., $n_o = 6$ in eqs.~\ref{eq:FD_expressions}, \ref{eq:FD_second_derivatives}, \ref{eq:FD_weights}) and diagonalization via CheFSI used filters of order $60$ to $80$. Determination of spectral bounds within the CheFSI method used MATLAB's \textsf{eigs} function \citep{stewart2002krylov} with a relatively loose tolerance of $10^{-2}$. For the nanotube simulations described here, we ensured a gap of $10$ to $11$ Bohrs between the atoms located within the fundamental domain, and the boundary surfaces $\partial R_{\text{in}}$ and $\partial R_{\text{out}}$, in order for the electron density and the wavefunctions to decay sufficiently in the radial direction\footnote{{We have carried out tests regarding the effect of the amount of vacuum padding (i.e., distance between $R_{\text {in }}$ or $R_{\text {out}}$ and the atoms of the structure) on the energies and forces,  and have observed the deviations in these quantities drop to $10^{-5}$ or so (in atomic units) at a vacuum padding of about $8$ Bohr, for the systems considered here. In our actual simulations, we use a somewhat larger vacuum  padding of $10-11$ Bohrs and the tube radii are also chosen accordingly.}}. Real space and reciprocal space discretization parameters were chosen on a case by case basis, as described later. 
\subsection{Materials Systems: Group IV Nanotubes}
\label{subsec:Materials_Systems}
Nanotubes and other similar systems are particularly well suited for study using the methods described in this work. We choose single walled nanotubes of carbon, silicon, germanium, and tin as materials systems for investigation here. These systems are used for carrying out numerical validation studies, and due to their technological importance, also for gaining insights into their properties by the use of our method. Such one-dimensional nanostructures from Group IV of the periodic table can be described in terms of a ``roll-up'' procedure \citep{evarestov2015theoretical}, starting from their two-dimensional sheet counterparts (i.e., graphene, silicene, germanene and stanene).  We collectively refer to these one- and two-dimensional materials as X (X = C, Si, Ge, Sn) nanotubes, and Xenes, respectively. Both these classes of materials have been intensely studied in recent years through both experimental and computational methods, due to their association with fascinating materials properties \cite{molle2017buckled, ni2011tunable, drummond2012electrically, balendhran2015elemental, zhu2015epitaxial, scalise2014vibrational, davila2014germanene, kara2012review, martel1998single, javey2003ballistic, popov2004carbon, gong2009nitrogen, park2009silicon, wu2012stable, park2011germanium, li2011controlled, zhao2006porous, xu2013graphene, bhimanapati2015recent, butler2013progress, naguib201425th, fiori2014electronics, koppens2014photodetectors, novoselov2004electric, novoselov2005two, geim2009graphene, iijima1993single,saito1998physical, sha2002silicon, wang2017band, blase1994hybridization, spataru2004excitonic, yang2000electronic, fagan2000ab, benedict1995static, zhang2003silicon, yang2005electronic, giovannetti2008doping, vogt2012silicene, davila2014germanene, zhu2015epitaxial, seifert2001tubular}. In particular, the electronic properties of deformed carbon nanotubes have received extensive treatment in the literature through theoretical and computational means \citep{yang2000electronic, ding2002analytical, yang1999band, najafi2016analysis, ding2003curvature, Dumitrica_Tight_Binding1, sreekala2008effect, rochefort1999electrical, heyd1997uniaxial, kane1997size}. Although a few computational studies on the electronic structure of the larger class of Group IV nanotubes are also available \citep{wang2017band, ghosh2019symmetry, abbasi2018structural, abbasi2019band, abbasi2018theoretical}, as far as we can tell, this work is the first to investigate from first principles, the behavior of these materials under torsional deformations, and to extend some well known results for carbon nanotubes to the broader class of Group IV nanotubes.

\begin{figure}[ht]
\centering
{\includegraphics[trim={3cm 3.5cm 2.0cm 3.5cm}, clip, width=0.80\textwidth]{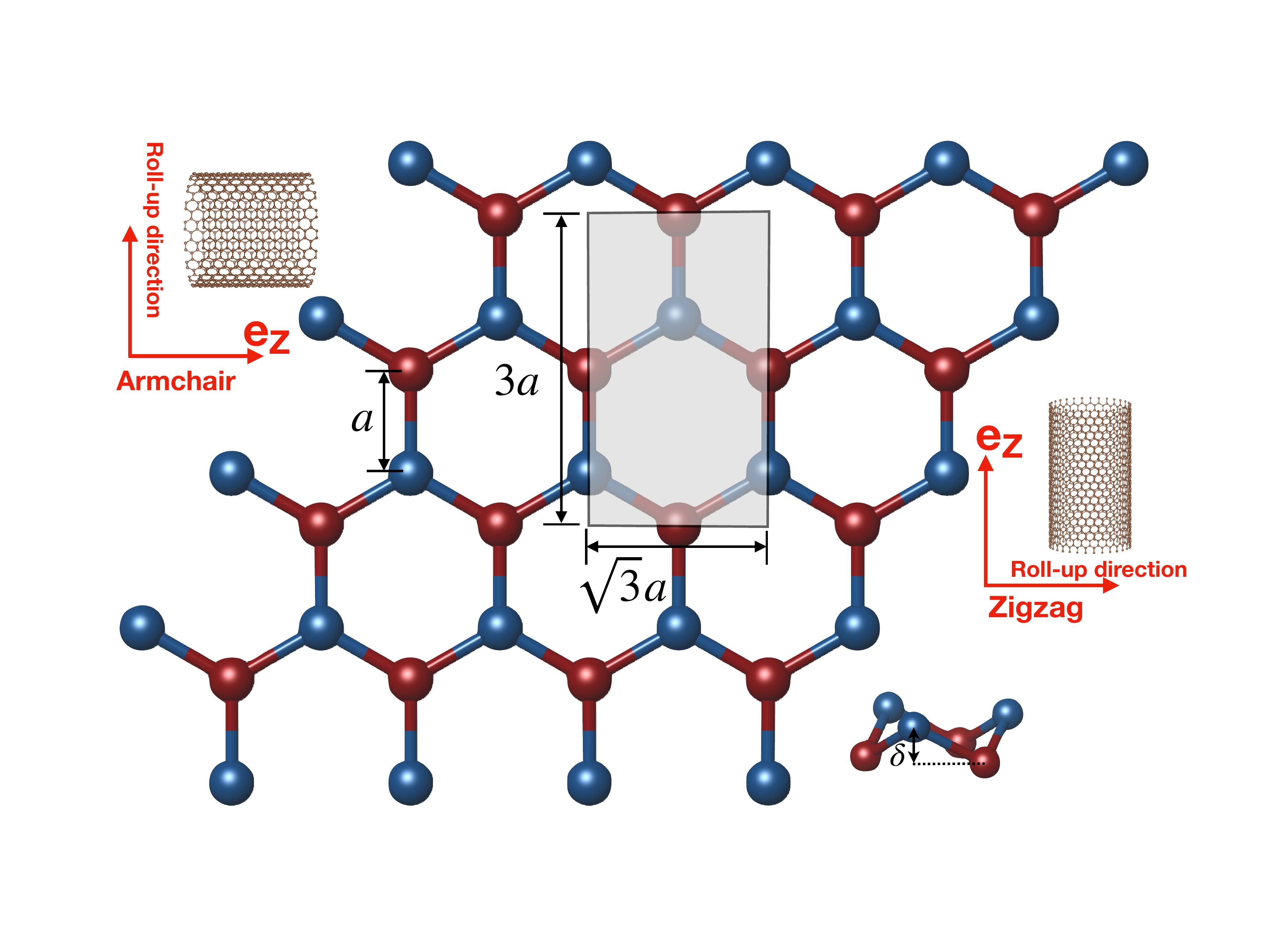}}
\caption{Roll-up construction of untwisted X nanotube, starting from Xene sheet. Atoms in the fundamental domain are shaded and are the same ones conventionally used for carrying out simulations of (planar) Xenes using orthogonal unit cells. The parameter $a$ represents the planar interatomic distance, $\delta$ represents the out of plane buckling parameter, and in-plane and out-of-plane atoms are shown in alternate colors.}
\label{fig:Xene_Roll_Up}
\end{figure}

By using the roll up construction on the Xene sheets (see Figure \ref{fig:Xene_Roll_Up}), we can represent untwisted tubes using just four atoms in the fundamental domain \citep{ghosh2019symmetry, James_OS, Dumitrica_James_OMD}, and a twist can be prescribed on the system by choosing a non-zero value of $\alpha$. Depending on the direction of rolling, the untwisted tubes can be classified as armchair or zigzag, and the fundamental period $\tau$ of the untwisted tubes in these cases are $\sqrt{3}\,a$ and $3\,a$ , respectively, with $a$ denoting the (planar) interatomic distance among the X atoms. Furthermore, the cyclic group order $\mathfrak{N}$ can be expressed in terms of the nanotube radius via the relation ${\mathfrak{N}}\,L = 2\pi R_{\text{avg.}}$. Here  $L = 3\,a$  and $\sqrt{3}\,a$, for armchair and zigzag cases, respectively, and $R_{\text{avg.}}$ denotes the average radial coordinate of the atoms in the fundamental domain. For subsequent simulations, we adopted the values of the parameter $a$, as well as the out of plane buckling parameter $\delta$, as reported in \citep{ghosh2019symmetry}. We include the values of the parameters in Table \ref{Table:LatticeParameters} for the sake of having a self contained presentation here.\footnote{To compute these parameters, the relaxed ground state structure of the Xene sheets (single layer) was computed using the plane-wave DFT code ABINIT \citep{Gonze_ABINIT_1, gonze2016recent}. The same pseudopotentials, exchange correlation functional and electronic temperature were used between ABINIT and Helical DFT. Energy cutoffs between $40$ and $60$ Ha, $30 \times 30 \times 1$ k-points, and  a cell vacuum of $25$ Bohr in the direction orthogonal to the sheets, were employed. At the end of the geometry relaxation procedure, the atomic forces and the cell stress were of the order of $10^{-5}$ Ha/Bohr  and $10^{-8}\;\text{Ha/Bohr}^3$, respectively. The agreement of these parameters with existing literature is quite good \citep{ghosh2019symmetry}, thus lending confidence to the physical properties of the X nanotubes as revealed via our simulations.}
\begin{table}[ht]
\centering
\begin{tabular}{c  c  c}
\hline
Material & $a$ (Angstrom) & $\delta$ (Angstrom)  \\   
\hline
Graphene  & 1.407 &  - \\
Silicene  &  2.200 & 0.404  \\
Germanene & 2.232 & 0.566 \\
Stanene  & 2.522 & 0.699 \\
\hline
\end{tabular}
\caption{Equilibrium lattice parameters of Xene sheets, as obtained from \citep{ghosh2019symmetry}, and used in subsequent Helical DFT simulations.}
\label{Table:LatticeParameters}
\end{table} 
\subsection{Convergence, accuracy and efficiency studies}
\label{subsec:convergence_accuracy_efficiency}
\begin{figure}[ht]
\centering
\subfloat[Convergence of ground state energy w.r.t. real space discretization]{\includegraphics[width=0.42\textwidth]{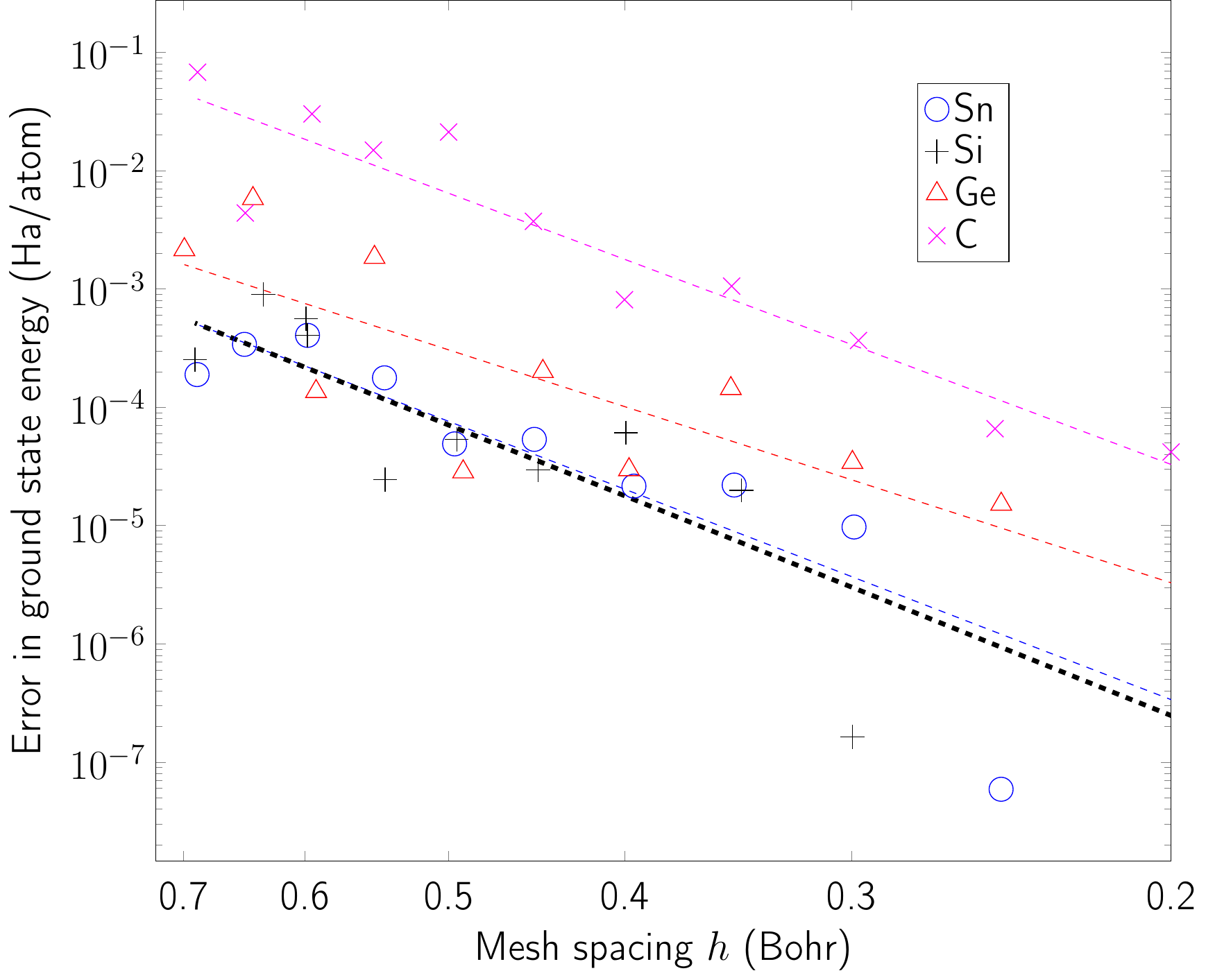}}\quad
\subfloat[Convergence of atomic forces w.r.t. real space discretization]{\includegraphics[width=0.42\textwidth]{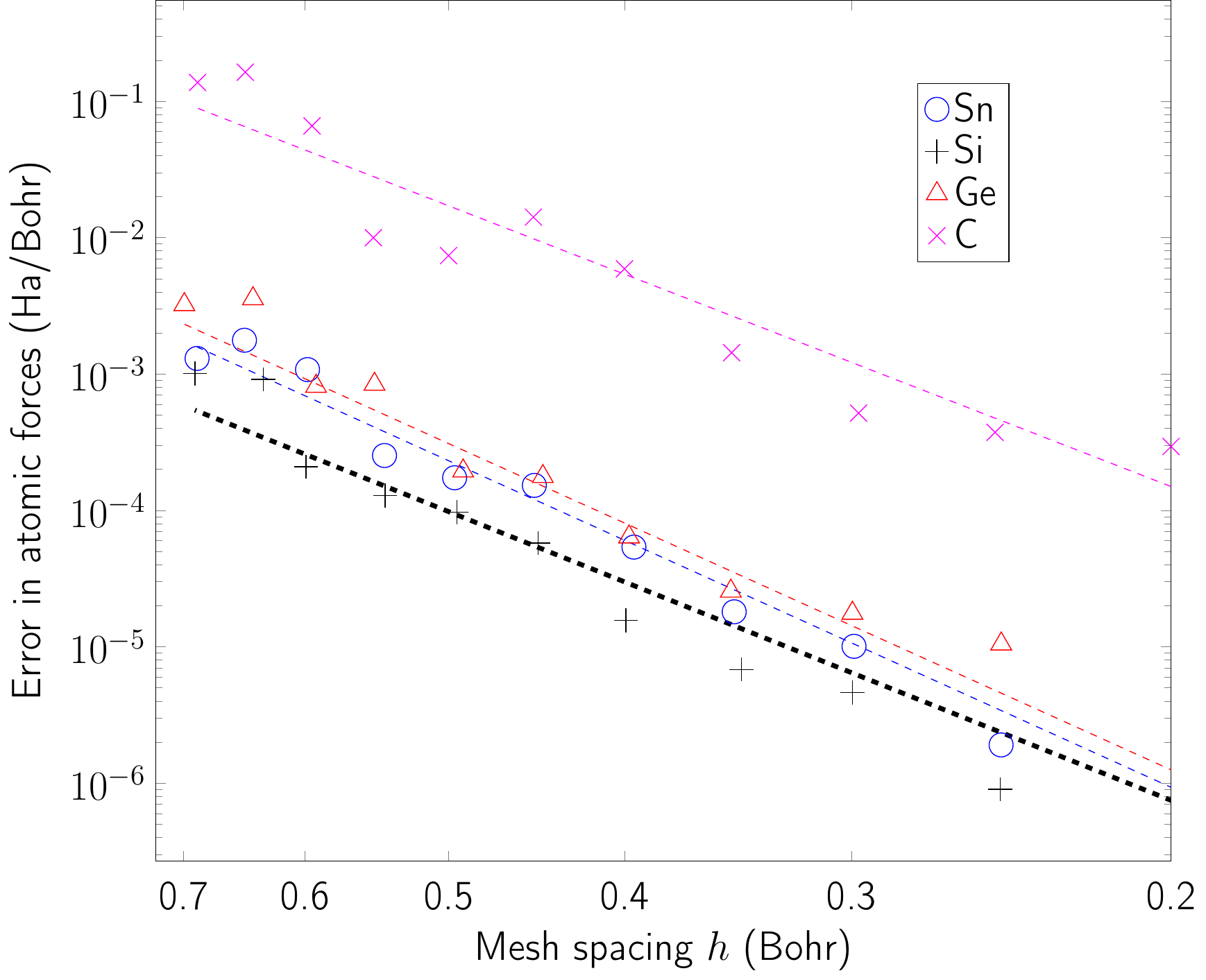}}\\
%%%%%%%%%%%%%%%%
\subfloat[Convergence of ground state energy w.r.t. reciprocal space discretization]{\includegraphics[width=0.40\textwidth]{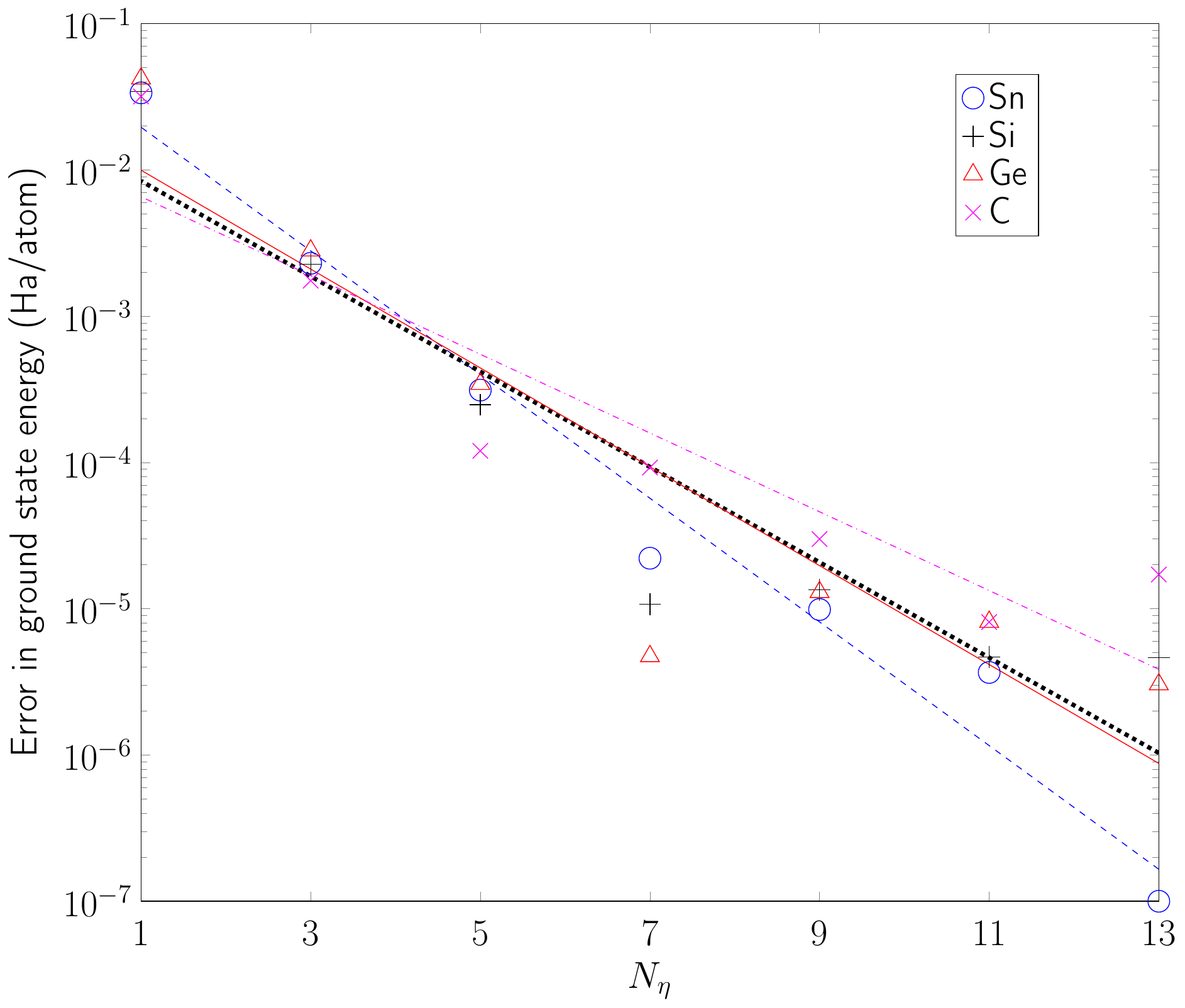}}\quad
\subfloat[Convergence of atomic forces w.r.t. reciprocal space discretization]{\includegraphics[width=0.40\textwidth]{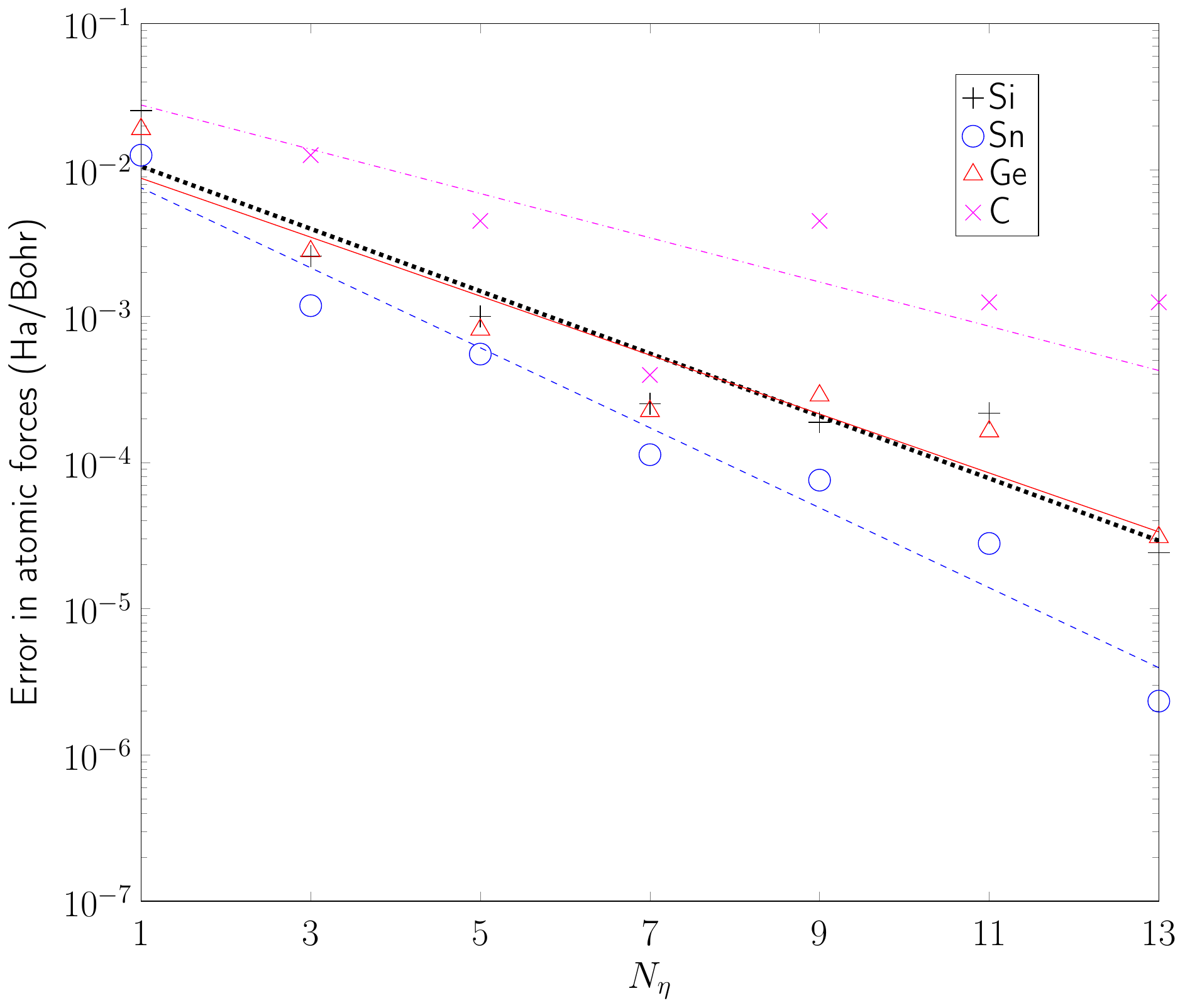}}
\caption{Convergence behavior of the numerical method for X nanotubes, with respect to real space and reciprocal space discretization parameters. The error in the forces is the magnitude of the maximum difference in all the force components on all the atoms. Dotted lines indicate straight line fits.}
\label{fig:Convergence}
\end{figure}

We begin with a discussion of the convergence properties of our numerical implementation {with respect to discretization parameters}. We choose armchair nanotubes of carbon (radius = $1.07$ nm, $\mathfrak{N} = 16$), silicon (radius = $0.97$ nm, $\mathfrak{N} = 9$), germanium (radius = $1.73$ nm, $\mathfrak{N} = 16$) and tin (radius = $0.99$ nm, $\mathfrak{N} = 8$), as example systems. We apply a twist to each of these systems by setting $\alpha$ between $0.003$ and $0.006$ (this corresponded to between $2.47$ and $8.86$ degrees/nm of imposed rate of twist). With all the other parameters of the computational method fixed to values described earlier, the only remaining quantities that can dictate the accuracy of the numerical solutions are fineness of the real and reciprocal space discretizations. Accordingly, we study the convergence behavior of the ground state energy and the atomic forces as a function of the mesh spacing $h$, and the number of reciprocal space points $N_{\eta}$ used in the calculations. The results are shown in Figure \ref{fig:Convergence}. For the mesh convergence study, we used $h = 0.15$ Bohr to evaluate the reference value while computing errors, while for studies involving convergence with respect to reciprocal space discretization, we used $N_{\eta} = 21$ as reference.

\begin{figure}[ht]
\centering
\subfloat{\includegraphics[width=0.55\textwidth]{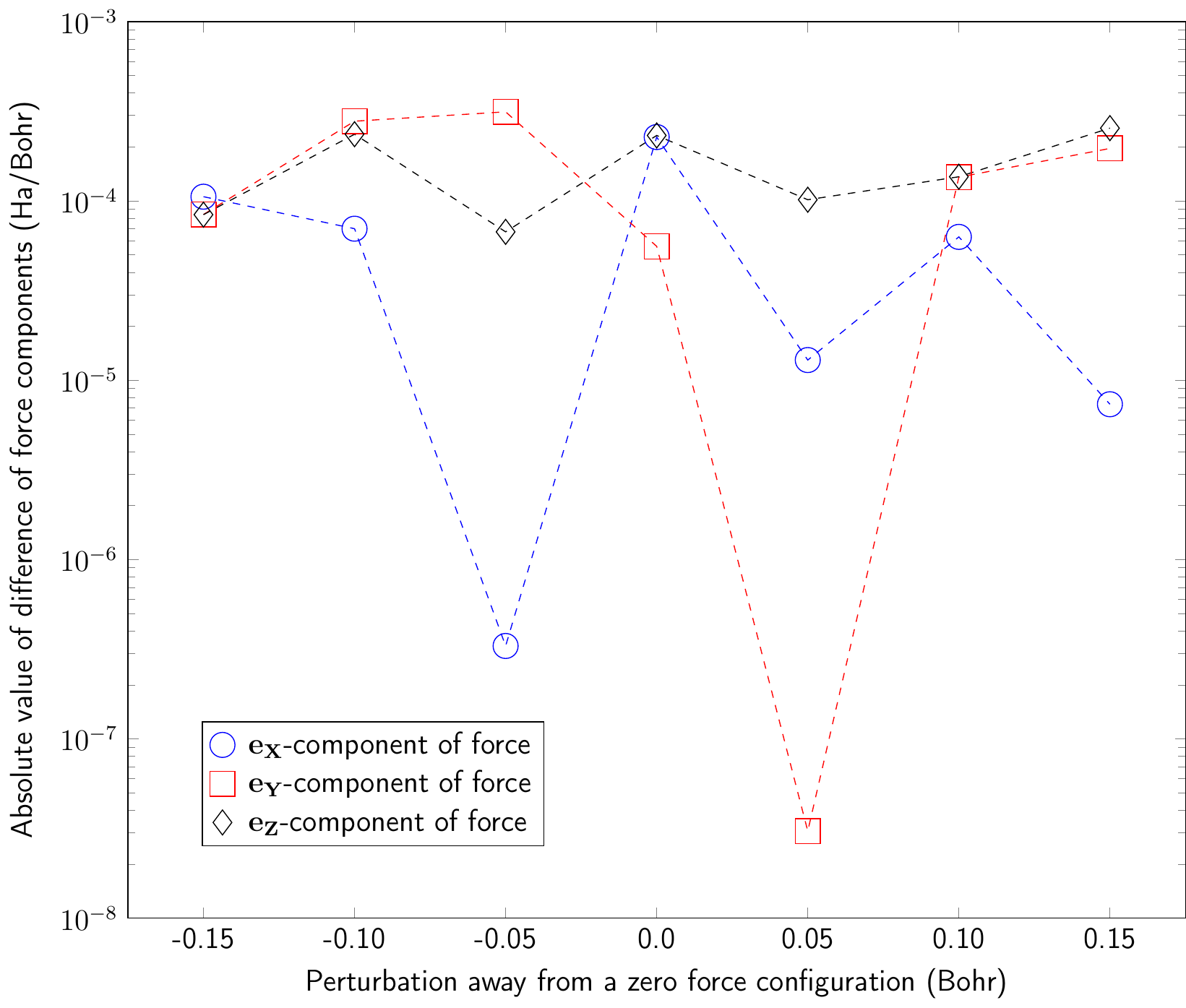}}
\caption{Consistency of the energies and forces as calculated by the Helical DFT code. For this test, a relaxed configuration of an armchair silicon nanotube (radius = $0.96$ nm) subjected to a rate of twist $= 5.67$ degree/nm was chosen. One atom in particular was then translated along $\bfe_{\bfX}, \bfe_{\bfY}$ and $\bfe_{\bfZ}$ directions (one direction at a time). The force components on the atom were obtained both via computing the derivative of a spline fit of the energy at each configuration, and direct evaluation of eq.~\ref{eq:Hellman_Feynman_Force}. The absolute value of the difference is shown in each case. The agreement is $\calO(10^{-4})$ Ha/Bohr or better in all configurations, giving us confidence the results produced by the code.}
\label{fig:energy_force_consistency}
\end{figure}

From the figures, we see that the numerical method converges systematically in each of the cases under study. By fitting straight lines to the convergence data with respect to $h$, we observed slopes between $5.5$ and $6.5$ which are somewhat lower than values observed for finite difference calculations using (untwisted) cylindrical coordinates \citep{ghosh2019symmetry}. We are also able to estimate that a mesh spacing of about $h = 0.3$ Bohr, and a value of $N_{\eta} = 15$ are more than sufficient to reach chemical accuracy thresholds in all cases (i.e., $10^{-3}$ Ha/atom in the energies and $10^{-3}$ Ha/Bohr in the atomic forces), and we used these discretization choices in structural relaxation calculations in subsequent sections. Figure \ref{fig:energy_force_consistency} shows the consistency of the forces and the energies as computed by Helical DFT at this level of discretization (i.e., numerical derivatives of the free energy per unit cell as computed via eq.~\ref{eq:electronic_free_energy}, yield the atomic force as computed via eq.~\ref{eq:Hellman_Feynman_Force}). To compute the energies and band structures of relaxed structures, we employed the finest discretization parameters that we could reliably afford within computational resource constraints. This corresponded to the choices $h=0.25$ Bohr and $N_{\eta} = 21$. 

Next, we come to a discussion of verification of the numerical method against results produced by standard, widely used plane-wave codes such as ABINIT \citep{Gonze_ABINIT_1, gonze2016recent}. As described earlier, this can be an arduous endeavor since such codes may require a very large number of atoms to be included in the periodic unit cell, in order to mimic the systems being simulated via Helical DFT. Moreover, in order to accurately accommodate the boundary conditions implemented in Helical DFT, a large amount of vacuum padding has to be often employed in the plane-wave code unit cell, and  nanotube-like structures tend to encase a large amount of vacuum as it is. These factors together can result in slow convergence of the electrostatics problem, as well as, poor conditioning of the systems of equations being solved by the plane-wave code. The latter issue, in turn, leads to SCF convergence problems which tend to worsen if calculations at high accuracies are required (i.e., upon using a large value of $E_{\text{cut}}$ for the plane-wave code). With these factors in mind, we chose the  armchair carbon and silicon nanotube systems described above for comparison against ABINIT. For the former, we did not prescribe any twist and use a $64$ atom unit cell. For the latter, we prescribed a twist of $\alpha = 0.1$, and used a $360$ atom unit cell. While dealing with these systems in ABINIT, periodicity was naturally enforced along the Z axis, Dirichlet boundary conditions were enforced along the X and Y axes by padding with a large amount of vacuum, and an SCF preconditioner ($\mathsf{diemac}$ option in ABINIT) was used to deal with instabilities associated with spatial inhomogeneities in the periodic unit cell. Helical DFT was made to use four atom unit cells for both examples. For each of these model systems, we observed that the energies (in Ha/atom) and the forces (in Ha/Bohr),  from ABINIT and Helical DFT agreed with each other to $\mathcal{O}(10^{-4})$, thus giving us confidence in the accuracy of the results produced by our method.\footnote{{Convergence and accuracy properties of the Helical DFT code have also been discussed in our earlier contribution \citep{banerjee2021ab}. However, the materials systems considered in this work are different from \citep{banerjee2021ab}, and so, we include this discussion here for the sake of a self-contained presentation. In particular, carbon is known to be associated with somewhat hard pseudopotentials and these studies helped us determine the appropriate discretization parameters for this element, so that numerically accurate predictions of electromechanical properties of carbon nanotubes could be made.}}

\begin{figure}[ht!]
\centering
\subfloat[SCF iteration wall times (normalized). Y-axis is logarithmic. ]{\includegraphics[width=0.99\textwidth]{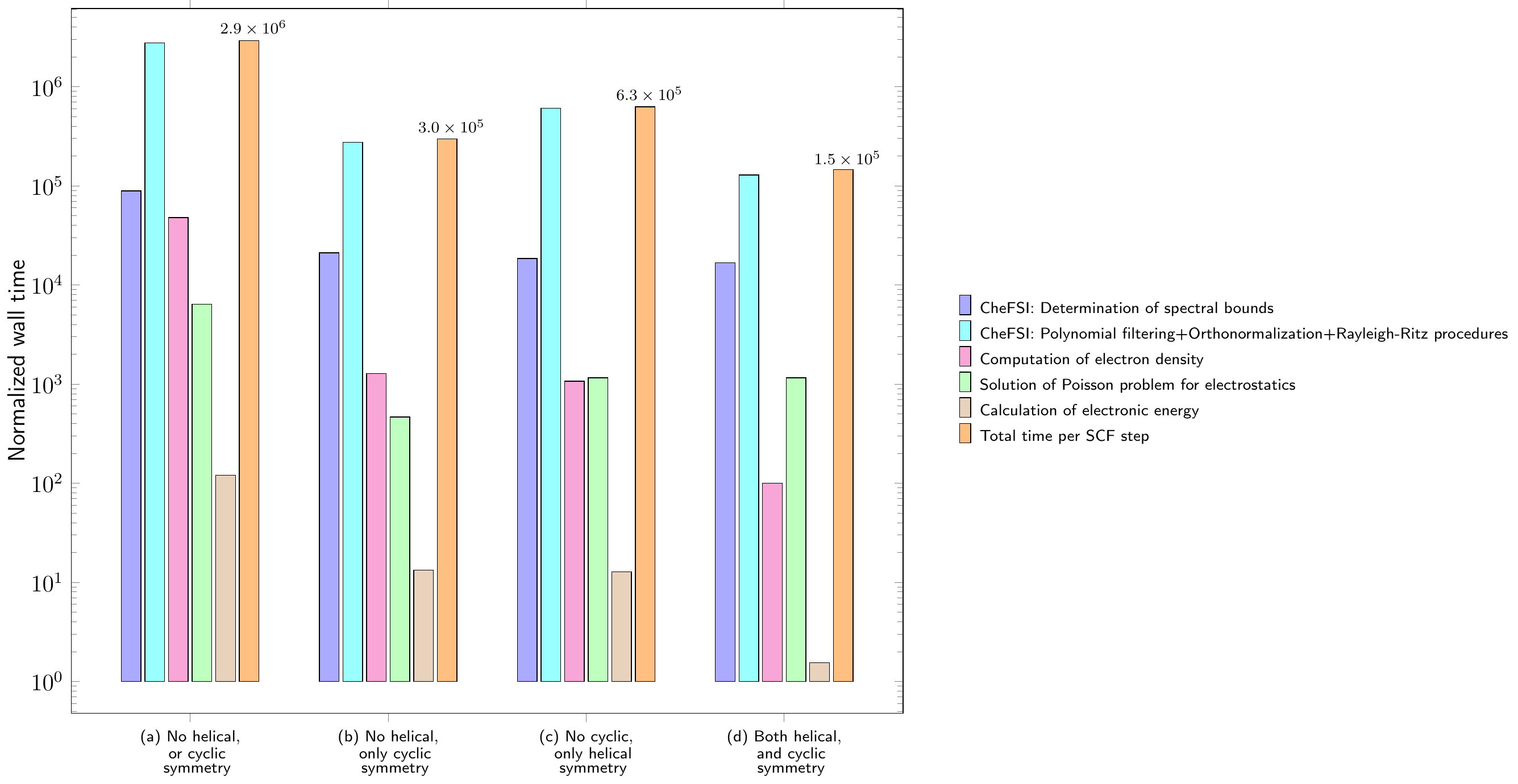}}\\
\subfloat[ Force calculation wall times (normalized). Y-axis is logarithmic. ]{\includegraphics[width=0.99\textwidth]{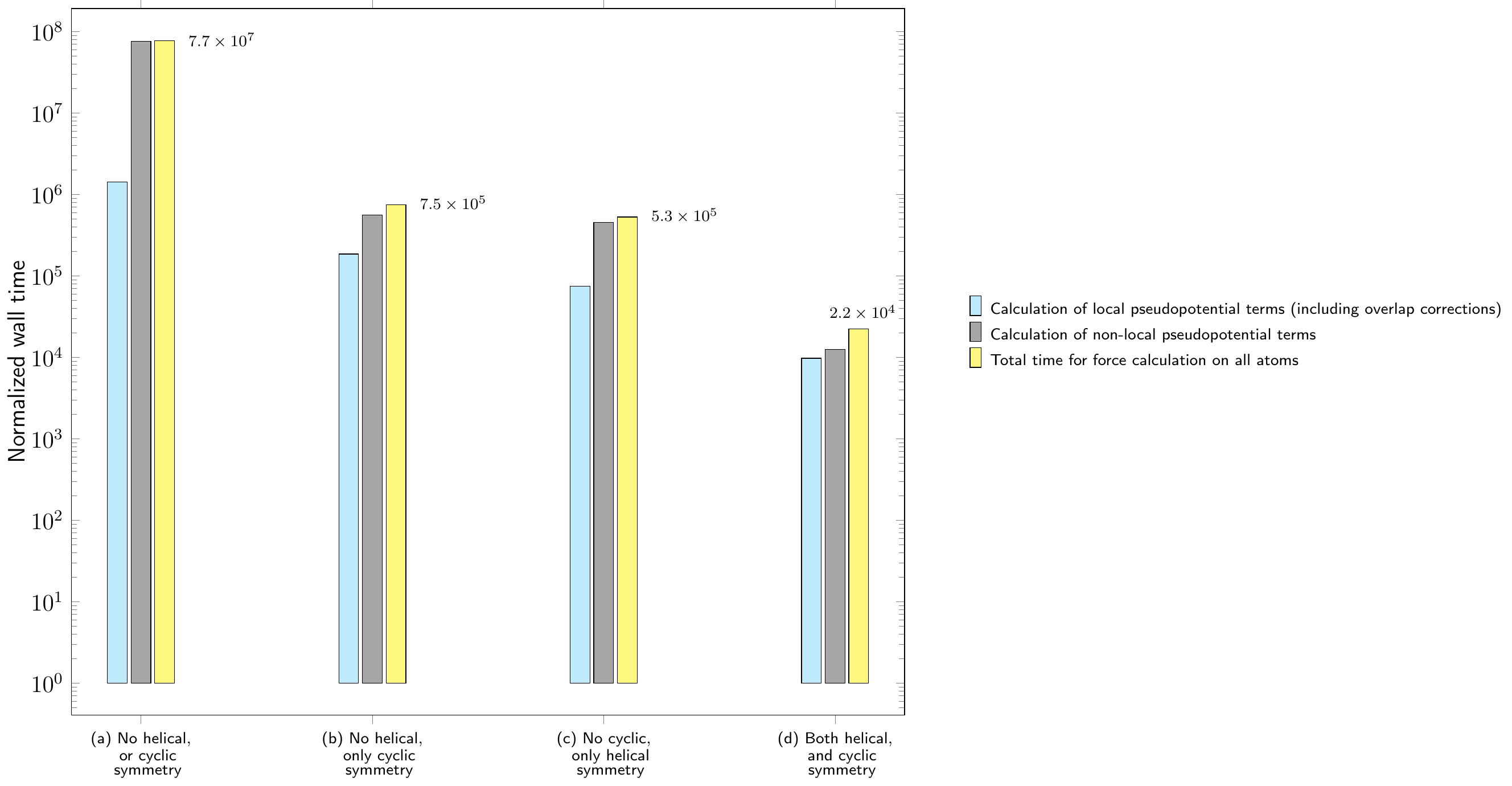}}
\caption{Influence of symmetry adaptation on computational wall times (single core). Numbers appearing in the plots above indicate the total time per SCF step and the total time for computation of the forces (both quantities normalized).}
\label{fig:symmetry_wall_times}
\end{figure}

Based on the above tests, we were also able to observe that even a well optimized  plane-wave code like ABINIT can take up to orders of magnitude more in simulation time (measured in c.p.u. hours) compared to Helical DFT, when simulations of nanotube structures (particularly, ones with imposed twist) are desired. This makes our computational method a particularly attractive choice in the first principles characterization of such systems. The relative efficiency of our method stems from the use of a coordinate system and a computational domain that are well adapted to the geometry of the twisted structure, and also from the appropriate use of symmetry. To highlight the latter aspect, we considered again the silicon nanotube system subjected to a twist of $\alpha = 0.1$. We used Helical DFT to calculate the ground state electronic structure of this system by considering the following four equivalent scenarios: 
\begin{enumerate}[(a)]
\item No helical or cyclic symmetries ($360$ atom unit cell with $\alpha = 0$ and periodicity along $\bfe_{\bfZ}$, $N_{\eta} = 1$ and only $\nu = 0$ considered).
\item Only cyclic symmetries ($40$ atom unit cell with $\alpha = 0$ and periodicity along $\bfe_{\bfZ}$, $N_{\eta} = 1$ and $\nu = 0,1,\ldots,8$ considered).
\item Only helical symmetries ($36$ atom unit cell with $\alpha = 0.1$, $N_{\eta} = 10$, and only $\nu = 0$ considered).
\item Both cyclic and helical symmetries considered ($4$ atom unit cell with $\alpha = 0.1$, $N_{\eta} = 10$ and $\nu = 0,1,\ldots,8$ considered).
\end{enumerate}
The single core wall times required for each SCF step, and computation of the atomic forces at the end of the SCF iterations are compared in Figure \ref{fig:symmetry_wall_times}. 

From these plots, it is clear that the SCF wall time is approximately $20$ times lower for the case with full symmetry adaptation, when compared to the case in which no cyclic or helical symmetries were used. Even more drastically, the computational wall time for the calculation of the force is about 3 orders of magnitude lower for the former case, when compared to the latter.  These computational advantages tend to be even more dramatic for simulations in which the angle of twist is relatively low (e.g. $\alpha = 0.0005$ to $0.005$), and such cases tend to arise routinely while probing the torsional response of the nanotubes in the linear elastic regime, as described in the next section. 

\begin{figure}
\centering
\subfloat{\includegraphics[width=0.60\textwidth]{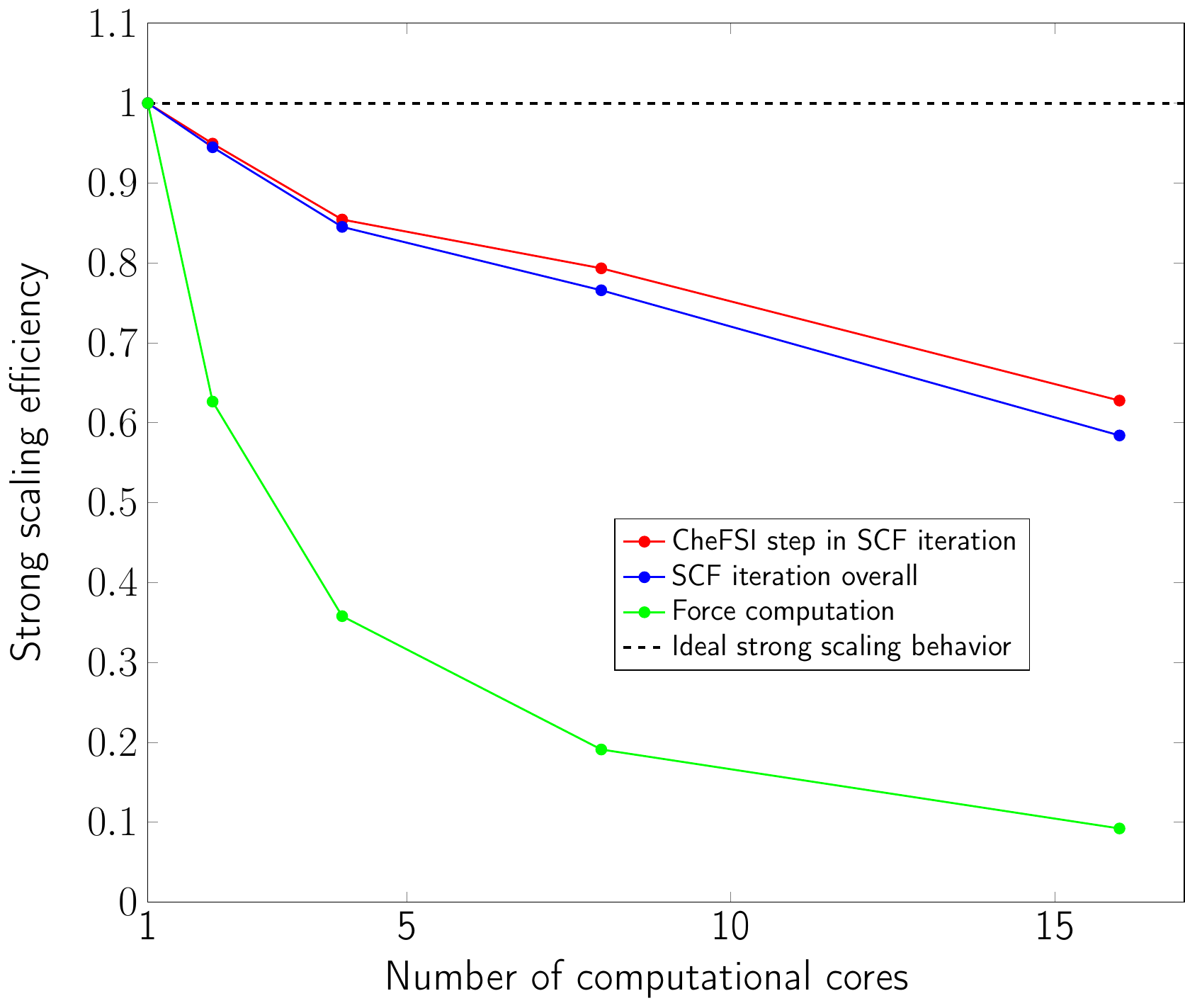}
}
\caption{\footnotesize{Strong scaling behavior of the Helical DFT code.}}
\label{fig:strong_scaling}
\end{figure}

Finally, we show in Figure \ref{fig:strong_scaling} the strong scaling behavior of the numerical implementation. We use case (d) described above for this study. We see that up to $16$ computational cores, the code has a strong scaling efficiency of about $60\,\%$. This follows the strong scaling efficiency of the CheFSI step closely, since this forms the dominant computational cost in every SCF step (see Figure \ref{fig:symmetry_wall_times}(a)). The scaling of the force computation step is far worse, dropping to about $10\,\%$ at 16 cores.  In general, the scaling behavior is expected to improve  for problems with a larger number of $\eta$ and $\nu$ points (e.g. for simulations of nanotubes of large diameter) since the current version of the code only uses parallelization over different values of $\eta$ and $\nu$. Improvement of the scaling behavior of the code, particularly by use of domain decomposition and band parallelization techniques in conjunction with the MATLAB Parallel Server framework (to enable deployment over distributed memory computers) is the scope of future work.\footnote{{As pointed out to us by an anonymous reviewer, these scaling performance figures suggest that the Helical DFT  code is heavily memory bound in the regime in which the data was collected, and therefore, perhaps a better metric might be to estimate the percentage of total peak performance. However, estimating this number involves calculation of the number of floating point operations performed during the operation of the code, and this can be somewhat challenging due to the use of both MATLAB and} {C source code. Furthermore, due to the lack of internal MATLAB routines for estimating flops, only tools developed by the MATLAB user community can be employed. We ran tests using the Lightspeed suite \citep{Lightspeed} and we focused only on one the core routines of the code, i.e., the matrix vector-product implementation. Our tests suggest that on the  $18$-core Intel Xeon Gold 5220 processor, the core routines reached about $10.7$ \% of the peak performance (peak performance data obtained from the Intel website \citep{Intel_Web}), which is not entirely unexpected due to the large amount of data movement operations associated with the calculation \citep{Lecture_Gropp}.
}}
\subsection{Computation of torsional stiffness from first principles}
\label{subsec:mechanical_simulations}
\begin{figure}[ht]
\centering
\subfloat{\includegraphics[width=0.60\textwidth]{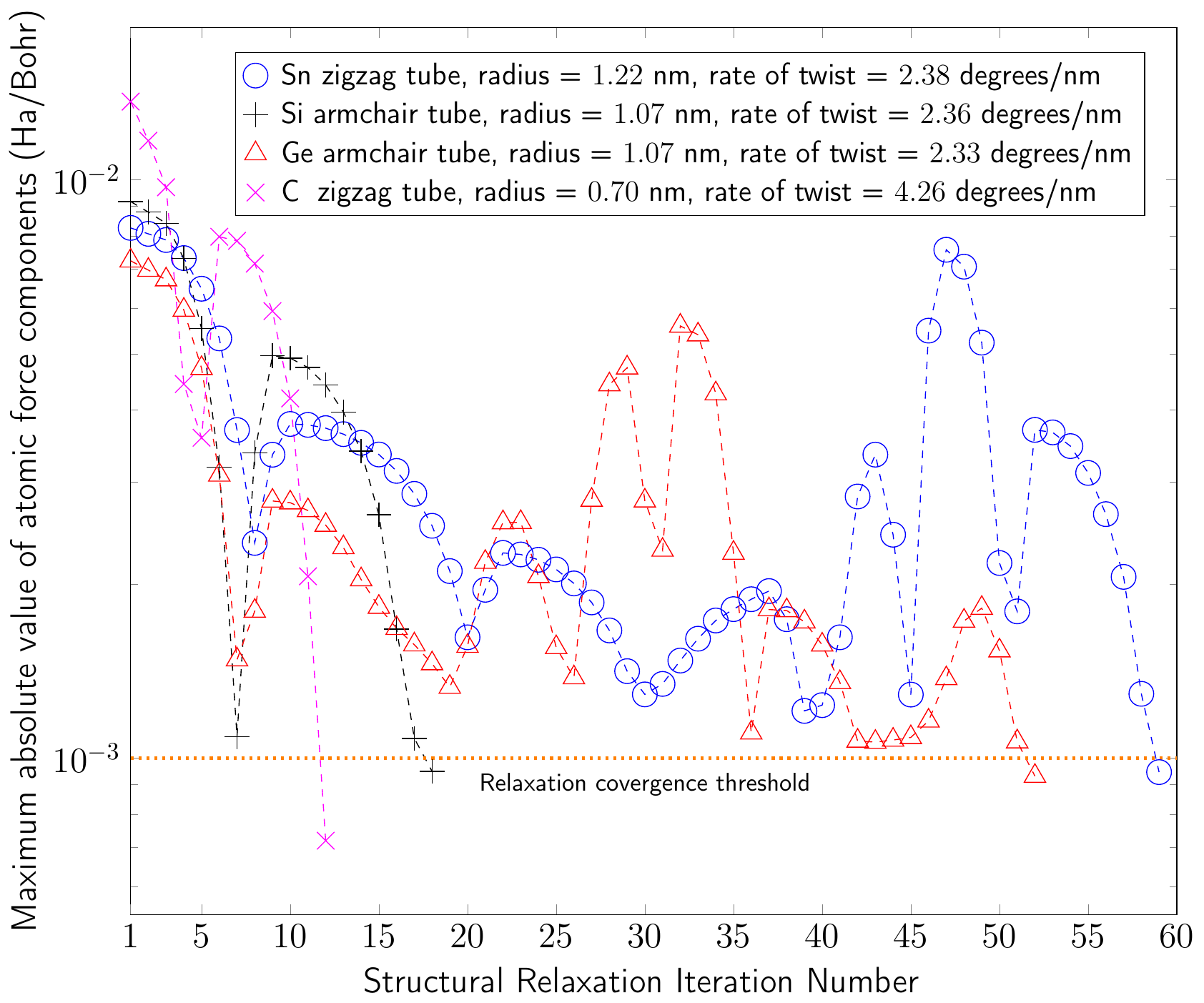}}
\caption{Examples of \textit{ab initio} structural relaxation of twisted structures using Helical DFT.}
\label{fig:Fire_Relax}
\end{figure}

\begin{figure}[ht]
\centering
\subfloat[Zigzag Sn nanotubes]{\includegraphics[width=0.48\textwidth]{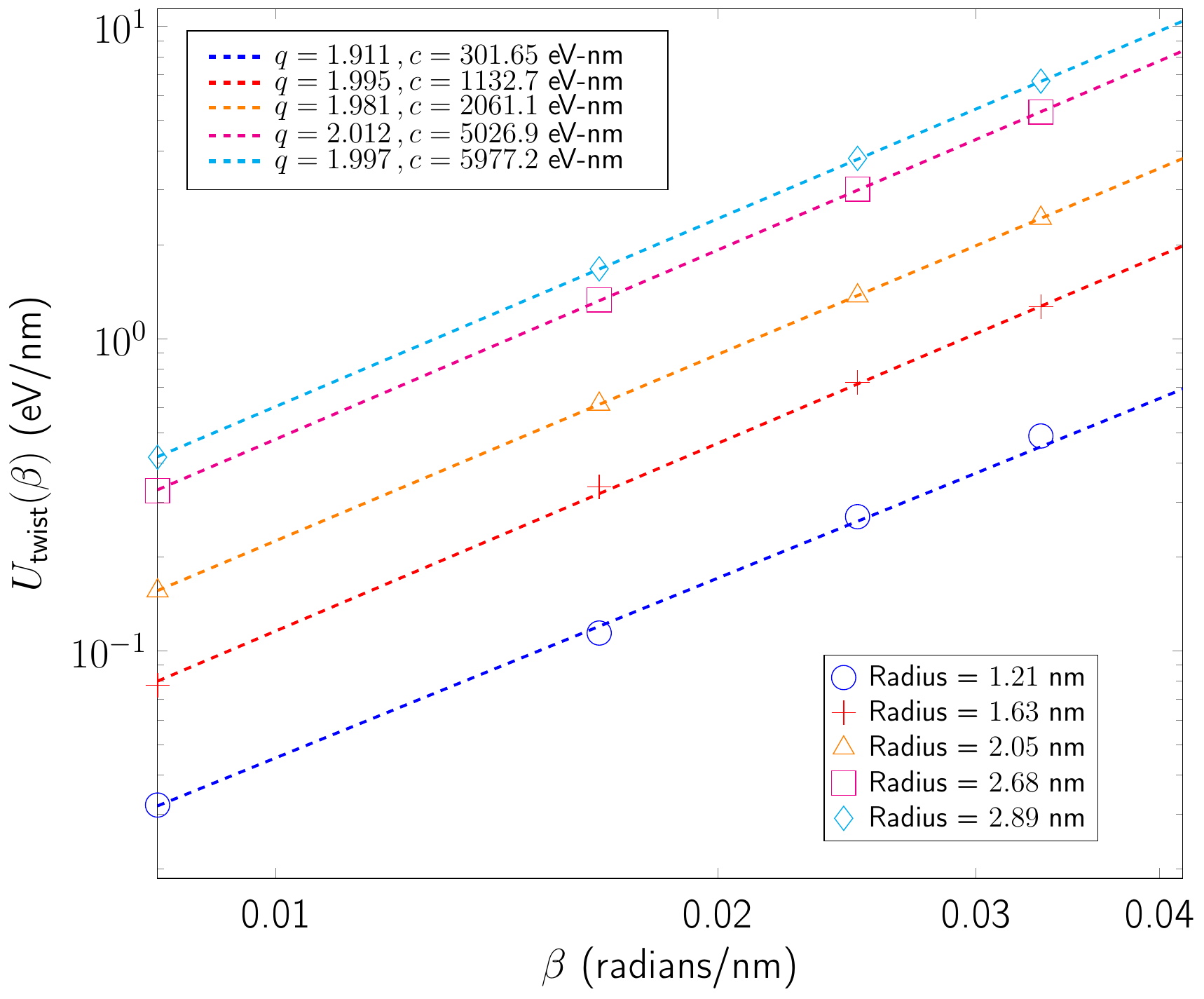}}\quad
\subfloat[Armchair C nanotubes]{\includegraphics[width=0.48\textwidth]{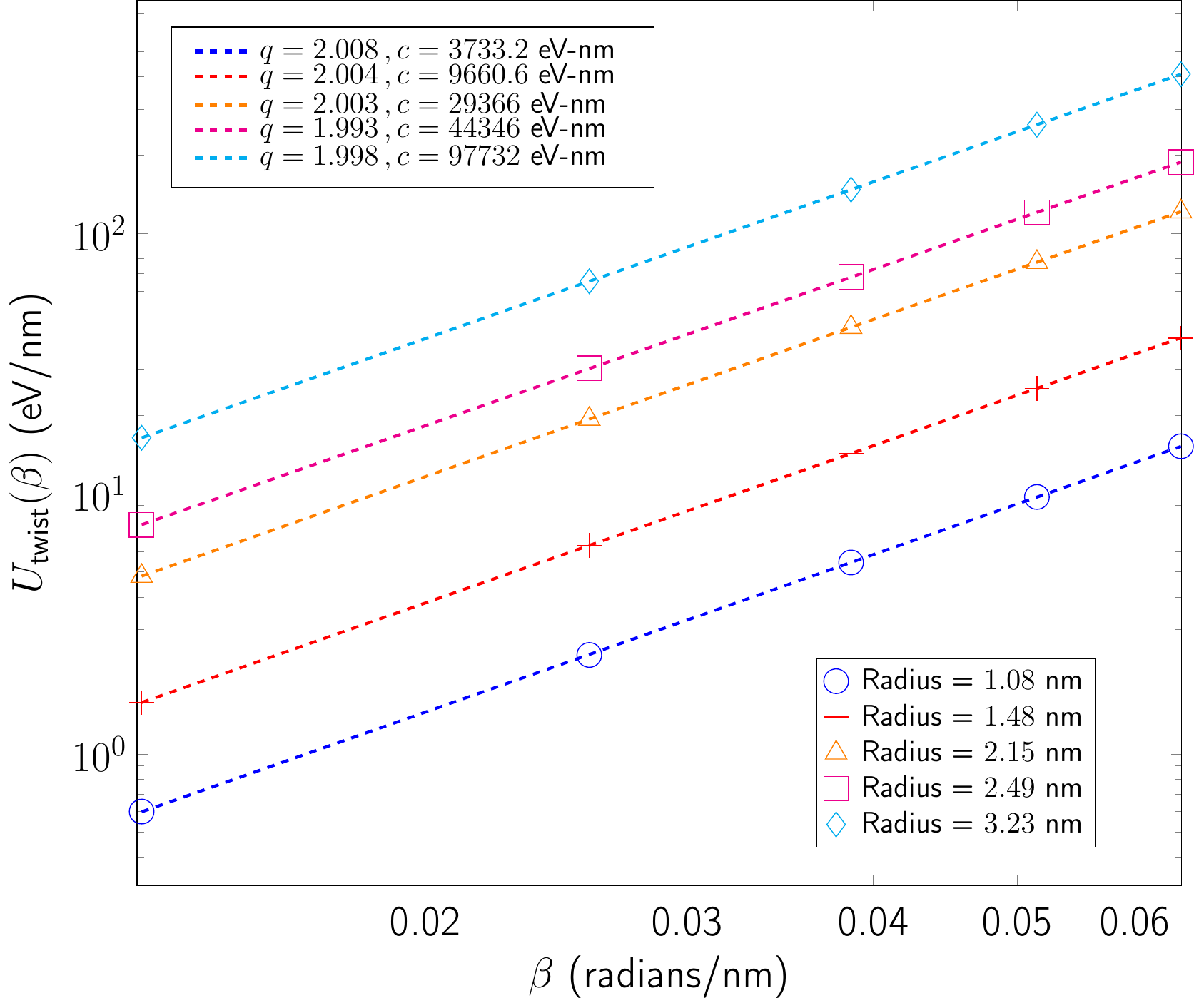}}
\caption{Dependence of twist energy per unit length on angle of twist per unit length (i.e., rate of twist)  for two representative classes of nanotubes. Dotted lines indicate straight line fits of the data to an ansatz of the form $U_{\text{twist}}(\beta) = c\times\beta^q$.}
\label{fig:linear_twist}
\end{figure}

\begin{figure}[ht]
\centering
\subfloat[Zigzag nanotubes]{\includegraphics[width=0.48\textwidth]{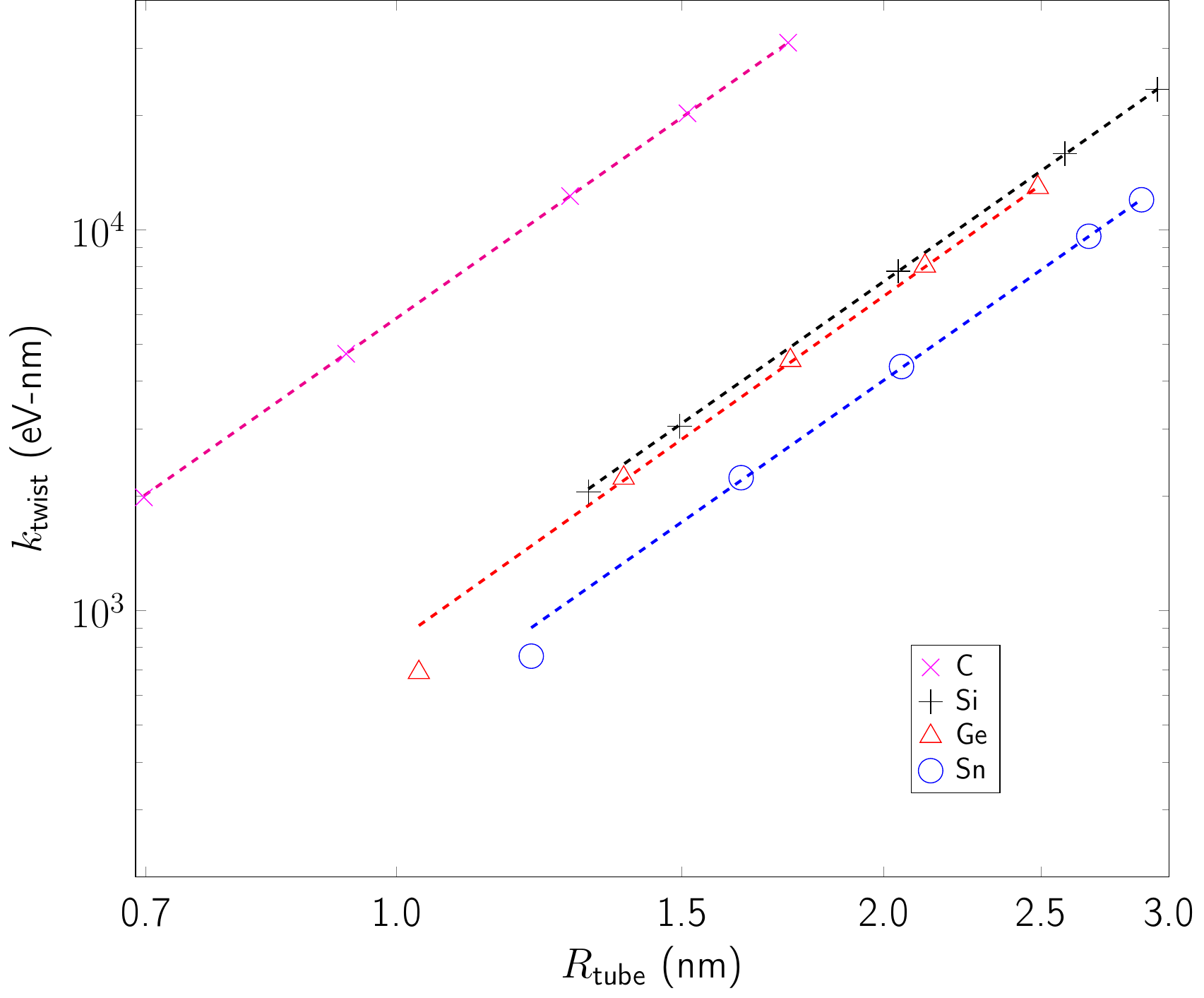}}\quad
\subfloat[Armchair nanotubes]{\includegraphics[width=0.48\textwidth]{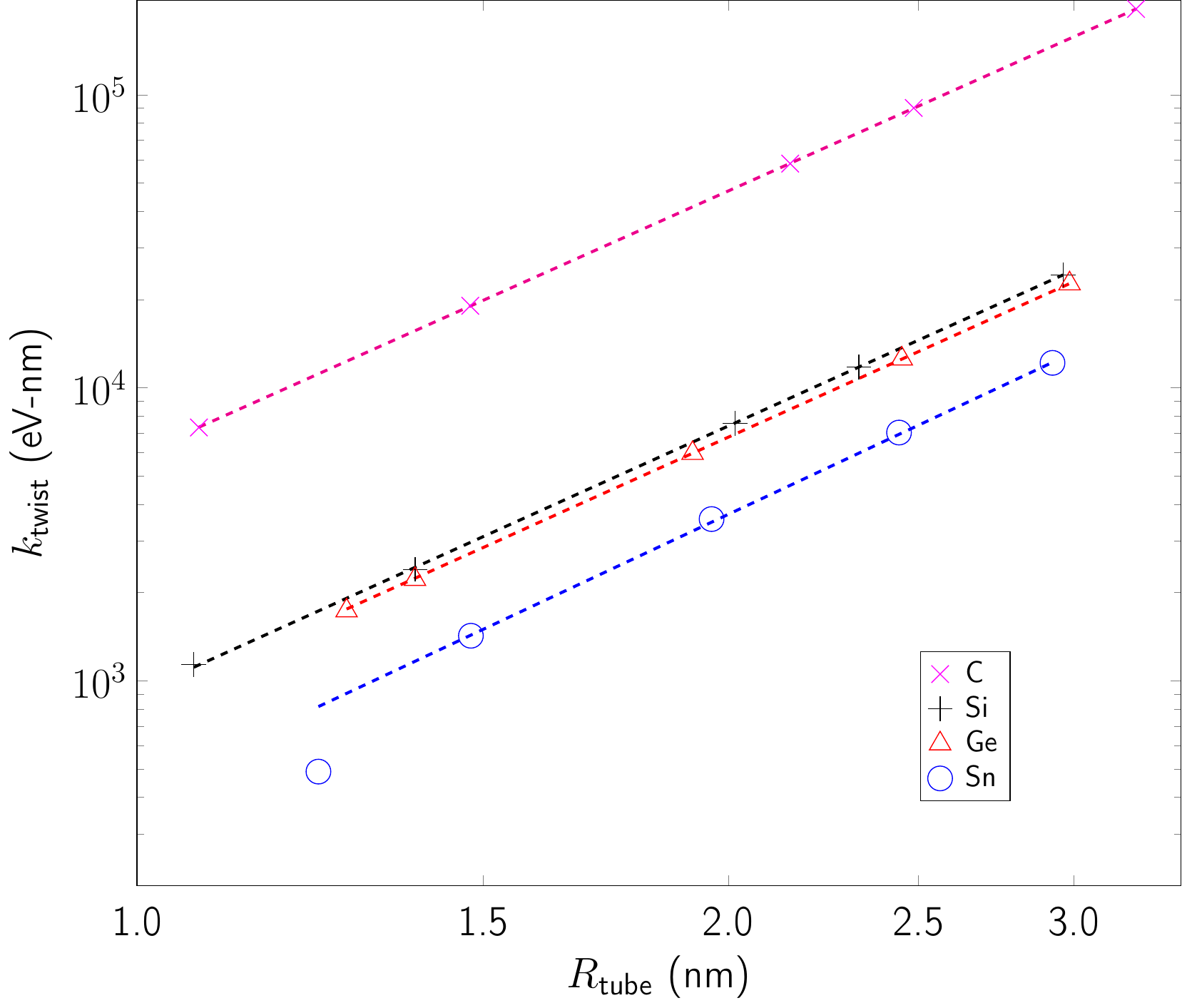}}
\caption{Variation of torsional stiffness $k_{\text{twist}}$ (eV-nm) with tube radius $R_{\text{tube}}$ (nm). Both axes are logarithmic. Dotted lines correspond to $k_{\text{twist}} = \kappa\times R_{\text{tube}}^p$ with fitted parameters for $\kappa$ ($\text{eV/nm}^2$) and $p$ (Table \ref{Table:kappa_p}).}
\label{fig:twist_radius}
\end{figure}

We now turn to demonstrations of the use of our computational method for evaluation of materials properties from first principles. We first concentrate on the mechanical response and evaluate the torsional stiffness of the X nanotubes in the linear elastic regime, \textit{ab initio}.  We choose $9$ to $10$ nanotubes of each material, about half of which are of zigzag type and the other half armchair. The nanotubes all had radii in the range $1$ to $3$ nm, approximately.  To carry out these simulations, we choose a four atom unit cell for the untwisted nanotube in each case, and perform structural relaxation using the FIRE algorithm \citep{bitzek2006structural} till all force components on all the atoms in the simulation cell dropped below $10^{-3}$ Ha/Bohr. We then successively increase $\alpha$ to impose twist, and in each case re-perform structural relaxation (see Figure \ref{fig:Fire_Relax} for some examples of the relaxation procedure). 
To avoid the appearance of torsional instabilities, we ensured that the prescribed rate of twist on the system was less than about $4.5$ degrees per nanometer \citep{Dumitrica_James_OMD}, and this corresponded to choosing $\alpha$ between $0.0005$ and $0.005$. We express the amount of applied twist per unit length of the tube (i.e., the \textit{rate of twist}) as $\displaystyle \beta = \frac{2\pi\alpha}{\tau}$, and compute the twisting energy per unit length of the structure as the difference in the ground state free energy per unit fundamental domain between the twisted and untwisted configurations (after atomic relaxation is carried out in both cases), i.e.:
\begin{align}
U_{\text{twist}}(\beta) = \frac{\mathfrak{N}}{\tau} \bigg({\calF}_{\substack{\text{Ground}\\ \text{State}}}(\calP^{**} , \calD, \calG|_{\beta}) - {\calF}_{\substack{\text{Ground}\\ \text{State}}}(\calP^{*}  , \calD, \calG|_{\beta=0})\bigg)\,.
\end{align}
Here, $\calG|_{\beta}$ and $\calG|_{\beta = 0}$ denote the symmetry groups associated with the twisted and untwisted structures, respectively. Also, $\calP^{**}$ and $\calP^{*}$ denote the collections of positions of the atoms in the fundamental domain, after relaxation in each case. For each of the nanotubes, we verified that mechanical response was in the linear regime, by fitting $U_{\text{twist}}(\beta)$ to a function of the form $U_{\text{twist}}(\beta) = c\times\beta^q$ and observing that $q \approx 2.0$ holds. We show a few examples in Figure \ref{fig:linear_twist}.
Next, using the above data, we estimated the twisting stiffness of each nanotube, defined as:
\begin{align}
k_{\text{twist}} = \hpd{U_{\text{twist}}(\beta)}{\beta}{2}\bigg\rvert_{\beta = 0}\,.
\end{align}
For each category of nanotube (i.e., armchair or zigzag, and type of material), we then studied the variation of $k_{\text{twist}}$ with the nanotube radius (computed as the average of the radial coordinates of all atoms in the fundamental domain), by using a fit of the form:
\begin{align} 
k_{\text{twist}} = \kappa\times R_{\text{tube}}^p\,.
\label{eq:stiffness_eqn}
\end{align}
The results from this procedure are shown in Figure \ref{fig:twist_radius} and the values of $\kappa$ and $p$ obtained in each case are displayed in Table \ref{Table:kappa_p}. Note that generation of this torsional response data required hundreds of individual simulations, which would not have been possible without the use of a specialized computational method such as the one presented here.
\begin{table}[htb]
\centering
\begin{tabular}{c c c c}
\hline
X & Type &  $\kappa$ ($\text{eV/nm}^{p-1}$) & $p$\\
\hline
C & Armchair  & $5905.76$ & $2.992$   \\ 
C & Zigzag  & $5859.81$ & $2.990$   \\ 
Si & Armchair  & $909.34$ & $3.026$   \\ 
Si & Zigzag  & $923.78$ & $2.986$   \\ 
Ge & Armchair  & $837.12$ & $3.018$   \\ 
Ge & Zigzag  & $829.57$ & $3.012$   \\ 
Sn & Armchair  & $418.23$ & $3.144$   \\ 
Sn & Zigzag  & $508.83$ & $2.984$   \\ 
\hline 
\end{tabular}
\caption{Torsional stiffness parameters for the $X$ nanotubes, with $k_{\text{twist}} = \kappa\times R_{\text{tube}}^p$}
\label{Table:kappa_p}
\end{table} 

A few comments are in order at this stage. First, we observe that the value of the exponent $p$ is nearly $3$ in every case. This suggests that the torsional response of the tubes is consistent with linear elasticity theory, in which $k_{\text{twist}}$ for a thin elastic tube with unit length, radius $R_{\text{tube}}$, thickness $t$, and shear modulus $G$ can be expressed as $Gt\pi R_{\text{tube}}^3$ \citep{timoshenko1968elements}. From this, it is possible to estimate the thickness-normalized shear modulus (i.e., $Gt$) of the Xene sheets as $\kappa / \pi$. Second, by comparing the different values of $\kappa$, we see that they span an order of magnitude across the different elements. In particular, for a given radius, $k_{\text{twist}}$ is the highest for carbon nanotubes and the lowest for those of tin, while nanotubes of silicon and germanium have intermediate values of this quantity close to each other. Third, for each material, the torsional response is quite similar in the armchair and zigzag directions with variations less than about $1.5\%$, except for the case of tin, in which case the variation is more substantial. This largely isotropic torsional response for the Xene nanotubes is quite distinct from the bending response of their sheet counterparts, which show strong anisotropic behavior that is correlated with the value of the normalized buckling parameter (i.e., $\delta/a$) for each material \citep{ghosh2019symmetry}. Our findings on the mechanical response of carbon nanotubes under torsion are broadly consistent with earlier studies for this material that used empirical potentials or tight-binding calculations \citep{Dumitrica_James_OMD, CNT_Dumitrica}, although the value of $\kappa$ reported here is lower from \citep{Dumitrica_James_OMD}, where Tersoff potentials were used \citep{tersoff1988new}.

Finally, we mention in passing, the effects of atomic relaxation. In general, if relaxation is not performed after the imposition of twist, the value of $k_{\text{twist}}$ for the system tends to be higher. The degree of variation can be quite different depending on the material involved. For carbon nanotube systems, we observed that $k_{\text{twist}}$ for an unrelaxed system was typically higher by a factor of about $1.08$, whereas for silicon nanotubes, this factor had the higher value of about $1.38$. Generally, these higher values of $k_{\text{twist}}$ also imply higher values of $\kappa$ by the same factors,  although the value of the exponent $p$ continues to be about $3$, when the fitting in eq.~\ref{eq:stiffness_eqn} is used.
\subsection{Investigation of electronic properties of nanotubes undergoing torsional deformation}
\begin{figure}[ht]
\centering
\subfloat{\scalebox{0.55}
{
\includegraphics[trim={2cm 1cm 3cm 2cm}, clip, width=\textwidth]{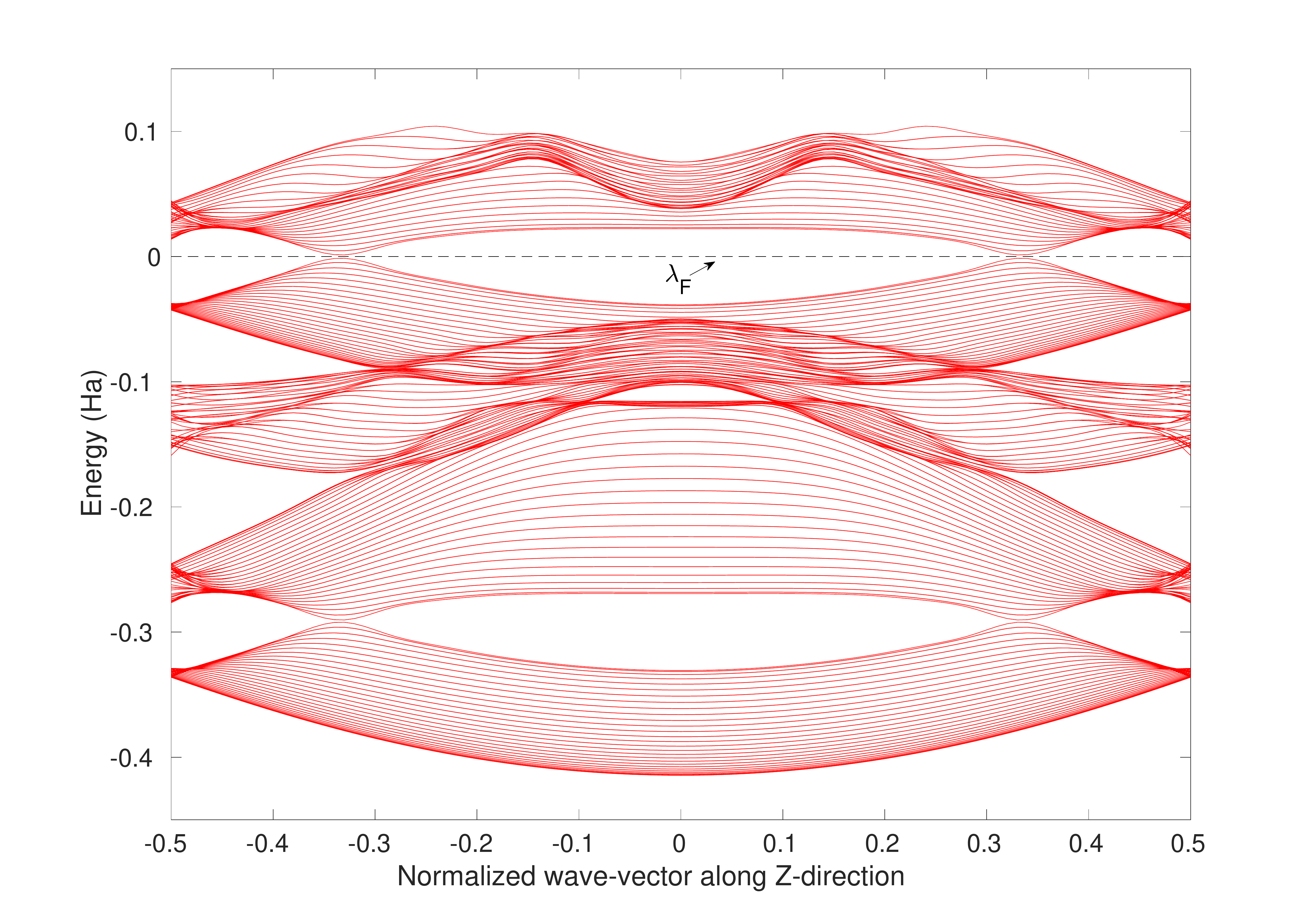}
}}
\caption{Conventional band diagram for an untwisted armchair Si nanotube of radius $2.96$ nm.}
\label{fig:band_diagram_conventional}
\end{figure}

\begin{figure}[ht]
\centering
\subfloat[2D surface plot of the eigenvalues $\lambda_j(\eta,\nu)$, for $j=8$.]{\scalebox{0.55}
{
\includegraphics[trim={1cm 1cm 0.5cm 1cm}, clip, width=\textwidth]{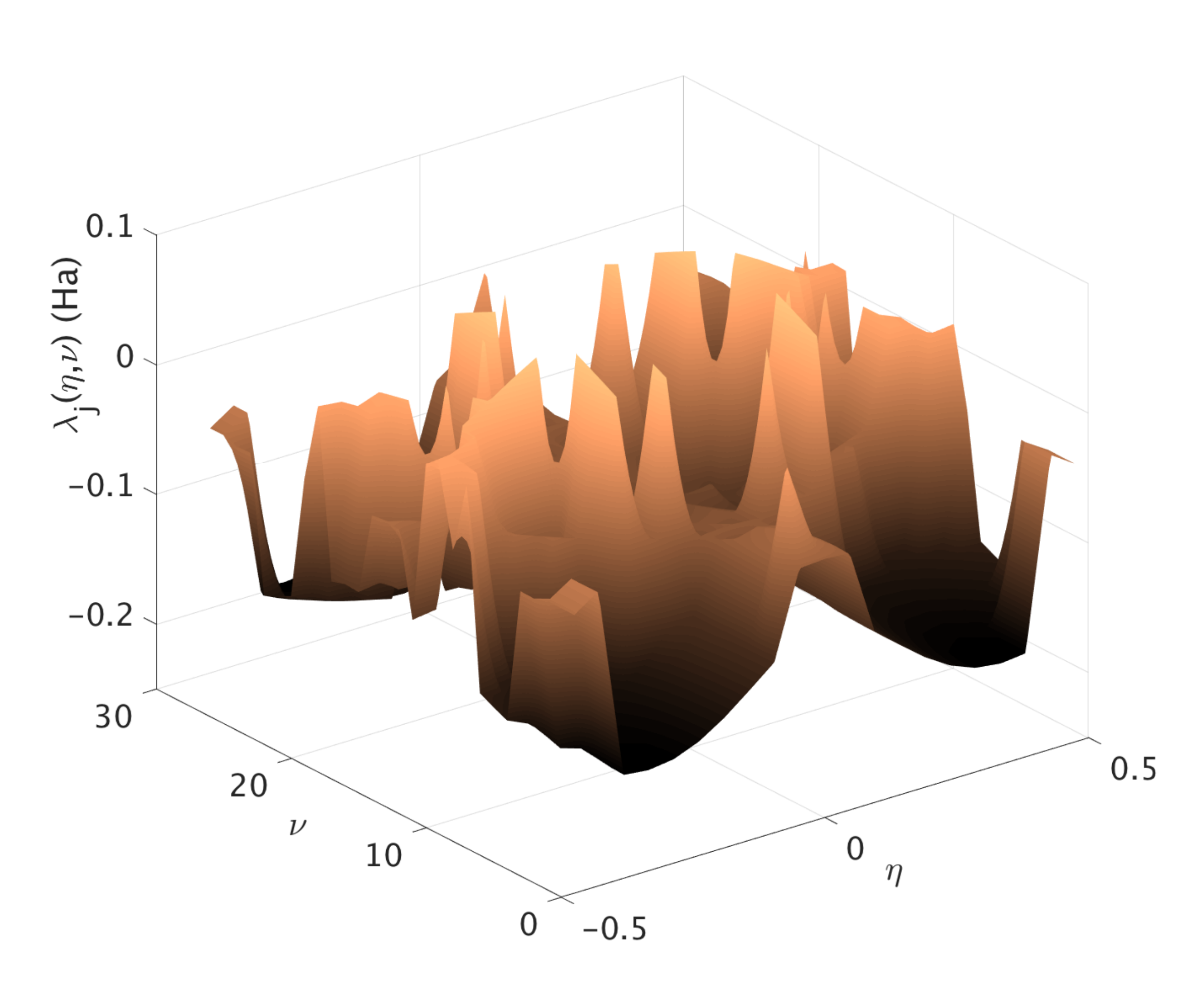}
}}\\
\subfloat[Symmetry adapted band diagram in $\eta$, along $\nu = 0$. Plot is symmetric about $\eta = 0$ due to time reversal symmetry.]{\includegraphics[width=0.48\textwidth]{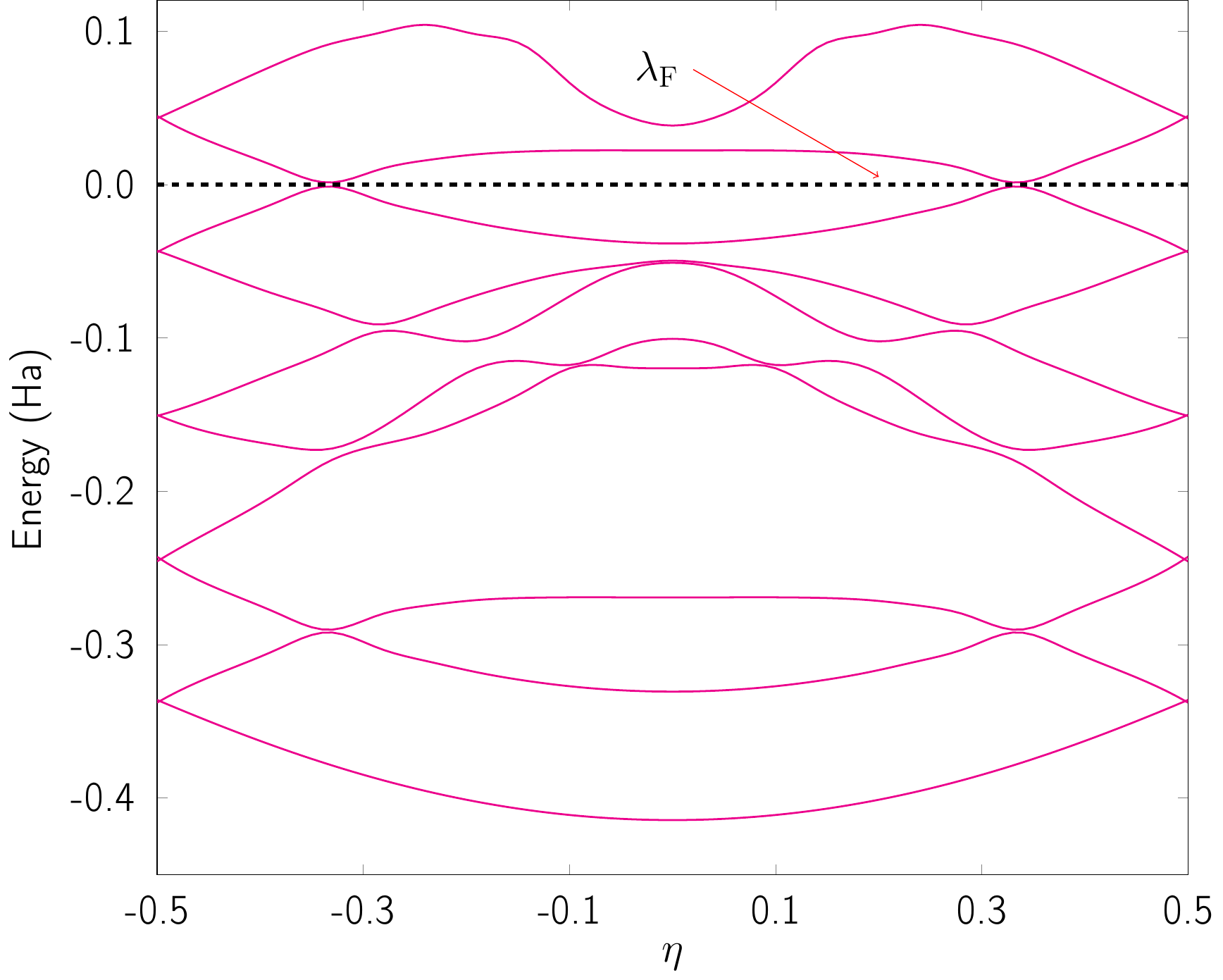} 
}$\;$
\subfloat[Symmetry adapted band diagram in $\nu$, along $\eta = 0$.  Plot is symmetric about $\nu  = 14$ due to time reversal symmetry.]{\includegraphics[width=0.48\textwidth]{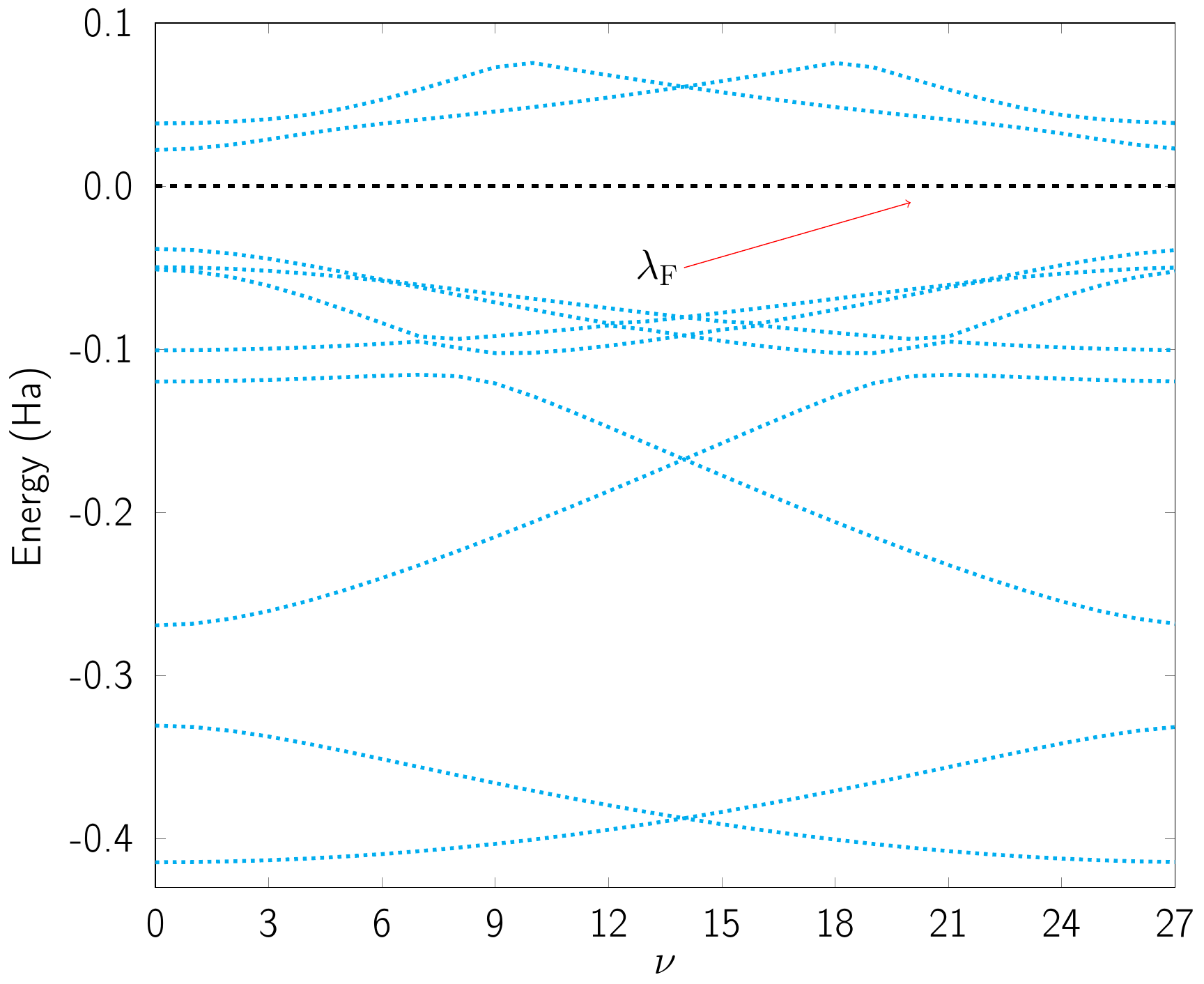}
}
\caption{Visualization of electronic states for the untwisted armchair Si nanotube (radius $2.96$ nm) using results from Helical DFT. Compare this to Figure \ref{fig:band_diagram_conventional}.}
\label{fig:electronic_states_plots}
\end{figure}

We now discuss the variation in electronic properties of nanotubes as they are subject to twisting. Due to the ability of our computational method to use symmetries connected with the system, electronic band-diagrams along both $\eta$ and $\nu$ can be obtained from Helical DFT. Moreover, the eigenvalues $\lambda_j(\eta,\nu)$ as $j$ is held constant and $\eta, \nu$ are varied, can be plotted as a two-dimensional surface. Since $\eta$ and $\nu$ serve to label the set of characters, and are natural quantum numbers for twisted structures, they serve to provide a clean and intuitive interpretation of the electronic states of the system, and allow easy identification of  the size and type of band-gaps. In contrast, the traditional band diagram for a quasi-one-dimensional system using a periodic method can be far more complicated, even for an untwisted structure. We show some examples of this contrast in Figures \ref{fig:band_diagram_conventional} and \ref{fig:electronic_states_plots}. 

\begin{figure}
\centering
\subfloat
{\includegraphics[width=0.55\textwidth]{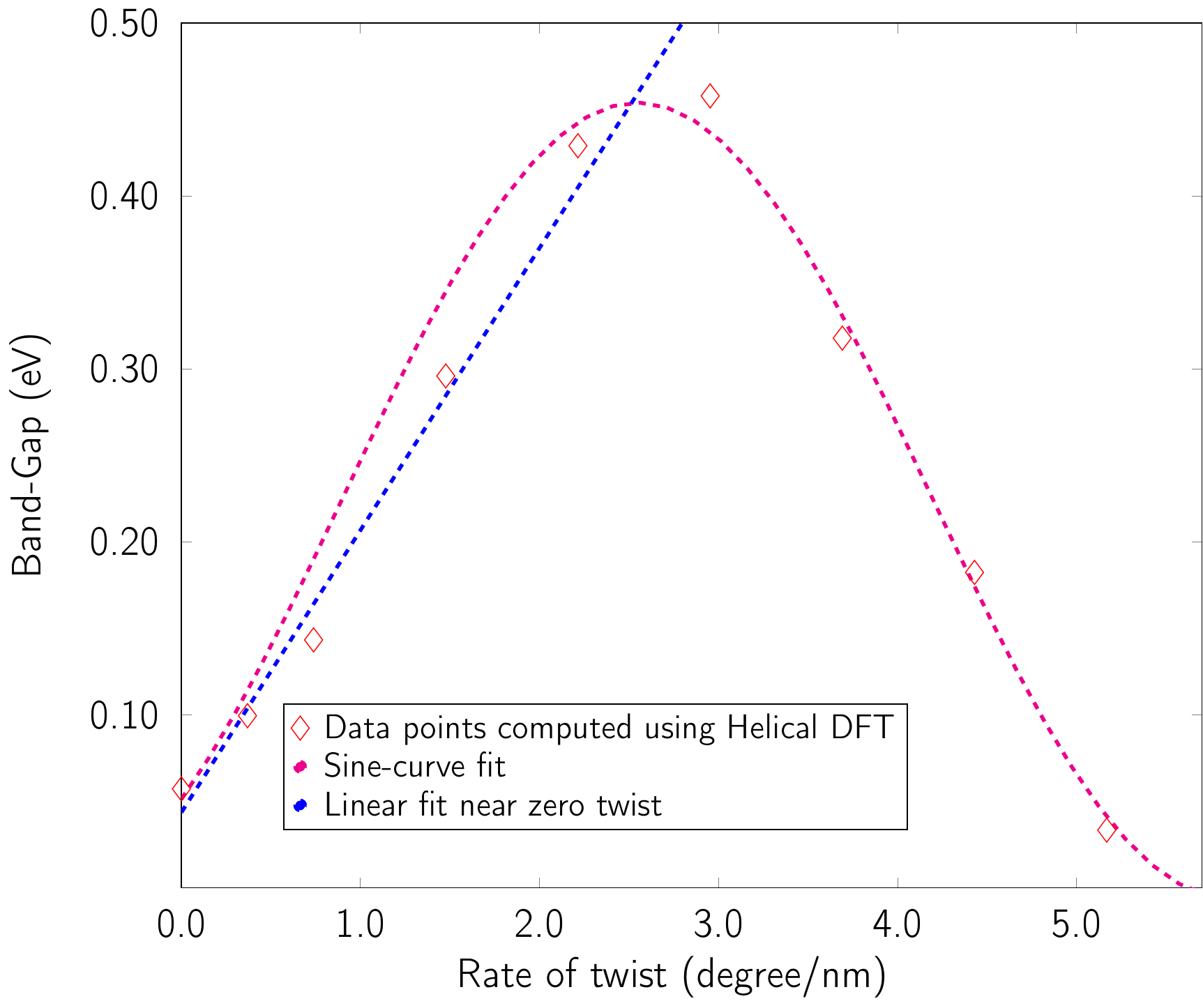}
}
\caption{Analysis of the variation of band-gap with applied twist, using an armchair carbon nanotube example (Radius = $1.07$ nm). The straight line fit near zero enables the evaluation of the tight-binding hopping parameter $t_0$, which comes out to be $2.897$ eV, in close agreement with \citep{Dumitrica_Tight_Binding1, correa2010tight, yang1999band}. The sine curve fit (in the non-linear response region) enables evaluation of the periodicity in the band gap variation and yields $\xi_{\text{period}}^{\text{fit}} =  0.1154$ rad/nm. The theoretical value from eq.~\ref{eq:period_BG} is $\xi_{\text{period}}^{\text{theory}} = 0.1217$ rad/nm, in close agreement.}
\label{fig:s2_t0_carbon}
\end{figure}

\begin{figure}[!ht]
\centering
\subfloat[Carbon nanotube, radius = $1.07$ nm]{\includegraphics[width=0.45\textwidth]{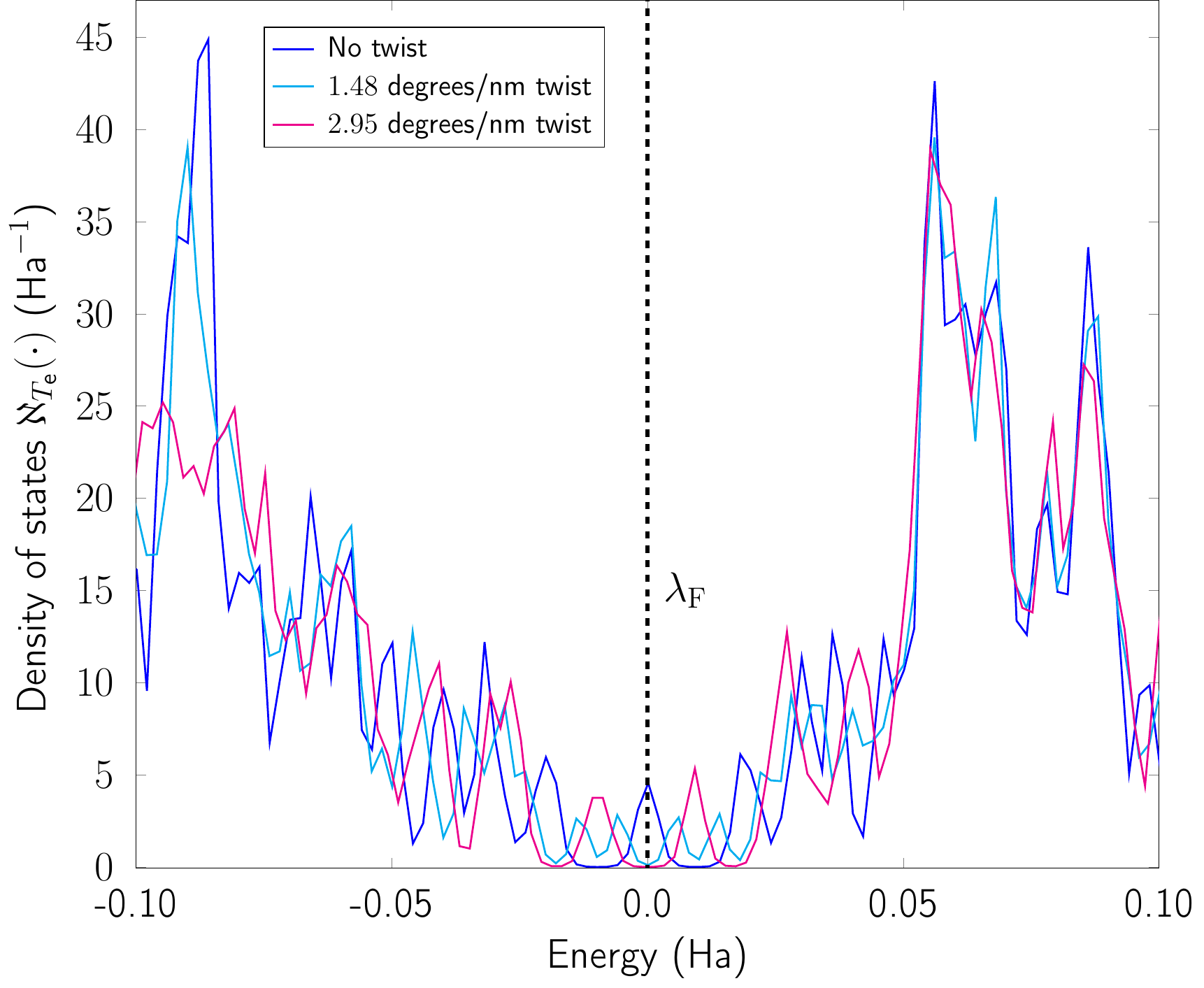}
}\quad
\subfloat[Silicon nanotube, radius = $2.96$ nm]{\includegraphics[width=0.45\textwidth]{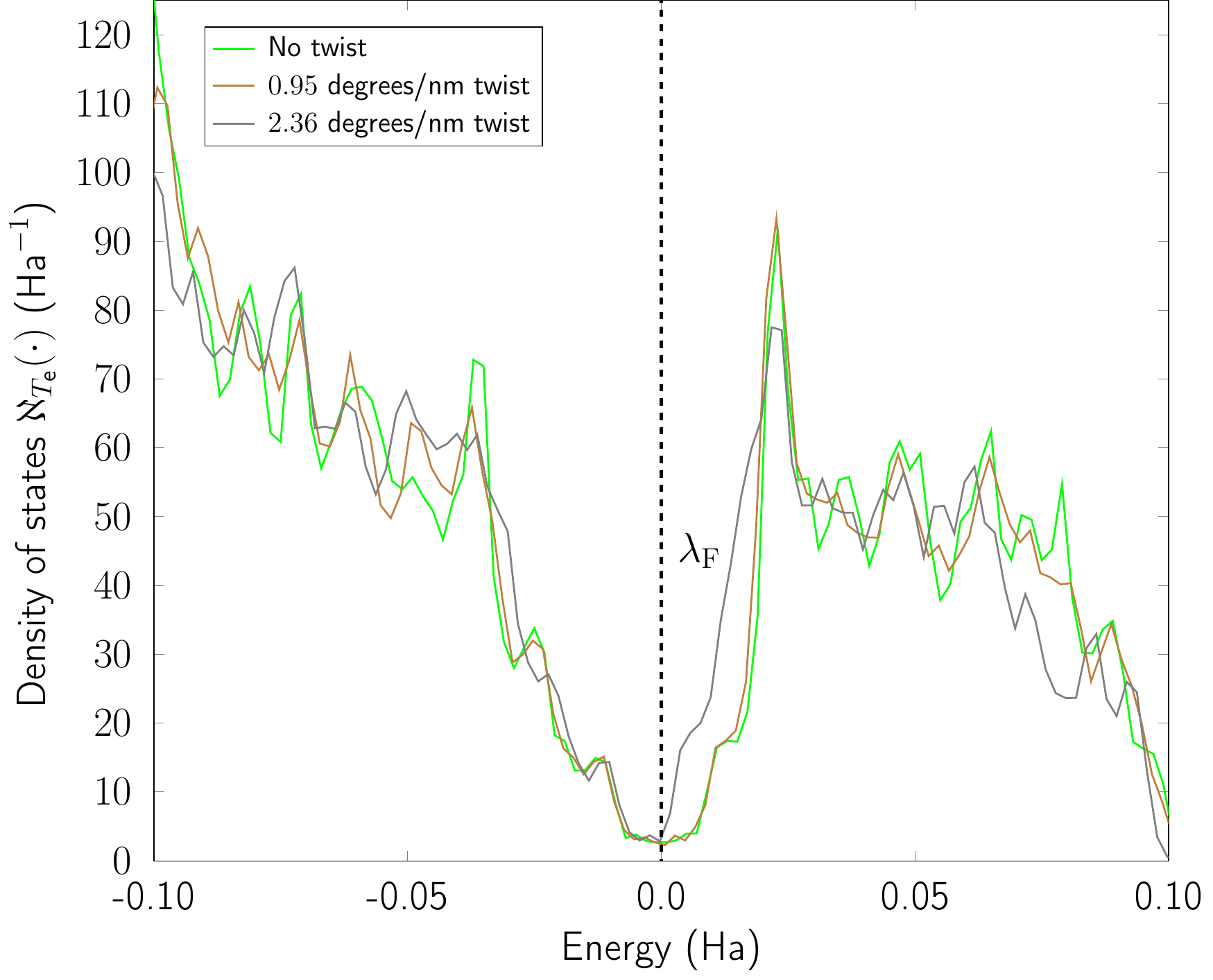}
}
\caption{{Variation in the electronic density of states near the Fermi level, for some armchair X nanotubes, when subjected to twist. The carbon nanotube undergoes a clear metal-to-semiconductor type transition upon twisting, as evidenced by the value of $\aleph_{T_{\text{e}}}(\cdot)$ falling to zero at the Fermi level. Other armchair Xene nanotubes (including the silicon nanotube shown here) do not show such stark variations, although changes in the electronic structure are clearly induced by the application of twist.}}
\label{fig:Density_of_States}
\end{figure}

Armed with the above tools, we study the variation in the bandgaps of nanotubes as they are subjected to twisting. For reasons explained later, we mainly concentrate on investigations related to armchair X nanotubes, although we also briefly comment on our findings related to zigzag X nanotubes subsequently. The behavior of carbon armchair nanotubes in particular, has received much attention in the literature \citep{yang2000electronic, ding2002analytical, yang1999band, Dumitrica_Tight_Binding1}, and serves as an important benchmark against which our results can be validated. Such nanotubes are known to be metallic \citep{yang1999band, Lieber_CNT, ding2002analytical} although in practical calculations, a vanishingly small bandgap at the location $\eta = \frac{1}{3}, \nu = 0$ (or equivalently, $\eta = -\frac{1}{3}, \nu = 0$) may be observed \citep{ghosh2019symmetry}. Upon twisting, armchair carbon nanotubes undergo a metal-to-semiconductor transition, with the characteristic feature that the bandgap-versus-rate-of-twist plot has a slope of  $3\,t_{0}\,R_{\text{tube}}$ in the linear regime (i.e., in the neighborhood of zero twist). Here $t_{0}$ is the tight-binding hopping parameter for carbon \citep{Dumitrica_Tight_Binding1}. Using armchair carbon nanotubes of radii $1.08$, $1.48$ and $1.88$ nm as examples, we used Helical DFT to compute the slope of the bandgap-versus-rate-of-twist plot in the linear regime and obtained values of $t_0$ between $2.6$ and $3.0$ eV (see Figure \ref{fig:s2_t0_carbon}). These agree well with the literature \citep{Dumitrica_Tight_Binding1, correa2010tight, yang1999band}, giving us confidence in the quality of our subsequent simulations. Upon twisting these nanotubes further, the band gap is known to further increase and then decrease, as the tube alternates between metallic and semiconducting states, and the period of oscillation (of the band gap versus rate of twist plot) is  theoretically known to be \citep{yang2000electronic, ding2002analytical, yang1999band, Dumitrica_Tight_Binding1}:
\begin{align}
\xi_{\text{period}}^{\text{theory}} = \displaystyle \frac{a}{R_{\text{tube}}^2}\,.
\label{eq:period_BG}
\end{align}
Here $a$ denotes the carbon-carbon bond length (see Table \ref{Table:LatticeParameters}). Using Helical DFT, we were able to compute the electronic density of states near the Fermi level and qualitatively verify the metal-to-semiconductor transitions in the armchair carbon nanotubes as they are twisted (see Figure \ref{fig:Density_of_States}(a)). To verify that Helical DFT also reproduces the quantitative aspects of the variation, we fit the band gap data from Helical DFT, to a general sine curve of the form:
\begin{align}
\text{band gap} = s_1 \sin \bigg(\frac{2\pi \alpha}{s_2} + \frac{2\pi}{s_3}\bigg) + s_4\,,
\label{eq:band_gap_sine_fit}
\end{align}
from which, the period of oscillation may be computed as:
\begin{align}
\xi_{\text{period}}^{\text{fit}} = \frac{2\pi s_2}{\tau} =  \frac{2\pi s_2}{\sqrt{3}a}\,.
\label{eq:gamma_s2}
\end{align}
We verified that $\xi_{\text{period}}^{\text{fit}}$ and $\xi_{\text{period}}^{\text{theory}}$ are in close agreement in all cases under study (see Figure \ref{fig:s2_t0_carbon} for a specific example). An alternate means of quantifying this agreement, following \citep{Dumitrica_Tight_Binding1}, is to equate $\xi_{\text{period}}^{\text{fit}}$ and $\xi_{\text{period}}^{\text{theory}}$, and estimate the bond length $a$, from this instead. In other words, by writing:
\begin{align}
s_2 =  \frac{\sqrt{3}\,a^2}{2\pi R_{\text{tube}}^2}\,,
\end{align}
or more generally, 
\begin{align}
s_2 = \sigma \times R_{\text{tube}}^{\mu}\,,
\label{eq:s2_scaling}
\end{align}
we may evaluate the exponent $\mu$ and the constant $\sigma$ from a plot of $s_2$ versus $R_{\text{tube}}$, and from this, we may further estimate the bond length as:
\begin{align}
a_{\text{fit}} = \frac{\sqrt{2 \pi \sigma}}{3^{\frac{1}{4}}}\,.
\label{eq:a_fit}
\end{align}
Using this procedure, {we arrived at} $\mu=-1.98$, and $a_{\text{fit}} = 1.37$ angstrom, both of which are very close to the expected values of $-2.00$ and $1.40$ angstrom, respectively. These results give us further confidence in the quantitative results obtained using Helical DFT.

\begin{figure}[!ht]
\centering
\subfloat[2D surface plot of the eigenvalues $\lambda_j(\eta,\nu)$, for $j=8$.]{\scalebox{0.55}
{
\includegraphics[trim={1cm 1cm 0.5cm 1cm}, clip, width=\textwidth]{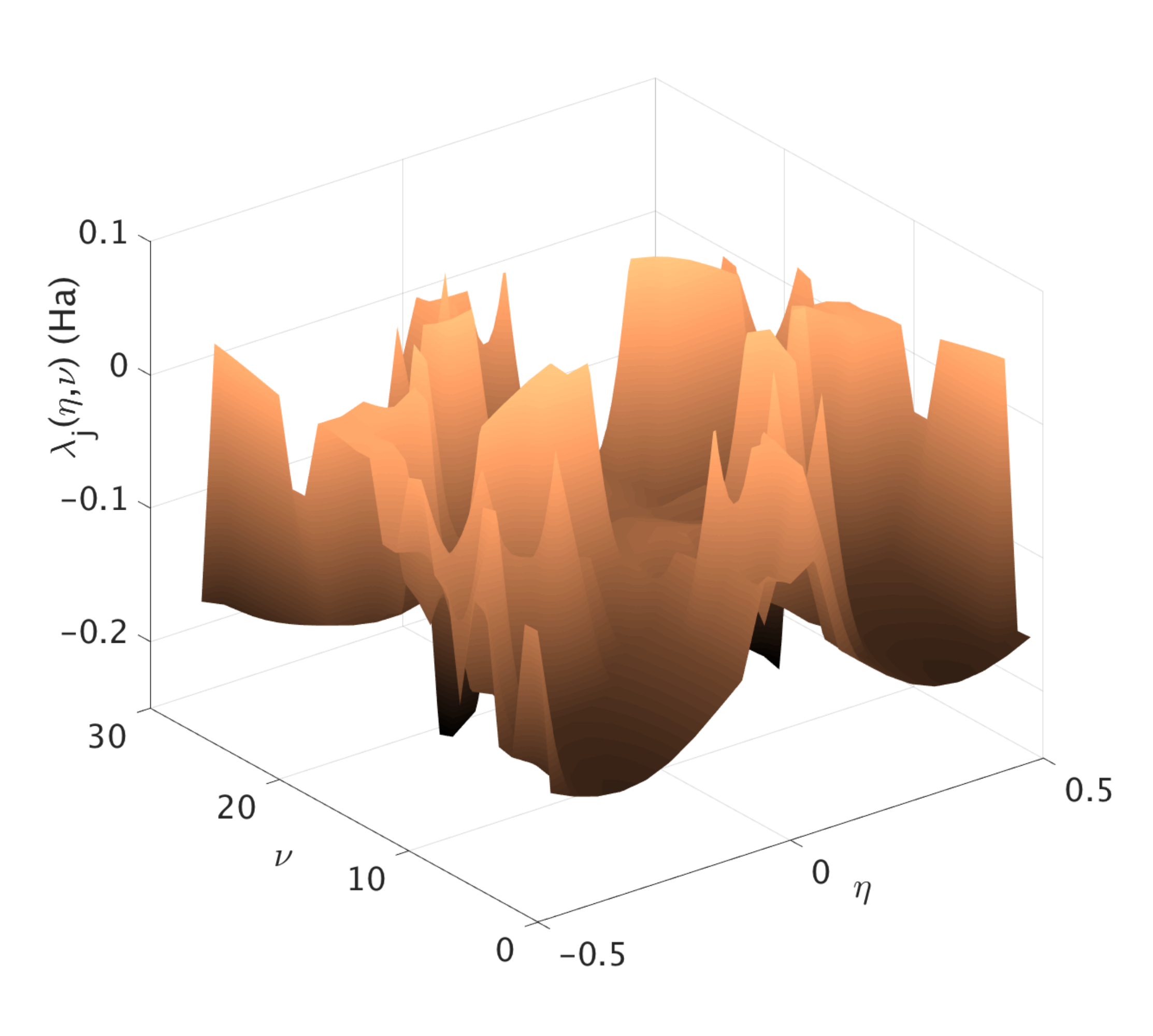}
}}\\
\subfloat[Symmetry adapted band diagram in $\eta$, along $\nu = 1$. Location of band gap  highlighted by blue rectangle.]{\includegraphics[width=0.45\textwidth]{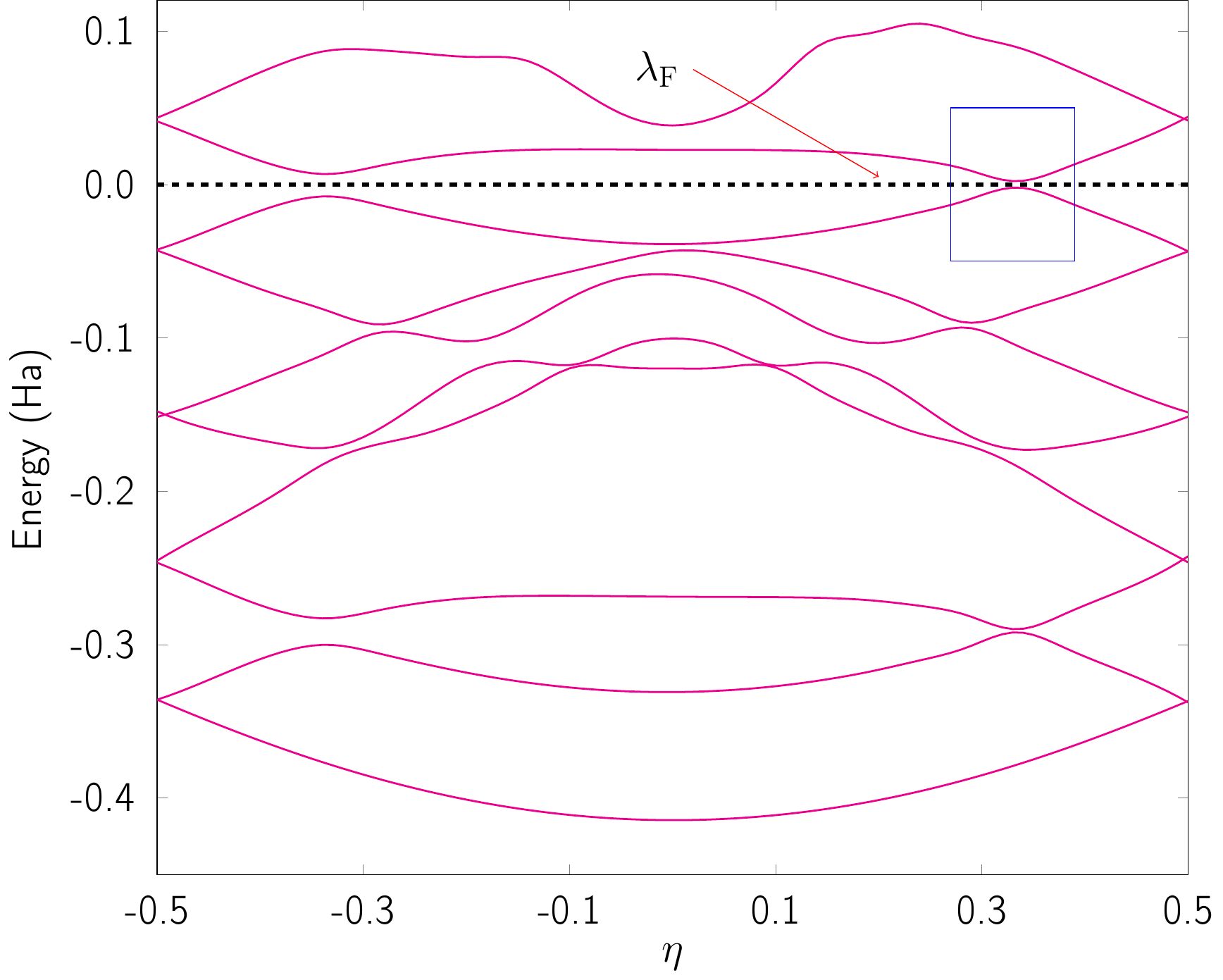}
}\quad
\subfloat[Zoomed in view of band gap and its location.]{\includegraphics[width=0.45\textwidth]{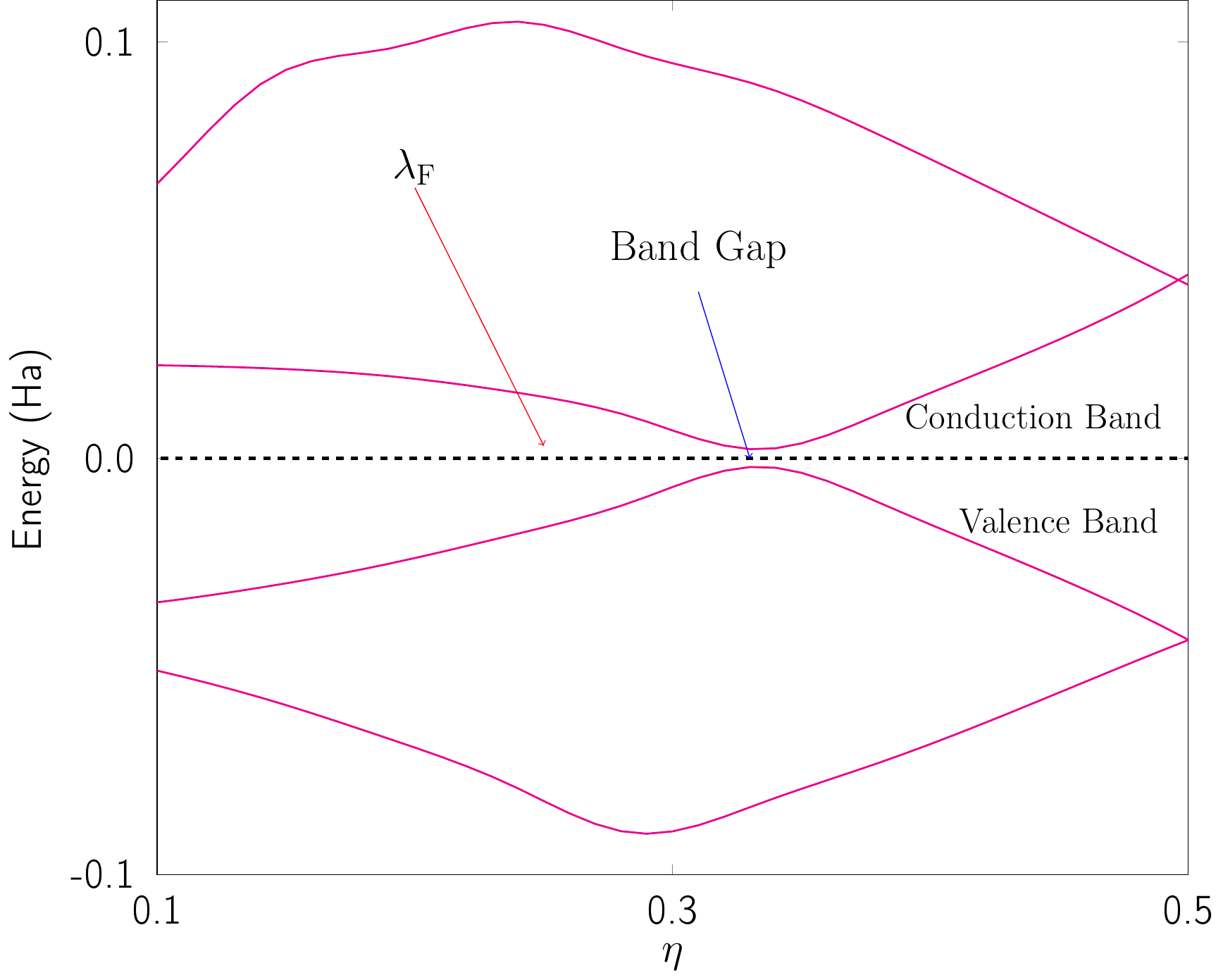}
}
\caption{Visualization of electronic states for twisted armchair Si nanotube (radius = $2.96$ nm), for $0.94$ degree per nanometer of applied twist. A small bandgap of about $0.11$ eV opens up in this case. Location of band gap ($\eta = 1/3, \nu = 1$) has been highlighted by blue rectangle in sub-figure (b) and a zoomed in view is available in sub-figure (c). The surface plot in sub-figure (a) also looks noticeably different from Figure \ref{fig:electronic_states_plots} (a).}
\label{fig:electronic_states_plots_again}
\end{figure}

\begin{figure}[!ht]
\centering
\subfloat[Silicon nanotubes]{\includegraphics[width=0.45\textwidth]{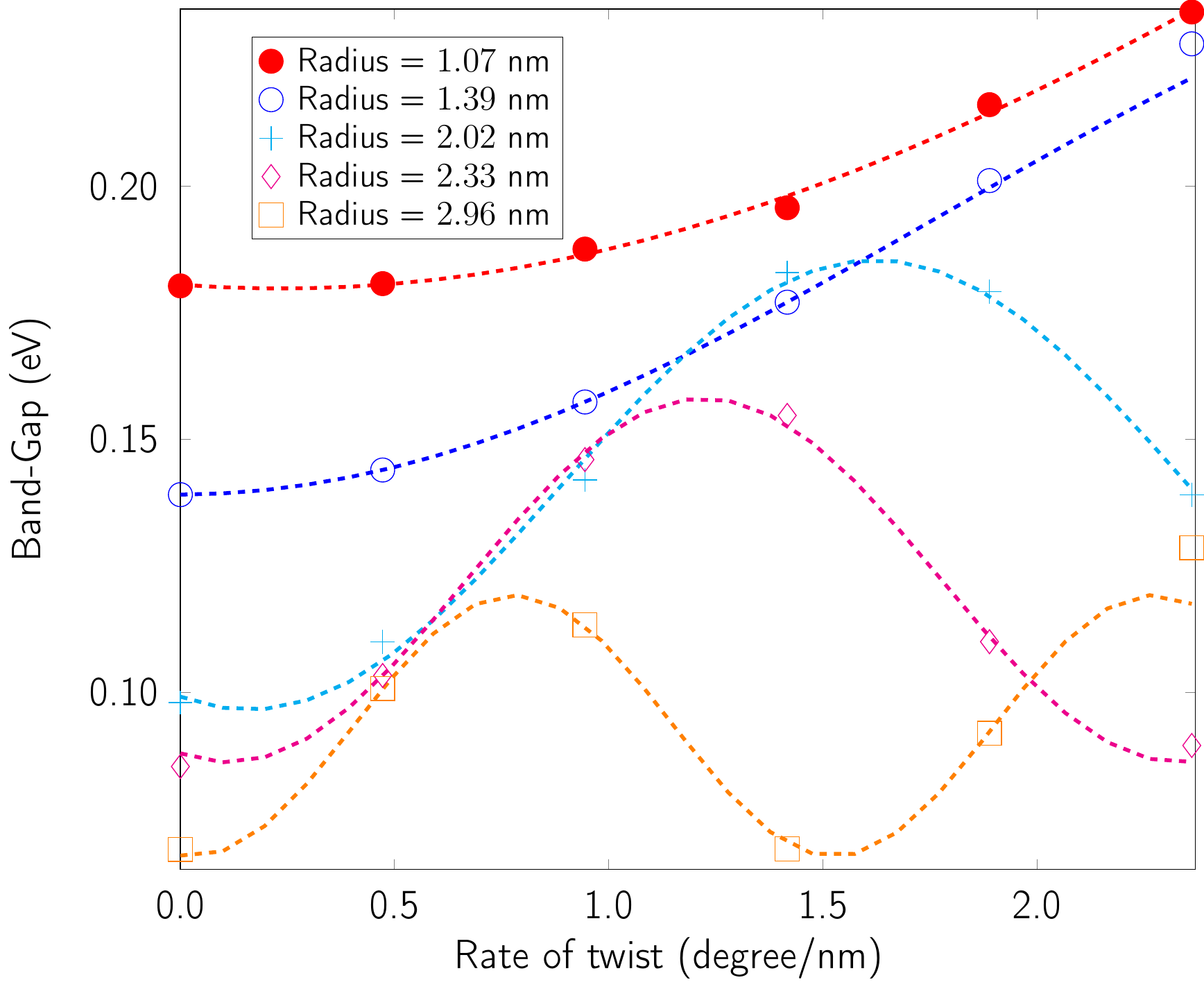}
}\quad
\subfloat[Germanium nanotubes]{\includegraphics[width=0.45\textwidth]{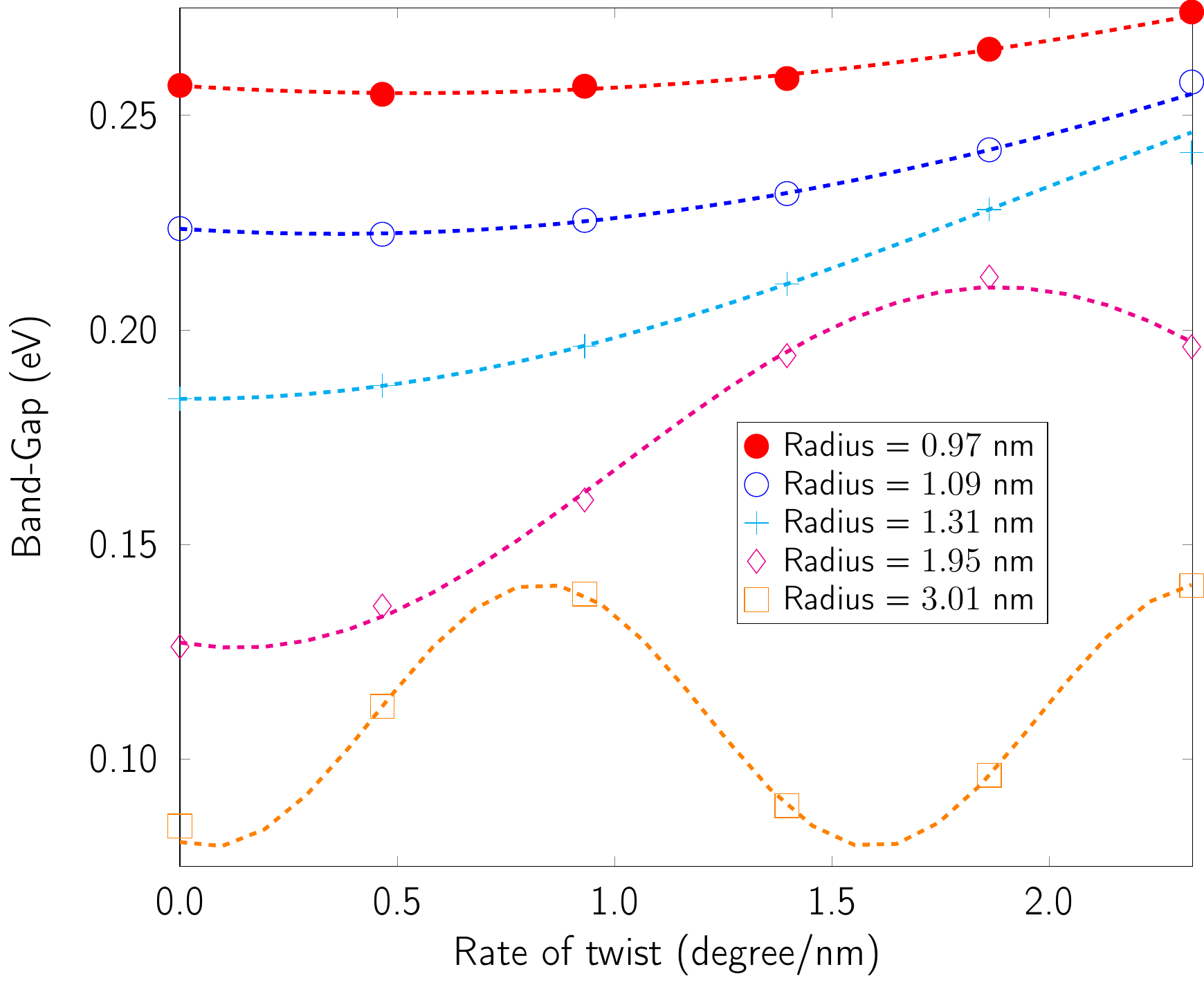}
}\\
\subfloat[Tin nanotubes]{\includegraphics[width=0.45\textwidth]{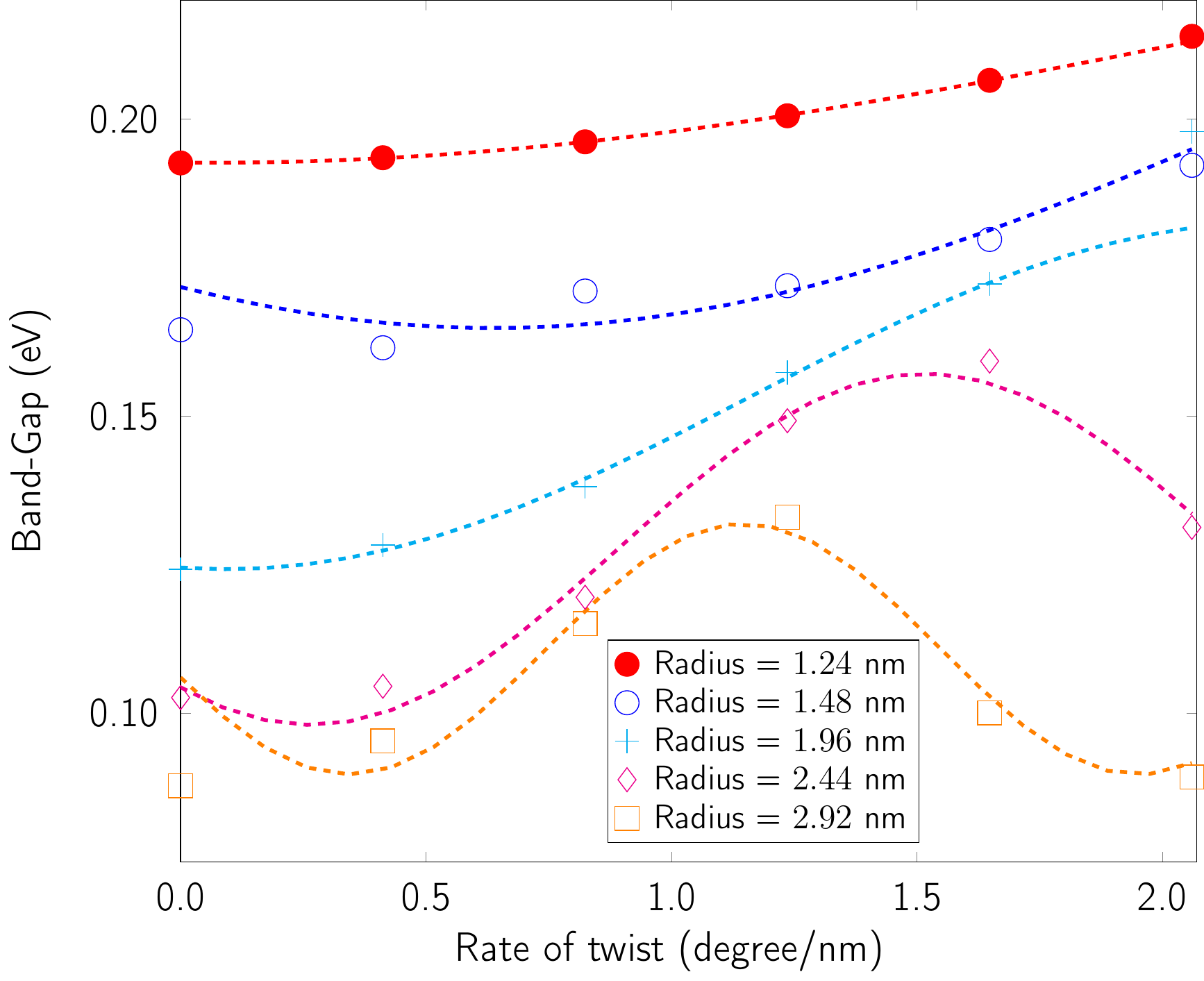}
}\quad
\subfloat[Investigation of scaling laws in periodicity of band gap variation]{\includegraphics[width=0.45\textwidth]{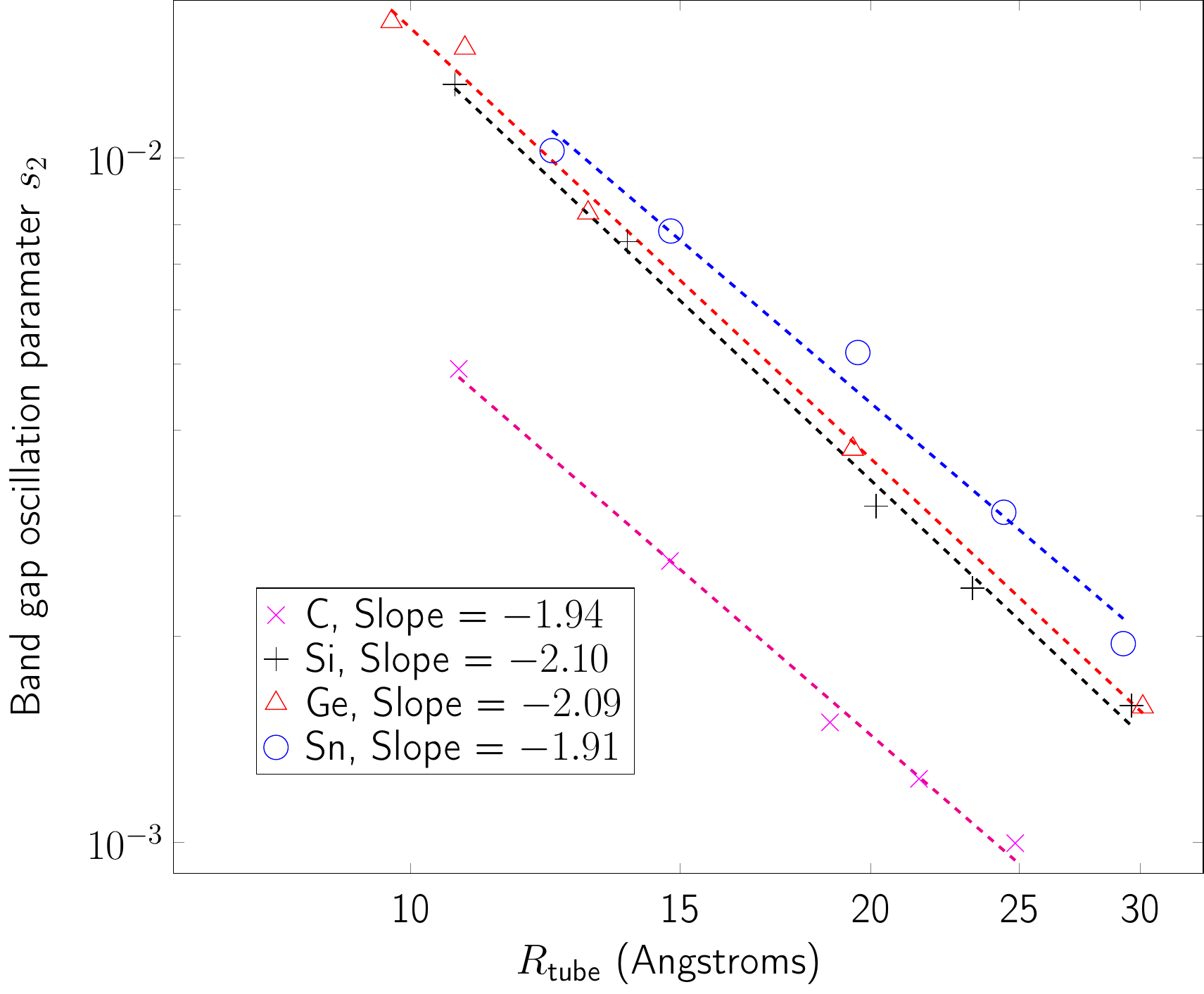}}
\caption{Variation of band gap with applied twist for armchair X (= Si, Ge, Sn) nanotubes. Sub-figures (a), (b) and (c) include data from Helical DFT, as well as sine curve fits (dotted lines) used to determine the band gap oscillation parameter $s_2$ (eq.~\ref{eq:band_gap_sine_fit}). Sub-figure (d) explores the variation of this parameter with the tube radius (eq.~\ref{eq:s2_scaling}). The slope of each of the straight line fits is close to $-2.00$, suggesting that the period of variation of the band gap scales in an inverse quadratic manner with the nanotube radius.}
\label{fig:s2_plots}
\end{figure}

Turning to the broader class of armchair group IV nanotubes (i.e., X = Si, Ge, Sn) we make the following observations using the data obtained from Helical DFT. In general, these nanotubes are semiconducting, with a direct band gap located at the same position as the armchair carbon nanotubes, i.e., $\eta = \frac{1}{3}, \nu = 0$ (or equivalently, $\eta = -\frac{1}{3}, \nu = 0$) for untwisted tubes. Upon twisting, these tubes also undergo periodic oscillations in their band gaps,\footnote{The location of the band gap initially continues to be the same as that of the untwisted tube, but then it transitions to small values in $\nu$ (i.e. $\nu = 1, 2,$ etc.), while remaining at the same location in $\eta$ (i.e., $\eta = \frac{1}{3}$). Thus, for relatively small twists, the band gap continues to be a direct one. Upon further application of twist however, the band gap becomes indirect and the eigenvalue just above the Fermi level is associated with a different value of $\nu$ as compared to the eigenvalue just below the Fermi level, although the value of $\eta$ associated with these eigenvalues continues to be $\frac{1}{3}$.} although the amplitudes of the oscillations are generally more muted than the case of armchair carbon nanotubes, and we did not observe metal-to-semiconductor type transitions for most tubes. For  tubes with larger radii however, the untwisted states can be associated with vanishingly small band gaps to begin with --- owing to the decay relations obeyed by the band gaps \citep{ghosh2019symmetry, wang2017band}, and these tubes are likely to be practically metallic at room temperature. Therefore, changes to the band gap upon application of twist can be more easily discerned (See Figures  \ref{fig:electronic_states_plots} and \ref{fig:electronic_states_plots_again} for an example involving an armchair silicon nanotube). To quantify the periodic changes in the band gaps, we obtained the period of oscillation in each case using the sine curve fitting procedure outlined above (eq.~\ref{eq:band_gap_sine_fit}), and computed the power law dependence of the period on the tube radius by means of eq.~\ref{eq:s2_scaling} (see Figure \ref{fig:s2_plots}). The values of $c$ and $\mu$ so obtained are shown in Table \ref{Table:s2_scaling}. 

The results are clearly suggestive of the fact that the period of variation of the band gap scales in an inverse quadratic manner with the tube radius for all armchair X nanotubes. We also observed that evaluation of eq.~\ref{eq:a_fit} using the values of $\sigma$ shown in Table \ref{Table:s2_scaling} leads to quantities that are fairly close to the values of $a$ shown in Table \ref{Table:LatticeParameters}, for each armchair X nanotube, suggesting that the theoretical relation in eq.~\ref{eq:period_BG} is generally valid for this entire class of nanotubes.
\begin{table}[ht]
\centering
\begin{tabular}{c  c  c}
\hline
Material & $\sigma$ ($\text{\AA}^{-\mu}$) & $\mu$  \\   
\hline
Carbon  & 0.52 &  -1.98 \\
Silicon  &  1.86 & -2.10  \\
Germanium & 1.91 & -2.09 \\
Tin  & 1.34 & -1.91 \\
\hline
\end{tabular}
\caption{Parameters for the scaling law $s_2 = \sigma \times R_{\text{tube}}^{\mu}$ for armchair X nanotubes. Here, $s_2$ is the bandgap oscillation parameter as defined in eq.~\ref{eq:band_gap_sine_fit}. The value of $\mu$ in each case is close to $-2.00$, suggesting that the period of variation of the band gap scales in an inverse quadratic manner with the nanotube radius for these tubes.}
\label{Table:s2_scaling}
\end{table} 
               
\begin{figure}[!ht]
\centering
\subfloat[Type II zigzag tin nanotubes]{\includegraphics[width=0.45\textwidth]{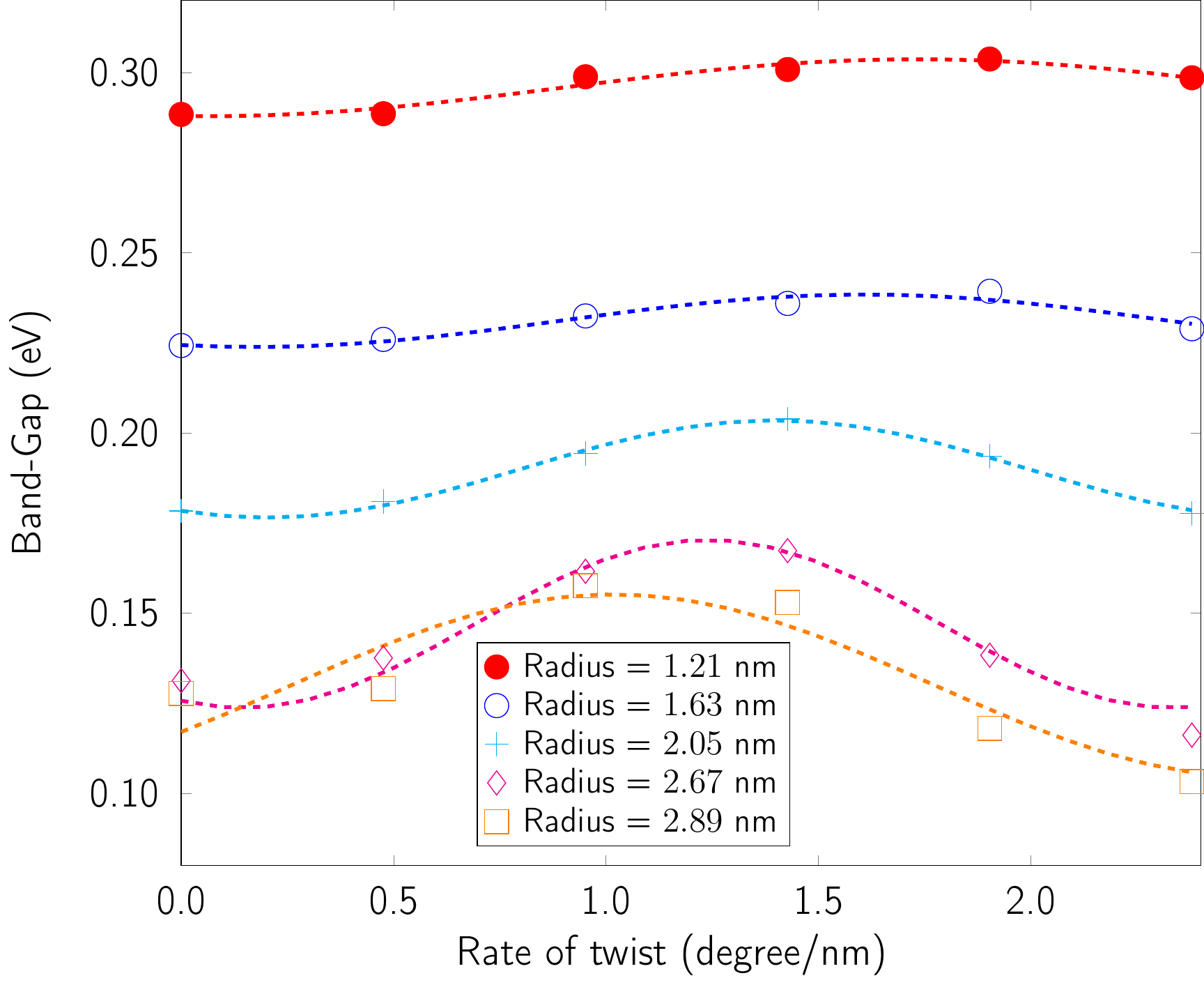}
}\quad
\subfloat[Type III zigzag carbon nanotubes]{\includegraphics[width=0.45\textwidth]{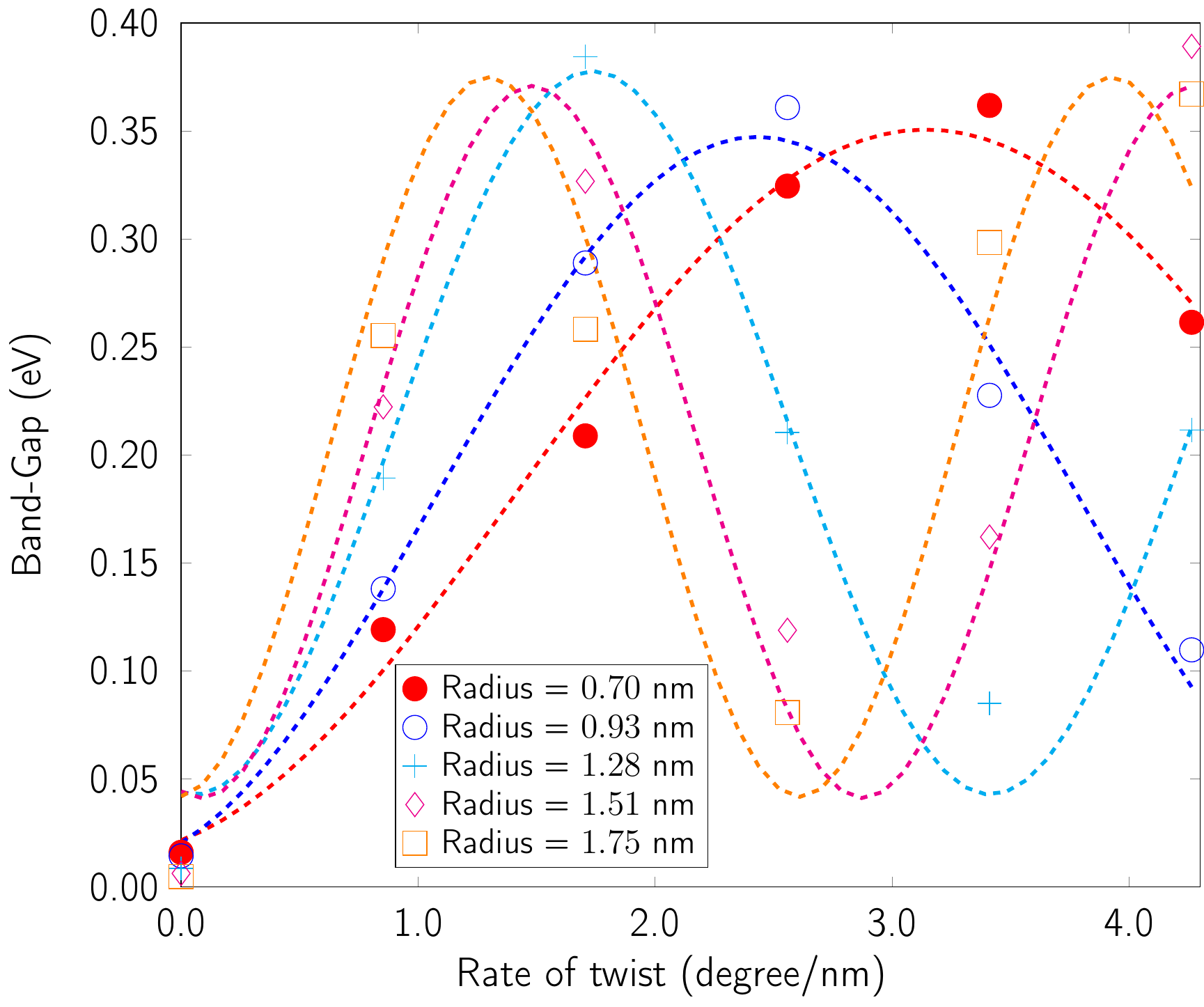}
}
\caption{Variation of band gap with applied twist for some zigzag nanotubes. Data from Helical DFT, as well as sine curve fits (dotted lines) are included. For most zigzag X nanotubes, particularly of Types I and II, the band gap changes little with twist. Sub-figure (a) shows examples of this using Type II tin nanotubes. In contrast, Type III zizag carbon nanotubes (sub-figure (b)) are metallic in the absence of twist and show more pronounced oscillatory changes between metallic and semiconducting states upon being twisted.}
\label{fig:zigzag_twist_data}
\end{figure}

Finally, we touch upon our investigations related to zigzag X nanotubes. These can be of different ``types'' \citep{ghosh2019symmetry, ding2002analytical}, i.e.,  Type I, II or III, depending on whether $\text{mod}(\mathfrak{N},3) = 1, 2$ or $0$. In general, zigzag X nanotubes, barring Type III carbon variants, are semiconducting \citep{ghosh2019symmetry, ouyang2002fundamental}, and 
the untwisted tubes have direct bandgaps located at the following values of $\eta$ and $\nu$ -- Type I carbon nanotubes: $\eta = 0, \nu = \frac{\mathfrak{N}-1}{3}$; other Type I nanotubes: $\eta = 0, \nu = \frac{\mathfrak{N}+2}{3}$; Type II nanotubes: $\eta = 0, \nu = \frac{\mathfrak{N}+1}{3}$; Type III nanotubes: $\eta = 0, \nu = \frac{\mathfrak{N}}{3}$. We  found that the band gaps of Type I and II zigzag X nanotubes tend to have a rather limited response to torsional deformations, consistent with earlier observations made regarding zigzag carbon nanotubes specifically \citep{yang2000electronic, ding2002analytical, yang1999band}. For most of these types of materials, the band gaps are non vanishing at zero twist for even relatively large radii tubes and the subsequent changes to their band gaps due to twisting are fairly small at the levels of torsional deformation we considered. This tends to cause issues in discriminating between actual changes to the band gaps due to deformation, and the numerical noise associated with the simulations. Therefore, although we did observe oscillatory patterns in the band gap versus rate of twist plots (see Figure \ref{fig:zigzag_twist_data} for an example) we found it difficult to extract scaling laws from this data unambiguously. Out of all the different zigzag X nanotubes however, the Type III variants of carbon are metallic, especially at larger radii (i.e., when curvature effects are minimal) \citep{ghosh2019symmetry, ouyang2001energy}, and we observed such tubes to be quite sensitive to torsional deformations. Similar to the case of armchair nanotubes, we observed these tubes to show oscillatory behavior between metallic and semiconducting states (see Figure \ref{fig:zigzag_twist_data}), and an analysis of the period of variation of the band gap (using eq.~\ref{eq:band_gap_sine_fit} and \ref{eq:s2_scaling}) yielded $\mu = -1.98$, thus suggesting an inverse quadratic dependence on the radius. A thorough re-investigation of scaling laws in the electronic response of zigzag X nanotubes, by making use of more accurate numerical techniques (based on spectral methods \citep{My_Shivang_HelicES_paper, agarwal2021spectral}, for instance) remains the scope of future work.
\section{Conclusions}
\label{sec:conclusions}
In summary, we have presented a computational technique that allows systems associated twisted geometries to be simulated efficiently and accurately from first principles. We have formulated the symmetry adapted governing equations, laid out numerical implementation strategies and detailed various aspects of our implementation. Our technique uses a higher order finite difference discretization scheme based on helical coordinates, employs ab initio pseudoptentials and can be used to simulate quasi-one-dimensional systems, as well as their deformations, conveniently and without needing major computational resources. As an application of our method, we have systematically studied the behavior of single wall zigzag and armchair group-IV nanotubes in the range of (approximately) 1 to 3 nm radius, as they undergo twisting. Through an extensive series of simulations, we have demonstrated how certain mechanical properties of these nanotubes can be extracted from first principles using our technique, and we have also elucidated different aspects of the variation in the electronic properties of these materials as they undergo torsional deformation. In particular, using our simulations, we have been able to extend some well-known features of the electro-mechanical properties of carbon nanotubes to the broader class of Group IV nanotubes.

As a follow up of this work, we aim to employ the computational technique discussed here for the study of other nanotube materials, including multi-wall elemental nanotubes, and those made from transition metal dichalcogenides. An efficient C/C++ implementation of the computational method which makes use of domain decomposition and band parallelization (in addition to the currently implemented parallelization in $\eta$ and $\nu$), to improve scaling and computational wall time performance is the scope of ongoing and future work. Concurrently, the development of an efficient spectral scheme \citep{My_Shivang_HelicES_paper, agarwal2021spectral} in the spirit of \citep{Banerjee2015spectral},  which overcomes some of the inherent limitations of the current finite difference technique is also an area of active investigation. Finally, a long term goal associated with applications of the current computational method involves the design and discovery of exotic materials phases which show strong coupling between mechanical deformations (such as twist and extension/compression) and other electronic/optical/magnetic/transport properties.
\section*{Acknowledgement}
ASB acknowledges startup support from the Samueli School Of Engineering at UCLA, as well as funding from UCLA's Council on Research (COR) Faculty Research Grant. ASB would like to thank Neha Bairoliya (Univ.~of Southern California) for providing encouragement and support during the preparation of this manuscript. HMY would like to thank Elliona Li for her help in preparing some of the figures in this work. ASB would like to thank Vikram Gavini (University of Michigan) and Swarnava Ghosh (Oak Ridge National Lab) for insightful discussions during the early stages of preparation of the manuscript. ASB and HMY would like to thank UCLA's Institute for Digital Research and Education (IDRE) for making available some of the computing resources used in this work.
%\section*{\refname}
\newpage
\bibliographystyle{elsarticle-num}
\begin{center}
---
\end{center}
\bibliography{main}
\end{document}